\documentclass[12pt]{iopart}

\usepackage{graphicx}

\usepackage[utf8]{inputenc}
\usepackage{siunitx}
\usepackage{caption}
\usepackage[numbers,square,sort]{natbib}

\begin{document}

\title[All the Fun of the FAIR]{All the Fun of the FAIR: Fundamental physics at the Facility for Antiproton and Ion Research}

\author{M.~Durante$^1$, P.~Indelicato$^2$,
B.~Jonson$^3$, V.~Koch$^4$, K.~Langanke$^{5}$,
Ulf-G.~Mei{\ss}ner$^6$, E.~Nappi$^7$, T.~Nilsson$^3$,
Th.~St\"ohlker$^8$, E.~Widmann$^9$, M.~Wiescher$^{10}$ }

\address{
$^1$ Trento Institute for Fundamental Physics and Applications (TIFPA), National Institute of Nuclear Physics (INFN),
Via Sommarive 14, 38123 Povo (TN), Italy; and Department of Physics, University Federico II, Monte S. Angelo, Via Cintia, Naples, Italy.\\
$^2$ Laboratoire Kastler Brossel, UPMC-Sorbonne Université, CNRS, ENS-PSL Research University, Collège de France, Case\ 74;\ 4, place Jussieu, F-75005 Paris, France\\
$^3$ Institutionen f\"{o}r Fysik, Chalmers Tekniska H\"ogskola, S-412 96 G\"oteborg, Sweden\\

$^4$ Nuclear Science Division, Lawrence Berkeley National Laboratory, 1 Cyclotron Road,
Berkeley, CA 94720, USA\\	
$^5$ GSI Helmholtzzentrum f\"ur Schwerionenforschung, Planckstr. 1,
D-64291 Darmstadt, Germany \\
Institut f\"ur Kernphysik, Schlossgartenstr. 2, D-64289 Darmstadt, Germany\\
$^6$ Helmholtz-Institut f\"ur Strahlen- und Kernphysik und Bethe Center for Theoretical Physics, Universit\"at Bonn,
D-53115 Bonn, Germany\\
Institute for Advanced Simulation, Institut f\"ur Kernphysik and J\"ulich Center for Hadronphysics, D-52425 J\"ulich, Germany\\
$^7$ INFN, Istituto Nazionale di Fisica Nucleare, Sez. Bari, Bari, Italy\\
$^8$ GSI Helmholtzzentrum f\"ur Schwerionenforschung, Planckstr.1, D-64291 Darmstadt, Germany\\
Helmholtz Institute Jena, D-07743 Jena, Germany\\
Friedrich-Schiller University Jena, Institute for Optics and Quantum Electronics, D-07743 Jena, Darmstadt\\
$^9$ Stefan Meyer Institute for subatomic Physics, Austrian Academy of Sciences, 1090 Vienna, Austria\\
$^{10}$ Department of Physics \& Joint Institute of Nuclear Astrophysics, University of Notre Dame, Notre Dame, Indiana 46556, USA
}
\ead{k.langanke@gsi.de}

\newpage

\begin{abstract}
	The Facility for Antiproton and Ion Research (FAIR) will be the
	accelerator-based
	flagship research facility in many basic sciences and their applications        in Europe for the coming decades.
        FAIR will open up unprecedented research opportunities in hadron
        and nuclear physics, in atomic physics and nuclear astrophysics
	as well as in applied sciences like materials research, plasma
	physics and radiation biophysics with applications towards novel
	medical treatments and space science.
	FAIR is currently under construction as an international facility
	at the campus of the GSI Helmholtzzentrum for Heavy-Ion Research
	in Darmstadt, Germany. While the full science potential of FAIR
	can only be harvested once the new suite of accelerators and
	storage rings is completed and operational, some of the
	experimental detectors and instrumentation are already available
	and will be used starting in summer 2018 in
	a dedicated research program at GSI, exploiting
	also the significantly upgraded GSI accelerator chain.

	The current manuscript summarizes how FAIR will advance our
	knowledge in various research fields ranging from a deeper
	understanding of the fundamental interactions and symmetries in
	Nature to a better understanding of the evolution of the
	Universe and the objects within.

\end{abstract}
	
	\maketitle

\section{The scientific motivation of FAIR}

Our world is governed by four fundamental interactions. The strong force
operates among quark and gluons at very short distances and is responsible for
the composition and dynamics of atomic nuclei. The electromagnetic force,
of inifinite range, governs atomic physics and chemistry, and hence
most of the phenomena of everyday life. The weak interaction, operating
like the strong force on very short distances, but with much smaller strength,
is the only force which can change electromagnetic charges and hence
is responsible for radioactivity observed in nuclear beta decay. Despite
being the smallest in strength of all interactions by far, gravity
nevertheless dominates the large scale structure of
our Universe.

Modern physics describes all four interactions by well-established
theories. Three of these (quantum chromodynamics (QCD) for the
strong force, and electroweak theory for the unified  electromagnetic and
weak forces) are based on the concept of quantum gauge theory, while
gravity is described in the framework of general relativity. The formulation
of a quantum theory of gravity and the unification of all forces within
one theoretical framework is one of the ultimate goals of physics.
Another driver of today's science is, on the one hand, to explore and understand
the plethora of structures and phenomena incorporated and predicted
by the fundamental theories, and, on the other hand, to test
their limitations and applicability
looking for potentially new physics beyond them.
Like no other place, the Facility for Antiproton and Ion Research (FAIR)~\cite{FAIR}
will contribute, directly
and indirectly, to this quest.

FAIR is the next-generation accelerator facility for fundamental and applied
research providing a worldwide unique variety of ion and antiproton
beams for a large variety of experiments.
FAIR is an international facility with 10 partner countries
currently under construction at the GSI Helmholtzzentrum
f\"ur Schwerionenforschung in Darmstadt, Germany. More than 2500 scientists
and engineers from more than 50 countries on 5 continents are involved in the
preparation and definition of the research at FAIR which is expected
to start full operation in 2025.

\begin{figure}[t]
\includegraphics[width=150mm, angle=0]{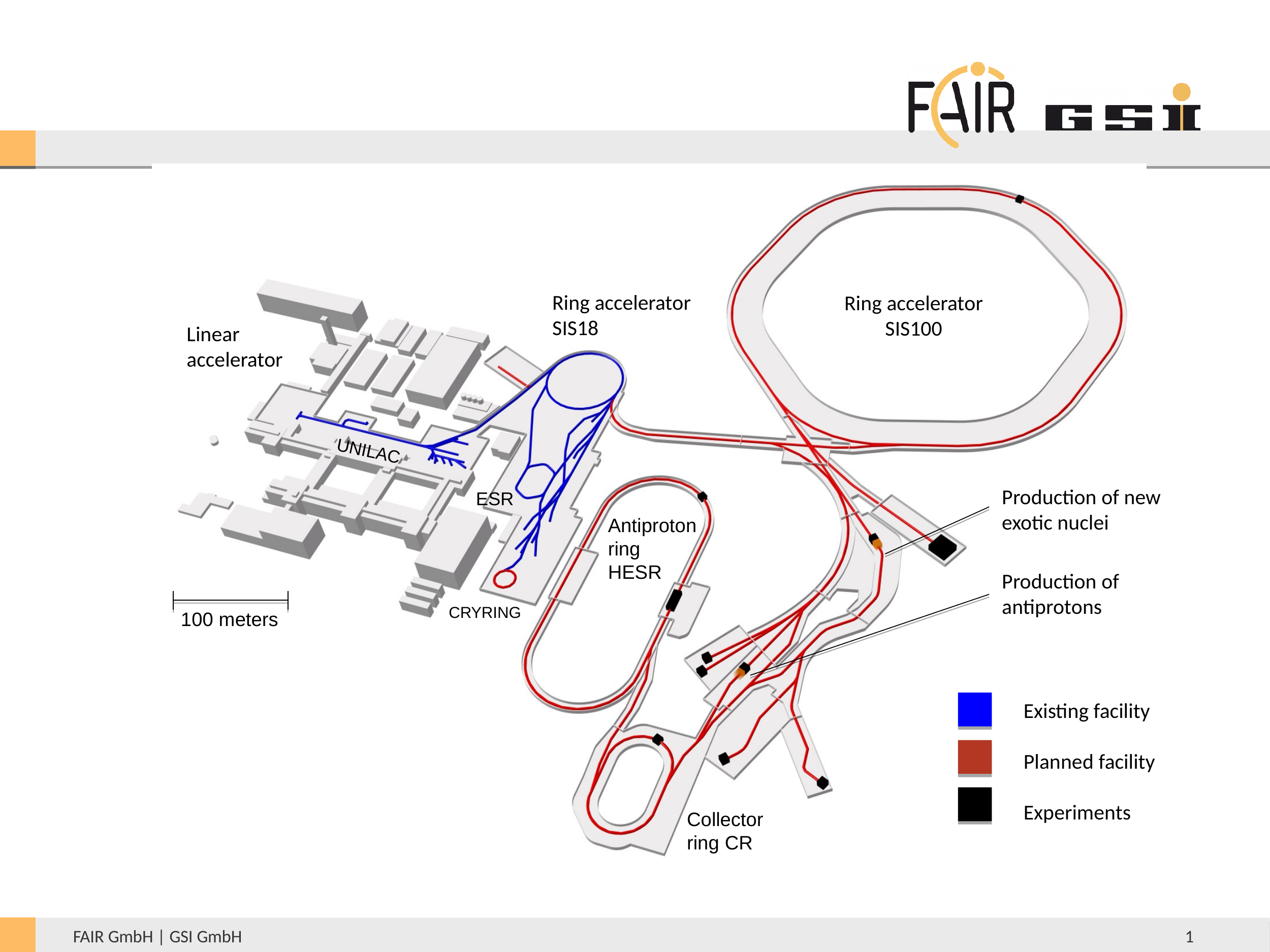}
	\caption{\label{fig:layout} The FAIR accelerator complex}.
\end{figure}

The heart of FAIR
is the superconducting, fast-ramping heavy-ion synchroton SIS100. This
high-intensity machine is supplemented by a proton linear accelerator
used for the production of antiprotons, a variety of rings for stored
cooled ions and antiprotons, and the Superconducting Fragment Separator (SFRS)
for the generation and clean identification of secondary beams of short-lived
ions.
The FAIR accelerator complex is unique by offering beams of all ion species
and antiprotons at high energies with unprecedented high intensities
and quality (i.e. with very precise energy and profile)
which are available at
several experimental areas with a suite of novel instrumentation
and detectors for fore-front research in hadron, nuclear, atomic and plasma
physics, and applications in material sciences and bio- and radiation physics.
The layout of FAIR is shown in Fig. \ref{fig:layout}.

The main thrust of FAIR research focuses on the structure and evolution
of matter on both the microscopic and the cosmic scale, bringing the
Universe in the Laboratory.
In particular, FAIR will expand our knowledge in various scientific
fields beyond current frontiers. To this end, researchers worldwide have organized themselves
in four international collaborations (PANDA, CBM, NUSTAR and APPA) addressing, respectively:

\begin{itemize}

\item
the investigation of the properties and the role of the strong force
in shaping and building  the visible world around us and its
role in the evolution of the universe;
\item
the phase diagram of quantum chromodynamics, in particular
at the high densities as they are encountered in neutron stars;
\item
tests of symmetries and predictions of the standard model of
		electro-weak theory, in particular in the range of
		strong-field quantum electrodynamics, and special relativity;
	\item
the properties of nuclei far off stability, in particular studying their role
for the origin of the elements in the universe;
\item
applications of high-quality, high-intensity beams in research areas
that provide the basis for, or address directly, issues of
applied sciences and technology.
\end{itemize}

In this manuscript we will discuss the outstanding research opportunities
at FAIR. In our layout, we follow the 2017 NuPECC long range plan, which
lists FAIR as a top priority in all 6 chapters of the report
\cite{NuPECC2017}. Chapter 2 deals with the quest to explore hadron structure and dynamics
exploiting proton-antiproton annihilation. Chapter 3 discusses how ultrarelativistic
heavy-ion collisions allow to study the
phase diagram of nuclear matter and the nuclear Equation of State. In chapter 4 we discuss
how FAIR enlarges the reach to exotic nuclei far-off stability. Chapter 5 is devoted to FAIR's 
unrivaled contributions to nuclear astrophysics, in particular to the origin of the elements
in the Universe and to the associated astrophysical objects which make them. In chapters 6-9
we summarize the many possibilities FAIR offers in atomic physics exploiting highly-charged ions,
to test matter-antimatter symmetries with low-energy antiprotons, and in plasma physics and biophysics
using heavy-ion beams.

\section{FAIR as a QCD machine}

\subsection{A short introduction to Quantum Chromodynamics}

The strong interactions, described by Quantum Chromodynamics (QCD)~\cite{Fritzsch:1973pi},
constitute the last frontier of the so tremendously successful Standard Model of particle physics.
It unifies the strong, the electromagnetic and the weak forces. All interactions are described by gauge field theories,
with spin-1 gauge bosons supplying the interactions. For the strong interactions, these are the gluons, that not only
couple to the matter fields, the quarks, but also to themselves. Furthermore, these gluonic self-interactions lead
to the remarkable running of the strong coupling constant with two very distinct properties. At large momentum
transfer or energy, the coupling constant is small and allows for precision tests of the theory based on
perturbation theory in deep inelastic scattering and related processes~\cite{Gross:1973id,Politzer:1973fx}. At small
momentum transfer, that is for energies or momenta of the order of 1~GeV or so, the coupling becomes so large that
perturbation theory can not be applied any more. Even more remarkable, the fundamental fields of QCD have never been
observed in isolation, in fact, they are {\it confined} within hadrons and nuclei. This is the  feature that
distinguishes QCD from its electroweak siblings and constitutes a major challenge to theory and experiment as
will be discussed below.

Let us pause and review briefly a few properties of QCD that will be of relevance to the following discussion. First, while
there are six quark flavors, these matter fields fall into two distinct categories. On the one side, there are
the {\em light} quarks up, down and strange and on the other side the {\em heavy} quarks, with charm, bottom and top flavor.
Here, light and heavy refer to the scale of dimensional transmutation, $\Lambda_{\rm QCD} \simeq 250\,$MeV. QCD
also features  a number of (broken) symmetries. In the light quark sector, the chiral symmetry is spontaneously
broken with the appearance of eight pseudo-Goldstone bosons, that play an important role in hadron structure and
nuclear dynamics. In the heavy quark sector, spin and flavor symmetries for the charm and bottom quarks
are intertwining the properties of seemingly unrelated bound states and interactions. We note that the top quark
decays too quickly so that it can not form strong interaction composites. In addition, there are a few anomalies
that feature prominently in the QCD dynamics but need not concern us for the following discussions.

\subsection{The mysteries of the QCD spectrum}

As stated above, QCD forms complex, colorless bound states in forms of hadrons and nuclei. Let us consider hadrons
first. For a long time and even before QCD was formulated, the quark model served as an efficient tool to bring
order into the so-called particle zoo (the large number of observed hadrons). In fact, most of the states found
experimentally can be categorized as belonging to the two simplest forms of colorless bound states made of constituent quarks,
namely mesons composed of a quark and an antiquark and baryons as three quark composites. But it was already
pointed out by Gell-Mann~\cite{GellMann:1964nj} that more complex systems  are also allowed by the
symmetries, such as tetra- or pentaquarks (compact bound states of two quarks and two antiquarks or of four quarks and one antiquark, respectively). With the formulation of QCD, other so-called exotic forms of matter
became possible, such as glueballs (bound states made of gluons only) or hybrids (bound states of quarks and gluons).
These type of states fall into two distinct categories. First, they can have quantum numbers also allowed for
quark model states. In this case, one expects sizable mixing effects, as  e.g. for the lightest calculated scalar
glueball  and the observed scalar mesons in the 1.5~GeV
mass range, and therefore an unambiguous identification will be extremely difficult. Second, they can carry quantum
numbers not allowed by the quark model, which offers the best possibility to detect this intricate type of matter.
In fact, these types of states for a long time defied any experimental effort to establish them. However,
there have always been a number of states in the meson and baryon spectrum which did not fit into the quark model,
notably the $\Lambda(1405)$ in the strangeness $-1$ sector of the baryon spectrum, predicted by Dalitz and Tuan
in 1959 \cite{Dalitz:1959dn} from coupled channel dynamics, or the light scalar mesons ($\sigma$, $\kappa$,
$f_0(980), \ldots$). It also became clear that the constituent quarks underlying the quark model are not the
basic blocks of QCD, these are the so-called current quarks. In fact, the constituent quarks in some way model the
complicated environment in which the current quarks reside and can be considered as effective degrees of freedom.

\begin{figure}[ht!]
	\begin{center}
		\includegraphics[width=0.55\textwidth]{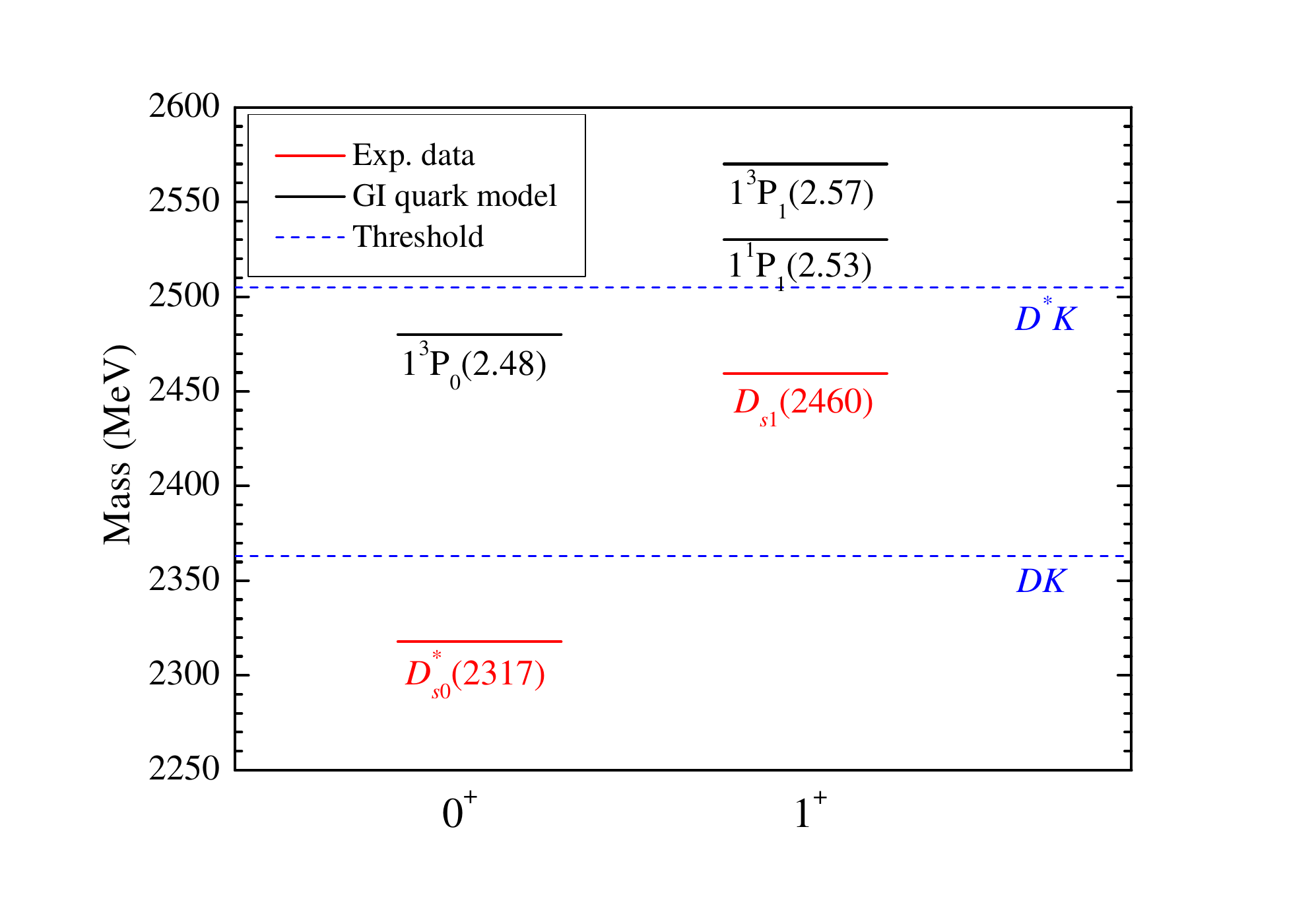}
        \vspace{-3mm}
	\caption{Spectrum of the mysterious charm-strange meson for given spin and parity (as indicated on the x-axes) compared
    to the otherwise successful quark model \cite{Godfrey:1985xj}. The pertinent $DK$      and $D^\star K$ thresholds are also shown.}
        \label{fig:Ds}
	\end{center}
\vspace{-5mm}
\end{figure}

Then, in 2003 with the first experimental findings of the positive parity charm-strange
mesons, the $D_{s0}^*(2317)$ at Babar \cite{Aubert:2003fg} and the  $D_{s1}(2460)$~\cite{Besson:2003cp} at CLEO, see Fig.~\ref{fig:Ds}
as well as the charmonium-like state produced in the exclusive decay
process $B^+ \to K^+ \pi^+\pi^- J/\psi$, the $X(3872)$, at Belle  \cite{Choi:2003ue},
a completely new area of hadron spectroscopy began, which can be dubbed
the ``XYZ-area''. A race at all colliders world-wide began, and many new states that did not fit at all in
the patterns predicted by the quark model were established, in particular also charged quarkonium states that
require at least two quarks and two antiquarks in their minimal flavor composition. Correspondingly, a flurry of theoretical
and lattice QCD studies tried to reveal the nature of these states and made predictions, some of which will be discussed
later. In addition, LHCb found indications for two pentaquark-type states with hidden charm~\cite{Aaij:2015tga},
which resurrected interest in exotic baryons after the very sad story of the light-flavored pentaquark $\Theta^+$, that
had unfolded in the beginning of the millenium~\cite{Hicks:2012zz}.
It should be noted that with the exception of the BESIII experiment
at the Beijing-Electron-Positron-Collider, with CLEO-c at Cornell  and COMPASS at CERN, most of the
experiments that contributed to this  spectroscopy revolution  (Babar at SLAC, Belle/BelleII at KEK, CDF and D0 at FNAL,
and LHCb, Atlas, Alice and CMS at CERN)  were {\it not} built for this purpose. This is also the reason why PANDA at FAIR can make
decisive contributions, as one of the main goals of PANDA is the spectroscopy in the charmonium region, as
will be discussed in more detail below.
While there is still no general consensus on the theoretical side about the nature of most of these states,
as a matter of fact a large number of these are located (very) close to two-particle thresholds. As examples consider
the $D_{s0}^*(2317)~[D_{s1}(2460)]$ that sits only 45~MeV below the $DK~[D^*K]$ threshold
or the $X(3872)$, that is essentially degenerate with the $D^0 \bar{D}^{0*}$ threshold. Therefore, such states are
premier candidates for hadronic molecules, that are loosely bound states of colorless hadrons much like the
deuteron in nuclear physics, which is bound by only 1~MeV per nucleon and fairly extended in space. It is important to realize that nuclei and hadron physics are just different manifestations of the emergence of structure in QCD and thus should not be considered
separately.
For recent reviews, see Refs.~\cite{Chen:2016qju,Lebed:2016hpi,Guo:2017jvc}.

As should be clear, these remarkable experimental findings have gone hand in hand with theoretical developments
that allow for model-independent studies of the spectrum or at least parts of it. Both in lattice QCD as well
as Effective Field Theories (EFTs) (or combinations thereof), much progress has been made towards an understanding
of the hadron spectrum. Let us first make some comments on lattice QCD. Here, the space-time is discretized
and the QCD path integral is solved by stochastic methods (Monte Carlo simulation) on high performance
computers. While this sounds simple, it requires a tremendous effort in devising fast algorithms and dealing
in a systematic fashion with the various artefacts introduced due to the finite volume, the finite lattice
spacing and the often unphysically  light quark (or pion) masses. A breakthrough was achieved in 2008, when the
first precision calculation of the light hadron spectrum with almost physical up and down quark masses
was performed~\cite{Durr:2008zz}. Since then, a cornucopia of results with light pion masses has appeared
and methods have been developed to compute particle decays into two and three particles, usually extending the
pioneering work of L\"uscher~\cite{Luscher:1986pf}. This is of utmost importance as most hadrons are resonances,
thus one must be able to compute their masses and widths. In addition, as one goes up in energy, more and
more two- and three-body channels open and further complicate the picture. It is fair to say that at present
only well separated resonances like the $\rho$-meson have really been calculated {\it ab initio} and
a few attempts have been made to look at more complicated systems with light and/or charm quarks
involving coupled channels, see e.g. Refs.~\cite{Feng:2010es,Mohler:2013rwa,Moir:2016srx}.
In addition, non-relativistic EFTs allow to address particular energy
regions, especially in the vicinity of two-particle thresholds or when heavy quarks are involved. Such a
framework allows one also to investigate kinematical structures like threshold cusps or triangle singularities, that
can mock up resonance signals in experiment. For a more detailed discussion, see~\cite{Guo:2017jvc}.

\subsection{Production and spectroscopy of baryon-antibaryon pairs}

In addition to the spectroscopy research discussed so far, PANDA at FAIR will also be able to contribute to the important interplay between the emergence
of structure in QCD and the interactions of the corresponding hadrons. This holds in particular
for the excitation spectrum of baryons with strangeness and the corresponding baryon-baryon
interactions, such a hyperon-nucleon or hyperon-hyperon forces. Indeed, surprisingly
little is known about the excitation spectra of (multi)strange and charm baryons. Since the cross sections
into baryon-antibaryon final states are large, the spectroscopy of excited hyperons is a compelling part
of the initial program at PANDA. Furthermore, the interactions between hyperons are another area where
the data situation is to say the least sparse. These interactions are not only interesting by themselves,
but are also known to play an important role for the physics of compact stars. In fact, with the new
area of gravitational wave astronomy~\cite{Abbott:2016blz}, the nuclear equation of state plays
an even more prominent role~\cite{Bauswein:2017vtn,Baym:2017whm}. Hyperon-hyperon and hyperon-nucleon
interactions can be obtained from the spectrum of single- and double-hypernuclei, that will be abundantly produced
at FAIR. On the theoretical side, effective field theory~\cite{Haidenbauer:2013oca}
and lattice QCD studies~\cite{Beane:2012ey,Inoue:2013nfe} are paving the way for a model-independent
analysis of these data. Another interesting feature is the possible appearance of  exotic bound states,
a field that was triggered by Jaffe's prediction of the H-baryon~\cite{Jaffe:1976yi}, an exotic six
quark ($\Lambda$-$\Lambda$ bound) state with strangeness $-2$, that has never been observed. However, this
state and other strange di-baryons are nowadays amenable to  lattice QCD~ \cite{Doi:2015oha}
as well as precise EFT studies~\cite{Haidenbauer:2011za}. This part of the FAIR hadron physics
program links directly to the superb nuclear structure and nuclear astrophysics investigations
that will be done by the NUSTAR collaboration at FAIR (see below). It is worth to emphasize again that hadrons and nuclei should be considered
together, as these are different manifestations of the structure formation in QCD.

\subsection{What will FAIR deliver?}

\begin{figure}[ht!]
	\begin{center}
		\includegraphics[width=0.55\textwidth]{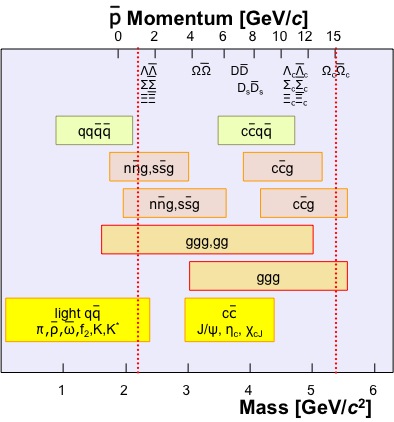}
	\caption{Overview of the hadron states accessible in antiproton-proton ($\bar pp$)
        annihilation as a function of antiproton momentum (upper scale) or center-of-mass energy (lower scale).                                        The red dashed lines indicate the range accessible with PANDA at the HESR. }
        \label{fig:ugm1}
	\end{center}
\vspace{-3mm}
\end{figure}

Here, we briefly discuss how the hadron and hypernuclear program with the PANDA experiment
at FAIR will advance our understanding of structure formation in QCD (for more details, see
the PANDA physics book~\cite{Lutz:2009ff}, being aware that there has been tremendous progress
in the field since then). All this is only possible because of the fine energy resolution
that can be achieved in proton-antiproton annihilation at FAIR's High-Energy Storage Ring HESR supplemented by modern
technologies incorporated in the PANDA detector. The accessible energy range of PANDA is depicted in Fig.~\ref{fig:ugm1}.

\noindent{\bf Gluonic excitations:}~The glueball spectrum in pure Yang-Mills theory is well
determined~\cite{Morningstar:1999rf}, with the lowest state a $0^{++}$ excitation with a mass below 2~GeV.
This state is expected to mix heavily with scalar mesons of a similar mass and no lattice QCD calculation
exists that accounts for this. Glueballs with higher spin and exotic quantum numbers in the mass range
from 2.5 to 5~GeV are only accessible with PANDA at FAIR. For hybrids, there are lattice QCD predictions
of hybrid meson states in the mass range from 2 to 2.5~GeV with exotic quantum numbers~\cite{Dudek:2010wm},
however, these calculations were done at rather large pion masses and did not properly include the
channel couplings alluded to before.  Non-relativistic Effective Field Theory (NREFT) also predicts hybrids in the mass region
between 4 and 4.5~GeV~\cite{Berwein:2015vca}, but it remains to be seen whether particular decay
properties that would allow to identify these states can be predicted. In any case, there is
a rich field for discovery here.

\smallskip

\begin{figure}[t!]
	\begin{center}
        \includegraphics[width=0.30\textwidth]{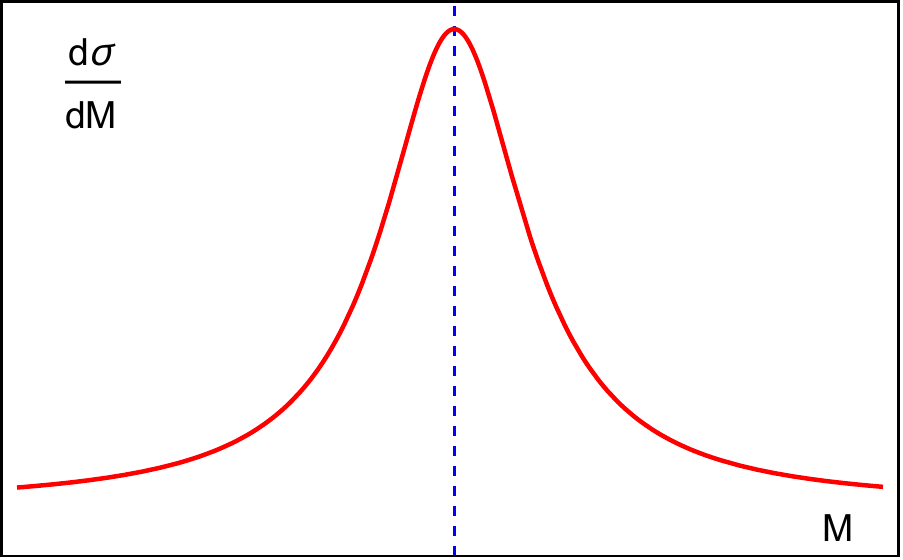} \hspace{0.95cm}
	\includegraphics[width=0.30\textwidth]{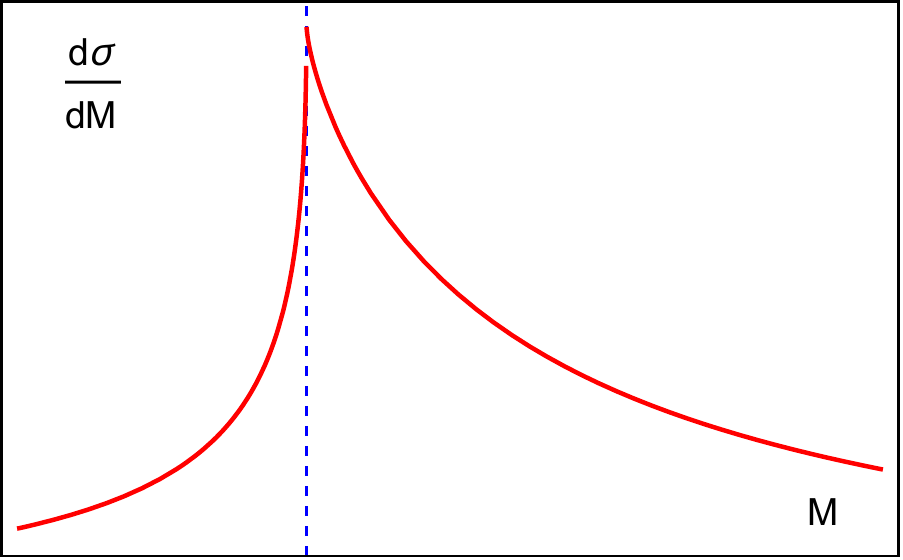}
	\caption{Behavior  of  near-threshold line shapes for
                compact (left panel) and molecular states (right panel) around a threshold
                indicated by the dashed perpendicular line. The x-axis
                shows $M = m_1 + m_2 + E$, with $E$ the energy and $m_{1,2}$ the masses of the
                two hadrons forming the molecular state.
                }
                \label{fig:ugm2}
	\end{center}
\vspace{-3mm}
\end{figure}

\noindent{\bf Charmonium spectroscopy:}~While many exciting results concerning the XYZ states have
been already obtained and more will be generated in the next years, PANDA will not only be a ``XYZ-factory'',
producing hundreds of these elusive states like the $X(3872)$ or $Z_c(3900)$ per day, the superb
resolution will also allow to make precision measurements of these particles, eventually revealing
their true nature. This is symbolically shown in Fig.~\ref{fig:ugm2}, where the line shape for a compact tetraquark state
is compared to a hadronic molecule with exactly the same mass  (and all other quantum numbers equal)
around a two-particle threshold. While the former is symmetric around the threshold energy, the latter is highly asymmetric and
further exhibits a visible  non-analyticity at the two-particle threshold. In addition, such line shape
measurements can also differentiate between bound and virtual states (for details, see the
review~\cite{Guo:2017jvc}). PANDA will be unique in performing such threshold energy scans and thus
will be able to determine the width of the $X(3872)$ (and other states) from the line shape, which is
not possible in $e^+e^-$ or $pp$ production experiments. So despite the
many results already obtained in this field, FAIR is in a unique position of solving a number of
remaining mysteries by producing data with an unprecedented accuracy.

 \smallskip

\begin{figure}[ht!]
	\begin{center}
        \includegraphics[width=0.50\textwidth]{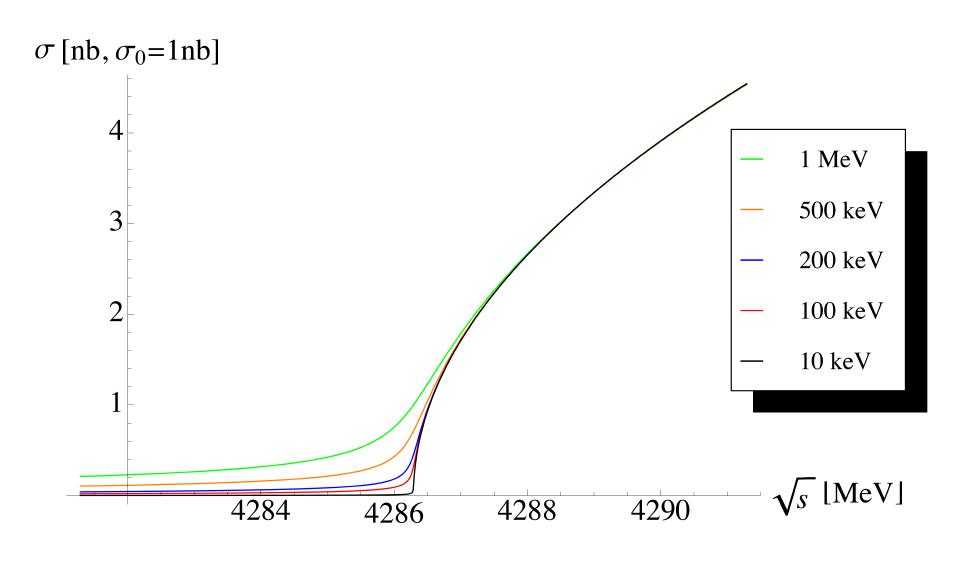}
	\caption{Excitation  function  to  simulated  scan  data  for a $D_{s0}^*(2317)$
                 with a width of 10~keV (lowest line) up to 1~MeV (upmost line).
                }
                \label{fig:ugm3}
	\end{center}
\vspace{-3mm}
\end{figure}

\noindent{\bf D-meson spectroscopy:}~Similar to the quarkonium sector, PANDA will also be able to
solve some puzzles in the heavy-light sector. Of particular importance will be a measurement of
the width of the $D_{s0}^*(2317)$, which decays isospin-violating to $D_s^+\pi^0$. In the quark
model, such isospin-violating decays are strongly suppressed and thus the width is calculated to
be 10~keV or less, see e.g.~Ref.~\cite{Godfrey:2003kg}, but larger than 100~keV if it is a molecule,
due to the channel couplings, see e.g.~~Ref.~\cite{Liu:2012zya}. The upper limit on its width is
presently 3.8~MeV~\cite{Patrignani:2016xqp}. Such a small width a predicted by the various
models is very difficult to measure directly. However,
the shape of this excitation function (the line shape) depends on the particle width,
thus the measurement of the shape can be used to deduce the particle width. A corresponding
simulation for PANDA is shown in Fig.~\ref{fig:ugm3}, indicating that the  width resolution of the
order of 100~keV is feasible~\cite{Mertens:2012kpa}.
This would be the most stringent test of the molecular scenario
so far and might lead to a new paradigm in heavy-light spectroscopy~\cite{Du:2017zvv},
similar to the light scalar mesons, were the quark-antiquark excitations do not form the
lowest multiplet but rather molecular states like the $\sigma(500)$, the $f_0(980)$ and so on.
This would be truely a remarkable achievement.

\smallskip

\noindent{\bf Strangeness physics:}~PANDA will also be a strangeness factory, because even in phase~1, hyperon-antihyperon pairs will be copiously produced. This will allow
to study excited states of double-strange hyperons ($\Xi^*$), triple-strange hyperons ($\Omega^*$), charmed
hyperons ($\Lambda_c^*, \Sigma_c^*$) as well as hidden--charm nucleons ($N_{\bar c c}$, pentaquarks).
In addition, antiproton annihilation on nuclei will give access to double hypernuclei to study
baryon-baryon interactions and possible di-baryons. This is a wide field with a large potential
for surprises and of high importance to nuclear astrophysics, e.g. neutron star structure. As is well-known, the nuclear equation of state is rather sensitive to
strangeness, as is also discussed in the upcoming chapters.

\smallskip

\noindent{\bf Electromagnetic processes:}~Another area where PANDA can make significant contributions
are processes involving real or virtual photons. In particular, our knowledge of the time-like
form factors of the proton can be advanced tremendously as individual measurements of the
electric and magnetic form factors as well as their relative phase  become possible at FAIR~\cite{Adamuscin:2006bk}.
In this way, one will be able to finally clarify whether strong enhancement in the cross section
at the two-nucleon threshold is due to a subthreshold resonance or generated by final-state
interactions~\cite{Haidenbauer:2014kja}.
Further, PANDA can make important contributions to the understanding of the proton's quark-gluon
substructure through measurements of transition  distribution amplitudes, generalized distribution amplitudes
as well as the Drell-Yan process.

\subsection{Outlook}

The hadron physics program at FAIR has been under severe criticism from the start. However,
over the years, there has been a tremendous revival of interest in spectroscopy due to the observation
of many unexpected states at colliders world-wide. A large number of fine data has been
collected and is being generated in the next few years at various laboratories world-wide,
but from the above discussion it should have become clear that there is a {\em very rich
discovery potential} with PANDA at FAIR. With the data collected over the years and the further
development of theoretical methods and calculations, we can look forward to a real understanding
of the emergence of structures in
QCD, which would constitute arguably the most  remarkable achievement of research at
the Facility for Antiprotons and Ions Research. The future for hadron physics
with/for FAIR looks very bright indeed.


\section{FAIR  - opening a new window into QCD matter}
Understanding the properties of QCD matter is not only of fundamental
interest but has also important implications for the world around
us. Ninety-eight percent of the mass of every atom is due to QCD
dynamics, with the remaining two percent coming from the
Higgs-mechanism. The structure of neutron stars as well as the
dynamics of the recently observed neutron star mergers are controlled
by the properties of QCD matter, in particular its equation of state
(EoS). More fundamentally, QCD matter will likely be the only
matter made from fundamental degrees of freedom of the standard model,
quarks and gluons, which can be actually
explored in the laboratory.  Understanding its phase
structure is therefore of central interest and it may inform us about
the structure of matter at the electro-weak and even higher energy
scales, and thus help us in constraining the dynamics of the early
universe right after the Big Bang.

In the laboratory QCD matter is produced by colliding heavy nuclei at
relativistic energies. Over the last decades, ultrarelativistic
collisions of heavy nuclei at increasingly higher center of mass
energies have been studied at the Relativistic Heavy Ion Collider
(RHIC) at BNL and at the Large Hadron Collider (LHC) at CERN. The
analysis of collected data provided detailed insights on the behavior
of the nuclear matter at very high temperatures and low net-baryon
densities, conditions very similar to those that prevailed some
microseconds after the Big Bang.  The experiments at RHIC and LHC have
paid off the huge investment in money and human resources made by DOE
and European Research Institutions by providing strong evidence of the
existence of the Quark-Gluon-Plasma (QGP), a scientific landmark of
paramount importance for understanding the properties of hadrons and
the evolution of the Universe. In particular is has been established
that the Quark-Gluon-Plasma is very opaque and, at the same time, it
behaves like a nearly perfect fluid, with a shear viscosity to entropy
ratio close to limit of infinitely strong coupling (for  reviews see
e.g. \cite{Braun-Munzinger:2015hba,Wang:2016opj}). Interestingly, the other
known substances exhibiting a similarly small shear viscosity to
entropy ratio are ultra-cold quantum gases with an interaction tuned
to be at the unitarity limit.  Experiments at RHIC and LHC have also
demonstrated that heavy ion collisions create a nearly thermal system
of strongly interacting matter where all possible states are created
according to their statistical weights. As a result even rather exotic
states such as the $\mbox{}^{3}_{\Lambda}H$ and
$\mbox{}^{3}_{\bar{\Lambda}}\bar{H}$
\cite{Abelev:2010rv,Adam:2015yta,Adam:2015vda}  have been observed
in these reactions. Therefore, given sufficient statistics, heavy ion
reactions may serve as sources for even more exotic multi-strange
nuclei which otherwise would be difficult to produce.

While most of the activity in heavy ion research has focused on the
highest energies to study high-temperature QCD matter, the interest in
collisions of heavy nuclei at lower energies has grown substantially
over the past years and has led to the design of a number of new
experiments at accelerator facilities under construction. The goal of
this novel strategy is to study QCD matter and its phase diagram at
large net-baryon density. Indeed, at moderate centre of mass energies,
the colliding nuclei create at the mid-rapidity region a medium with
finite net-baryon density which increases with decreasing beam energy,
with a maximum around 30 AGeV \cite{Randrup:2006nr}, where a density
of up to ten times that of the nuclear ground state is reached, i.e.,
a density which we believe exists in the cores of neutron stars.

Clearly the Compressed Baryonic Matter Experiment, CBM, at FAIR in
Darmstadt is well positioned to explore
many facets of QCD matter, find new exotic states (in conjunction with
the complementary PANDA experiment). In addition it will be able to
explore the equation of state at densities and temperatures close to
those probed by neutron star mergers. For a comprehensive review of
the CBM physics program see \cite{Friman:2011zz}.

\subsection{Exploring the QCD phase diagram}
\begin{figure}[ht!]
	\begin{center}
		\includegraphics[width=0.50\textwidth]{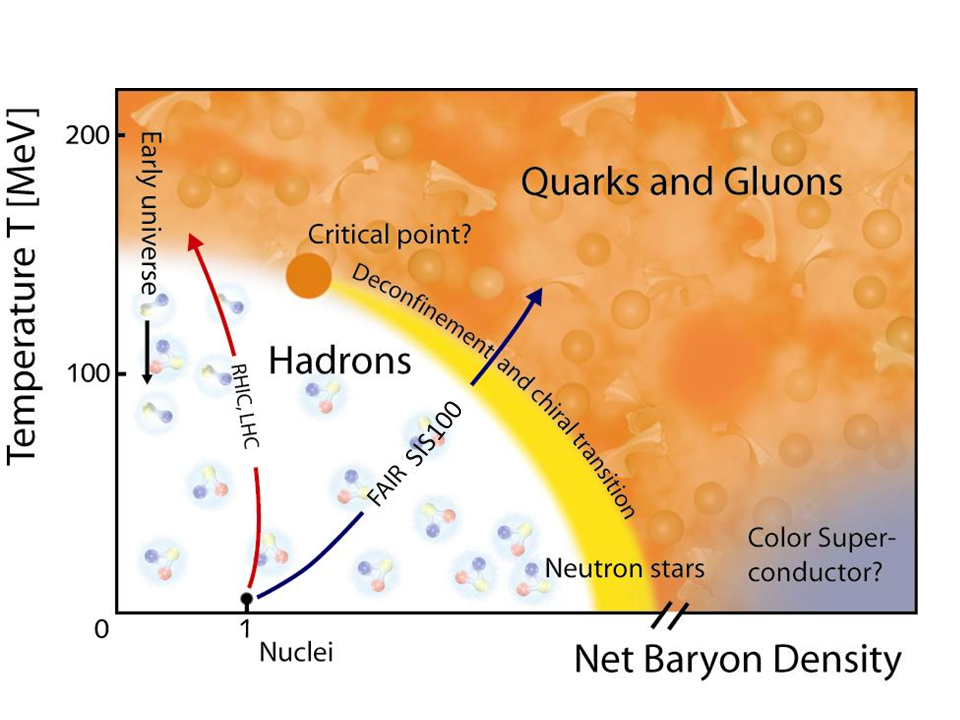}
	\caption{Sketch of the QCD phase diagram including a critical
          point and a first order co-existence region. Also shown are
          typical trajectories of heavy ion collisions at LHC, RHIC
          and FAIR. \label{fig:phase_diagram} }
	\end{center}
\vspace{-3mm}
\end{figure}

QCD matter at rather low net-baryon density has been extensively
studied in experiments at RHIC and LHC, where the existence of the QGP
and its unique properties -- near perfect fluidity and large
opaqueness -- have been established. At the same time Lattice QCD
calculations have determined that at vanishing net-baryon density the
transition from a hadronic system to the QGP is an analytic cross-over
\cite{Aoki:2006we}.  Also, experimentally no fluctuation signals
associated with a phase transition have been observed. However, the
situation at large net-baryon density may very well be different. Many
model calculations predict a first order phase co-existence region at
large densities (see e.g. \cite{Stephanov:2004wx}). This co-existence
region will end in a critical point, which, if it exists, would
constitute a landmark not only in the QCD phase diagram but in the
standard model. A sketch of the QCD phase diagram is shown in
Fig.~\ref{fig:phase_diagram}. Therefore, not surprisingly, parallel to
the study of high-temperature nuclear matter, interest in collisions
of heavy nuclei at lower energies has grown substantially over the
past years.

Unfortunately the region of the QCD phase diagram at large net baryon
densities and low temperatures cannot be addressed theoretically by
first-principle calculations like lattice QCD. Instead, one has to
rely on effective models and in particular on experiment and a
coherent theory effort. Since a relativistic heavy ion collision is a highly dynamical process, one of the most important tasks is the development of a dynamical framework which reliably propagates the signals for example of a critical point to the final state. At the highest energies viscous hydrodynamics combined with the equation of state from lattice QCD serves as such a framework. However, at the energies relevant for FAIR this framework needs to be improved and modified in many important aspects, for example: What are the initial conditions? What is the equation of state at large baryon densities? How does one faithfully propagate fluctuations resulting from a phase transition? Is hydrodynamics still valid? If not, what is the alternative? 
In addition to such a dynamical framework, it is, of course essential for this program, to have well defined
observables which are sensitive to possible structure in the QCD phase
diagram. One such observable are the cumulants of the net-proton
distribution which is expected to show a strong maximum at the
critical point \cite{Stephanov:2008qz,Stephanov:2011pb}. And indeed
preliminary data by the STAR collaboration from the first beam energy
scan at RHIC show a strong increase of the fourth order cumulant at
the lowest energies accessible at RHIC
\cite{Luo:2015doi,Luo:2017faz}. This increase towards lower energies
and especially the magnitude of the fourth-order cumulant at the
lowest energy of $\sqrt{s}=7.7$ GeV  cannot be explained by presently
available models, which all predict a decrease of the cumulants
\cite{He:2017zpg,Bzdak:2016jxo}. Therefore, if the preliminary STAR data are
confirmed in the second phase of the RHIC beam energy scan, one may
very well see the first hints of the QCD phase transition or critical
point. And since one has not yet seen the maximum of the cumulants,
FAIR and CBM may very well be in position to actually discover the QCD
critical point.

Of course it will require more than one observable to establish the
existence of a critical point. In addition to large fluctuations one
expects a softening of the equation of state and long lifetime of the
system. The former can be accessed by measuring the collective
expansion or flow of the system. The latter is best probed by the
measurement of the dilepton yield. In addition dileptons measure the
spectral function of the vector current and thus provide valuable
insights about the effective degrees of freedom. If, for example, the
observed new phase is the so called quarkyonic matter, where quarks
are still confined but chiral symmetry is restored
\cite{McLerran:2007qj}, one would expect to see significant changes in
the dilepton invariant mass spectra. Naturally, as more insights are gained e.g. from the RHIC beam energy scan, new ideas and and possible better observables will likely emerge. 
At the same time, it will be necessary to develop a theoretical framework, which allows for a quantitative comparison with the data.

\subsection{Strange matter and states }
QCD is not confined to the light flavors
which make up the matter of the Universe as we know it. A very
interesting question concerns the properties of matter where one or more
up or down quarks are replaced by strange quarks. Besides its
theoretical appeal this question has a direct impact for our
understanding of objects like neutron stars. 
For example, depending on the interaction of hyperons with neutron matter, 
it is conceivable that the inside of a neutron star is made out of hyperon matter 
\cite{Glendenning:1984jr}.
The observation of
single and doubly strange hyper-nuclei tell us already that matter
with strangeness is stable with respect to the strong interaction. But
what is the limit of this stability? Is it possible to have nuclei with
more than two units of strangeness? Here again heavy ion reactions
especially in the energy regime accessible by FAIR provide a unique
tool.

It is by now well established that the final state in a heavy ion
reaction is very close to thermal. Thus all possible states are
accessible, including ``exotic'' ones such as multiply strange
hyper-nuclei. To some extend this has been demonstrated by both the
STAR and ALICE experiment, where states such as the hyper-triton,
$\mbox{}^{3}_{\Lambda}H$, and anti-helium, $\mbox{}^{4}\bar{He}$, have
been observed in high energy collision
\cite{Abelev:2010rv,Adam:2015yta,Adam:2015vda}. In particular the
measurement of the $\mbox{}^{3}_{\Lambda}{H}$, which is bareley bound
by $2.39$  MeV and has a $\Lambda$ separation energy of only
$0.13$ MeV, demonstrates that even the most weakly bound systems can
be observed in spite of the violence of the collision. In addition to
hyper-nuclei these reactions may also be used to test the existence of
other objects, such as the Di-$\Omega$ state that has recently been
predicted to exist by Lattice QCD calculations \cite{Gongyo:2017fjb}.

At FAIR energies the conditions are ideal for the production and
observation of hyper nuclei, or other multi-strange metastable
objects (MEMOS) \cite{Steinheimer:2008hr}. The energy is sufficiently high to produce strangeness
in abundance and, at the same time, it is low enough so that the number
of baryons at mid-rapidity is high. Thus the probability for multiple
strange quarks to attach themselves to baryon and thus a nucleus is
rather large. In addition, the available high beam intensity together
with the capability of CBM to take an unprecedented data rate allow
for the detection of extremely rare states. This environment is also well suited
to clarify controversial issues like the  existence of anti-kaonic
nuclear bound states.

Together with the PANDA experiment and advances in Lattice QCD
which soon will allow for the calculation of multi-strange nuclei at
the physical pion mass, FAIR is very well positioned to advance our
knowledge of the strongly interaction matter with strangeness. It will, therefore, not only help us to answer important questions about strongly interacting many body systems in general, but also further our understanding how the strong interaction controls the properties of astrophysical objects, such as neutron stars.

\subsection{Equation of state }

\begin{figure}[ht!]
	\begin{center}
		\includegraphics[width=0.40\textwidth]{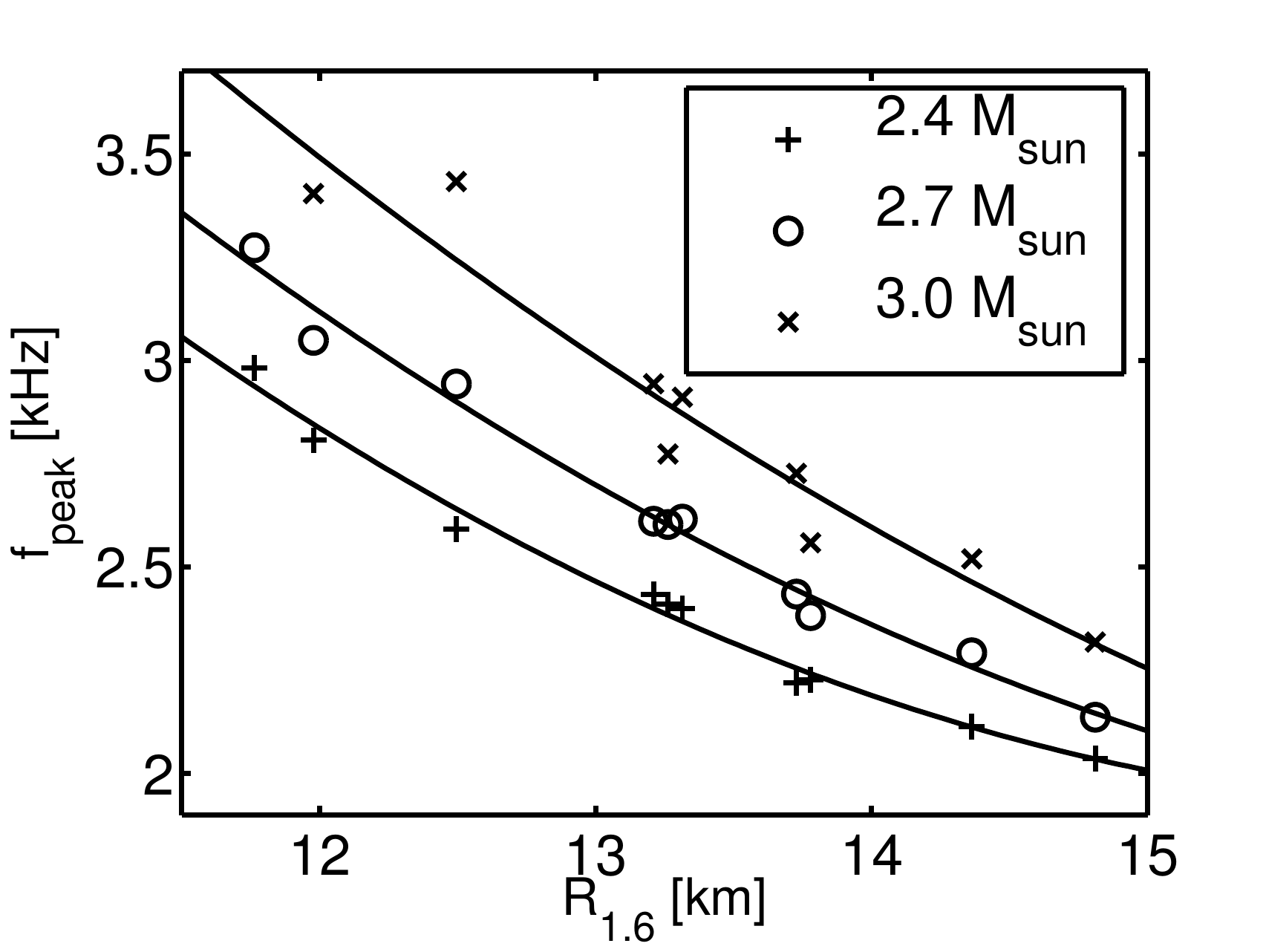}
        \includegraphics[width=0.40\textwidth]{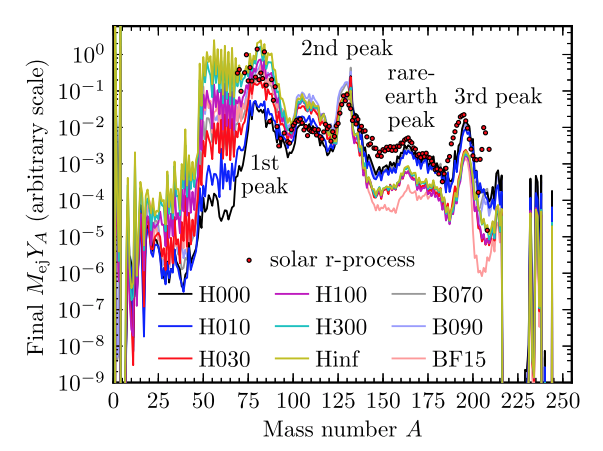}
	\caption{Left: Dominant post-merger gravitational wave frequency, $f_{peak}$ as a function of the radius $R1.6$ of a nonrotating neutron star with a gravitational mass of $1.6 M_{sun}$. Figure adapted from \cite{Bauswein:2015vxa}. The three symbols and curves correspond to different total binary masses. Right: Abundance of r-process nuclei produced in a neutron star merger for different lifetimes on the merged neutron star before collapse. Figure represents Fig.~4 of \cite{Lippuner:2017bfm}. \label{fig:NS_merge_EOS} }
	\end{center}
\vspace{-3mm}
\end{figure}

Beyond its importance for the understanding of fundamental properties
of QCD, the study of highly compressed nuclear matter is of high
interest in the context of astrophysics. The discovery of neutron
stars with masses of about two solar masses and, more recently, the
observation of a gravitational wave from neutron stars merging impose
severe constraints on the nuclear EoS, since the appearance of
hyperons softens the EoS to an extent that such heavy neutron stars
may collapse into black holes. While the structure of neutron stars is
governed by the EoS at small temperature, present simulations of
neutron star merges indicate that the shocks which eject the matter
relevant for the r-process reach temperatures of $T\sim 100$ MeV and
densities of several times nuclear matter density
\cite{Barnes:2013wka,Fernandez:2016sbf,Kasen:2017sxr}. 
These condition are very similar to those
which are responsible for the collective expansion of the matter in
heavy ion collisions available at FAIR. 
Therefore, a comprehensive analysis of collective flow, particle production etc  within the aforementioned dynamical framework, which will constrain the equation of state, informs the input simulations of neutron star mergers.   
Of course the matter governing
neutron stars and their mergers is more neutron rich than the typical
matter of a heavy ion collision. However, by selecting different
collision systems and possibly selecting events with a large isospin
asymmetry in the pion sector one may be able to study or at least
constrain the leading isospin dependence of the EoS.  These
measurements will then inform and constrain calculations of the
neutron star structure and hence complement the direct observation of
neutron star radii with improved resolution. Also, these measurements
will provide important inputs for the simulation of neutron star
mergers, which arguably are an important site for the r-process,
i.e. the production of heavy elements (see section 5). The sensitivity of the dynamics of neutron star mergers to the EoS is shown in Fig.\ref{fig:NS_merge_EOS}. In the left panel we show the 
dominant post-merger gravitational wave frequency, $f_{peak}$ as a function of the radius $R1.6$ of a non-rotating neutron star with a gravitational mass of  $1.6 M_{sun}$ for three different binary masses. The effect of the EoS is encoded in  the radius $R1.6$ such that a stiffer EoS results in smaller $R1.6$ (for detail see \cite{Bauswein:2015vxa}). On the right panel we show 
the yield of r-process nuclei produced in a neutron star merger for different lifetimes of the merged neutron star before collapse. This lifetime depends of the EoS and thus the production of r-process nuclei depend, among other, also on the nuclear EoS (for details see \cite{Lippuner:2017bfm}). Therefore, in conjunction with the NUSTAR program  FAIR  will enable to significantly advance our
understanding of the origin of the elements.

\subsection{The CBM experiment at FAIR}
FAIR will provide primary beams from the SIS100 synchrotron with
energies up to 29 GeV for protons, up to 11 AGeV for Au, and up to 14
AGeV for nuclei with $Z/A = 1/2$. At this facility, CBM will be the
leading project for carrying on the nuclear collision program at
SIS100, with well-defined upgrade options for exploiting, when it will
be available, the second FAIR accelerator, which will extend
the energy range to about 45 GeV per nucleon. Ten meters long, the CBM
experiment will measure both hadronic and leptonic probes with a large
acceptance in fixed-target mode. Its layout comprises a large-aperture
superconducting dipole magnet and seven subsequent detector systems
providing tracking and particle identification. A forward calorimeter
and the Projectile Spectator Detector will allow measuring the
centrality and the reaction plane. Measurements will be performed in
nucleus-nucleus, proton-nucleus, and - for baseline determination -
proton-proton collisions at various beam energies (for an overview of
the CBM experiment see \cite{Senger:2017oqn}).

A particular challenging feature of CBM is its capacity to cope with
the complex event  topologies typical for heavy-ion reactions at very
high interaction rates of up to $10^{7}$ collisions per  second, which is
three orders of magnitude higher than the rates reached in other
high energy heavy-ion experiments. The unprecedented high rate
capability of CBM combined with the high-intensity beams of FAIR will
allow overcoming the limitations in statistics suffered by current
experiments. These features will provide worldwide unique conditions
for performing high-precision measurements of multi-differential
observables and of extremely rare diagnostic probes which are
sensitive to the dense phase of the nuclear fireball. The yields and
flow  of identified anti-baryons, in particular multi- strange
hyperons, intermediate-mass lepton  pairs, and particles containing
charm quarks will allow a comprehensive  study of QCD matter at the
highest net-baryon  densities achievable in the laboratory to be
performed.

To achieve this ambitious goal, rigorous R\&D on innovative
technologies for detectors, electronics and data acquisition system
with trigger-less electronics and shipping time-stamped raw data to an
online computing farm was performed over the last decade, with the
emphasis on fast and radiation-hard detectors and electronics.  Owing
to the large amount of raw data delivered by the CBM detector set-up,
data processing and event reconstruction will be performed in
real-time.  The rather complex signature topology, which characterizes
most of the rare key observables, does not allow the use of
conventional hierarchical trigger schemes.  This necessitates very
fast and highly parallel algorithms, which reconstruct 4D event
topologies (i.e., in space and time), coping with the temporal event
overlap at high rates. Therefore, online reduction of the data rate by
more than two orders of magnitude when running at the highest
interaction rates is mandatory.  For the time being, the installation
and commissioning of CBM is planned during 2021-2024.

The physics programme  of CBM will be enriched and complemented by systematic studies of  electron-positron pair emission from vector meson decays performed by the upgraded set-up of the High Acceptance DiElectron Spectrometer HADES.

HADES\cite{HADES:2017oqn} is a large acceptance fixed target experiment formerly designed to investigate the virtual photon emission and the role of baryonic resonance decaying into intermediate $\rho$-mesons and into final states with open strangeness at the energies of the synchrotron facility SIS-18 of GSI\cite{HADESSIS18:2017oqn}.

 Once upgraded and moved to the FAIR tunnel, HADES will share with CBM the same SIS-100 beam up to kinetic beam energies per nucleon of 8 GeV thus bridging the knowledge gap with SIS-18. Whereas, in the mass range in which both experiments are sensitive, dilepton spectra will be measured independently thus minimizing systematic uncertainties.

\subsection{Outlook}
The comprehensive understanding of the fundamental properties of
QCD matter is among the major scientific goals at the intensity
frontier, where physicists use intense particle beams and highly
sensitive detectors to observe rare processes in search of new
phenomena.

In this context, the combination of the high-intensity beams of
heavy-nuclei at the forthcoming facility FAIR with the
"state-of-the-art" experiment CBM will provide unique conditions for
the study of the QCD phase diagram in the region of baryon-chemical
potentials larger than 500 MeV.

The very rich physics program of CBM covers the measurement, with
unprecedented precision and statistics at reaction rates up to 10 MHz,
of multi-differential observables and of rare diagnostic probes, such
as multi-strange hyperons, charmed particles and vector mesons
decaying into lepton pairs, which are sensitive to the dense phase of
nuclear matter.

These data will allow CBM to successfully explore the onset of chiral symmetry restoration 
and its effect on dilepton production, exotic forms of (strange) QCD matter and the
equation-of-state at condition close to those encountered in  neutron star mergers.

\section{Exotic nuclei and exotic beams at FAIR}

On October 16, 1967, at the CERN laboratory in Geneva, the first dedicated experiment
to produce and study atomic nuclei with very unusual neutron to proton ratios was performed~\cite{Hansen69}.
This experiment became the beginning of a new era in nuclear physics giving access to
large numbers of earlier unexplored radioactive nuclei. The experiment was named
ISOLDE, an acronym for Isotope Separation On Line. In short, the ISOL technique  
allows direct  separation of radioisotopes, produced by irradiating a target 
directly linked to the ion-source of an isotope separator. In this way one gets access to isotopes with 
half-lives down to the millisecond region. The nuclei with very unusual proton to neutron ratio
are referred to as {\it exotic nuclei.} The ISOL method has, over its fifty years of existence,
been developed into a very high level of sophistication and is one of the leading experimental
approaches for production and studies of nuclei over broad regions, far out towards the proton 
and neutron driplines, of the nuclear chart. 

The ISOL method was without real competition until the beginning of the nineties. Then
a new method to produce exotic nuclei was demonstrated at the Bevalac accelerator
at LBL in Berkeley.  With energetic heavy ions, irradiating a light target, it was found 
that the projectile fragments contained chains of new exotic nuclei. With a $^{48}$Ca beam
(at 220 MeV/u) one could, for example, produce 14 new neutron-rich isotopes~\cite{Westfall79}.
The first experiments using radioactive beams were performed at LBL in 1985 with beams
of radioactive Li isotopes with an energy of 790 MeV/u. The measured interaction 
cross-sections showed a remarkable increase for the last bound Li isotope, $^{11}$Li~\cite{Tanihata85},
an increase that soon was interpreted as the occurrence of a nuclear halo structure~\cite{Han87}.
The discovery of nuclear halos has had an unprecedented attraction towards the entire field
of exotic nuclei and radioactive beams~\cite{Jon04}.

Since then the ISOL and the heavy-ion production methods have, hand in hand, developed
into a worldwide undertaking. Today, when we write 2018, we dare to say that there has never 
before, in the history of  nuclear science, been such an intense effort in constructing and building 
new or upgraded radioactive beam facilities. The future FAIR facility is but one example. 

One of the main pillars under the proposal for a major upgrade of GSI into FAIR was the
experimental programme that over the past 25 years has shown the uniqueness of 
relativistic radioactive beams for investigations of exotic nuclei. FAIR will open the
door to new frontline experiments profiting from primary heavy ion beams
with highest energies and intensities up to the element Uranium.

\subsection{Some highlights from 25 years of exotic beams at GSI}

In the late eighties and early nineties there were hectic activities worldwide at the laboratories 
where exotic nuclei could be produced. The halo hypothesis was in focus and scrutinised in
different experiments that could give signals supporting the presence of the new structure. Experiments 
at ISOLDE, GANIL, Dubna, MSU and RIKEN showed bit by bit the presence of
halos in, for example, $^{8}$He, $^{11}$Be and $^{11}$Li.

A very timely coincidence was the start of the GSI high-energy programme allowing to
perform the first dedicated experiment to study light neutron-rich 
halo nuclei. It was realised that performing experiments with Radioactive Ion Beams (RIBs) 
at relativistic energies poses a number of challenges as well as advantages; the kinematical 
focusing of the messengers from the reaction yields excellent solid angle coverage. Furthermore, 
the heavy remaining fragments will  leave the reaction vertex at almost the same velocity as the 
incoming beam, permitting further tagging and characterisation. 

Today, a quarter of a century later, we can look back on a string of different experiments, over
the entire region of the nuclear chart, where the uniqueness of the RIBs has been the key to
the success. It is also these experiments that have given inspiration and guiding for the
developments of new, modern equipment for the future FAIR physics. We give here a few
selected highlights from these experiments at GSI.

The first dedicated study of halo nuclei at GSI started in 1992. With a 340 MeV/u $^{11}$Li
beam break-up reactions in C, Al and Pb targets were studied. The transverse and longitudinal
momentum distributions of the $^{9}$Li fragments were measured and the data revealed the 
expected narrow width of the momentum distributions, which is a signal of the large spatial  
distribution of the halo wave function~\cite{Humbert95}. 

The nucleus $^{11}$Li is the last particle-bound Li isotope. Its lighter neighbour $^{10}$Li
is, however, unbound towards neutron emission. Even if $^{10}$Li is
unbound it is still of scientific relevance since it can be shown to possess clear and distinct 
quantum properties. The most precise data have revealed a virtual $s$-state as ground 
state, with a scattering length of $a_s$=-22.4 fm, and with an excited $p$-wave resonance at 
0.566 MeV ($\Gamma$=0.548 MeV)~\cite{Aksyutina08}.

The very exotic nuclear systems $^{12}$Li and $^{13}$Li could be shown to  
belong to the family of unbound nuclei in an experiment where a 304 MeV/u
$^{14}$Be beam was directed towards a liquid hydrogen target for studies of the
$^1$H($^{14}$Be,$2pn$)$^{12}$Li and $^1$H($^{14}$Be,$2p$)$^{13}$Li
reactions~\cite{Aksyutina08}. The relative-energy spectrum $^{11}$Li+$n$ showed 
a virtual $s$-state, with a scattering length of -13.7 fm, that
could be identified as the $^{12}$Li ground state. This result contradicts 
shell-model calculations and is one example of the disappearance of 
$N$=8 as a magic number in the dripline region. Going one step further
the $^{11}$Li+2$n$ data gave evidence for a resonance at 1.47(31) MeV,
a resonance that will be on the agenda for further investigations at the
future FAIR.

The Borromean\footnote{A bound three-body system where none of 
the binary subsystems are bound is referred to as Borromean.} nucleus $^{6}$He 
has a very simple three-body structure, $\alpha$+$n$+$n$,
and has been studied in a number of different experiments. In one of these experiments 
the breakup of a 240 MeV/u $^{6}$He beam in carbon and lead targets was studied and
the inelastic nuclear and electromagnetic excitation spectra were obtained~\cite{Aumann99}. 
The deduced $E1$ strength was found to exhaust both the energy-weighted cluster sum rule and the
non-energy weighted one when integrating the strength up to 10 MeV excitation.
The non-energy weighted sum rule is related to the distance between the centre of mass
of the two neutrons and that of the whole nucleus and thus to the geometry of $^{6}$He.
From the data the  root-mean distance between the $\alpha$ core and the two valence neutrons
could be derived to be $r_{\alpha-2n}$=3.2 $fm$.

Another example of an unbound nucleus of special interest is
the $^{13}$Be, bridging the gap between the deformed
nucleus $^{12}$Be and the Borromean halo nucleus $^{14}$Be, where
the structure is under debate. $^{13}$Be ($^{12}$Be+$n$) 
has been found to have several excited states, which decay by 
neutron emission to the ground state and also excited states in 
$^{12}$Be. Complementary experiments in inverse and complete kinematics, 
populating states in  $^{13}$Be after neutron knockout from $^{14}$Be~\cite{Aksyutina13} and 
proton knockout from $^{14}$B have been performed.  An interesting low-energy 
structure with two overlapping broad $s$- and $p$-states have been 
established but there are still open questions concerning, for example, 
excited $d$-wave states.

By bombarding a carbon target with a $^{17}$Ne beam at an
energy of 500 MeV/u, a rather remarkable discovery was made. It was
found that  the two-neutron knockout channel  showed a clear resonance 
structure, corresponding to the nucleus $^{15}$Ne, which was found to be 
unbound by 2.522(66) MeV~\cite{Wamers14}. This is the lightest observed nucleus with isospin 
($T,T_z$)=(5/2,--5/2). The $^{13}$O+$p$+$p$ relative-energy spectrum and
the fractional energy spectrum showed that the $^{15}$Ne ground state has
a configuration with the two protons in the ($sd$) shell with a 63 \% 
($1s_{1/2})^2$ component.

The lightest nucleus exhibiting two-proton radioactivity is $^{19}$Mg, which 
for the first time was produced in neutron knockout from a 450 MeV/u
 $^{20}$Mg beam~\cite{Mukha07}. This experiment demonstrated the benefits of a
measurement performed in-flight with precise tracking of all fragments.
A half-life of 4.0(1.5) ps  and a $Q$-value for two-proton emission of 0.57(5) MeV 
were deduced from the data. A theoretical description of the data in a 
three-body model with $^{17}$Ne$\otimes$($\pi$d$_{3/2})^2$ is debated
since the $^{17}$Ne is a loosely bound nucleus with a possible two-proton
halo structure, which naturally would give the structure as a $^{15}$O core
with the four protons in the $(sd)$ shell.

One beautiful example on the production capacity of relativistic heavy ions
is given by an experiment utilising a 750 MeV/u $^{238}$U beam bombarding
a Be target. The produced fission fragments were analysed with the fragment
separator FRS and the production cross sections were measured. This led to
the discovery of 58 earlier unknown isotopes ranging from $^{54}$Ca to
$^{124}$Pd~\cite{Bernas97}. The location of the new isotopes in the chart of nuclides 
is very interesting since they closely follow the expected $r$-process path.
The data also included the doubly-magic nucleus $^{78}$Ni, a nucleus that had 
been discovered in an earlier experiment at GSI. Theory suggests that the double-magicity
is preserved for $^{78}$Ni ground state but also that a prolate band 
may be present at low energy. 

The doubly magic nucleus $^{100}$Sn was first observed as a particle-stable
nucleus in 1994 in parallel experiments at GSI and GANIL. In a new experiment~\cite{Hinke12}
performed in 2012 the $^{100}$Sn isotopes were produced using a 1 GeV/u $^{124}$Xe
beam bombarding a Be target. The produced Sn isotopes were implanted into a stack
of silicon strip detectors surrounded by an array of gamma detectors.
With a total of 259 produced $^{100}$Sn nuclei some very interesting results could be 
obtained. A half-life of 1.16(20) s could be deduced from the time distribution of the
decaying nuclei. Further a total of five gamma lines, corresponding to transitions 
between states in $^{100}$In fed in the $\beta^+$ decay, could be identified. The
end-point of the  $\beta^+$ spectrum was found at  3.29 MeV. From the data a log($ft$) value 
of 2.62 was obtained, which is the smallest  value for this quantity found for any nucleus so far. 

The nuclear equation of state (EoS) of neutron-rich matter is essential for the understanding of 
many phenomena both in nuclear physics and astrophysics. The electric dipole ($E1$) response of 
nuclei and, in particular, its dependence on the neutron-to-proton asymmetry, is governed by the 
symmetry energy and its density dependence. In an experiment 
the $E1$ strength distribution for the neutron-rich nucleus $^{68}$Ni~\cite{Rossi13} was  
studied in a kinematically complete experiment  in inverse kinematics. This was the first demonstration 
of a measurement of the dipole polarisability for a neutron-rich radioactive nucleus.

Precision mass and $\beta$ half-life measurements were performed with the storage ring 
ESR. The relativistic radioactive isotopes were produced by projectile fragmentation and fission 
reactions and then separated with the fragment separator FRS. They were subsequently injected
 into the cooler-storage ring ESR. This powerful experimental method gives access to all fragments 
 with half-lives down to the sub-millisecond range~\cite{Bosch13}.

\subsection{FAIR and exotic nuclei: \\The instrumentation to pick the most exotic 
flowers in the nuclear landscape.}

The vast experimental programme conducted at GSI during the past 25 years has successfully
demonstrated the power of secondary radioactive beams for studies of the most exotic nuclear 
reactions and of nuclear structure all the way out to the driplines and beyond. The possibilities 
offered at the new FAIR complex~\cite{Gutbrod06} is a real challenge for the 
scientific architects designing the  future equipments. 

\begin{figure}[t]
\begin{center}
\includegraphics[width=0.9\textwidth]{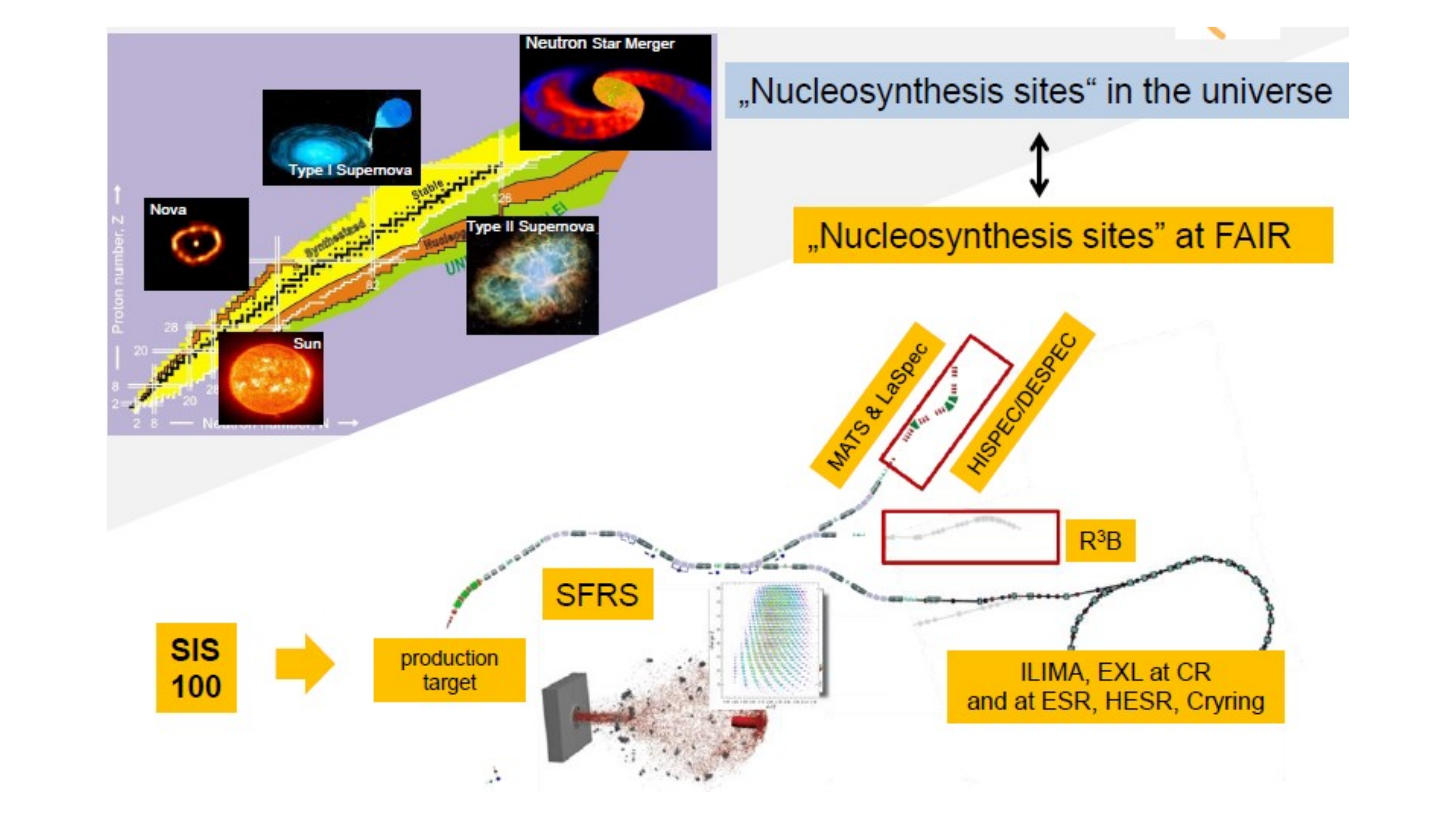}
\caption{\label{fig:nucl-sites}Layout of the nuclear physics RIB experiments at FAIR
with the fragmentation target, the Super FRS and the High-Energy, Low-Energy and Ring Branches (see text). As indicated by the nuclear chart,  exotic nuclei are crucial for astrophysical nucleosynthesis processes: neutron-deficient nuclei for novae and x-ray bursters and exotic nuclei with strong neutron excess for supernovae and neutron-star mergers. As discussed in the next section, FAIR will allow to produce many of these exotic nuclei and to determine  their properties.}
\end{center}
\end{figure}

The heart of the FAIR installation is the SIS-100 heavy-ion synchrotron, which will deliver
ion beams of unprecedented energies, intensities and quality. The RIBs used for the nuclear
physics research are produced in fragmentation reactions in a primary production target. 
The resulting cocktail of different energetic isotopes
is then  separated to deliver a specific isotope, or several, to
the experimental equipment. For this a new fragment separator has been designed, a separator
that can cope with the high-energy and high-intensity beams, which may reach the energy region of
around 1.5 GeV/u, and provide high separation power for elements up to Uranium ($Z$=92). 
The fragment separator is equipped with superconducting magnets assuring an efficient use of relativistic 
secondary beams characterised by a large phase-space.   
The new fragment separator is equipped with a total of 33 focusing and correction multiplets
and it is referred to as the Super-FRS. The large acceptance of Super-FRS, together with the 
primary-beam intensities for uranium ions from the SIS-100, will result in a gain in secondary-beam 
intensities of rare, radioactive isotopes of about three orders of magnitude.
There will also be a scientific programme directly linked to the Super-FRS; search 
for new isotopes and investigations of rare decay modes like multiple-proton or 
neutron emission, are planned.  For the other exotic nuclear 
physics programmes the Super-FRS can be said to be portal into the exotic nuclear landscape, 
since all secondary beams used in the experiments have to be selected by and delivered from it. 
From the Super-FRS there will be three individual beam lines to serve three different 
experimental areas: (i) the High-Energy Branch, (ii) the Low Energy Branch, and (iii) the Ring Branch (see Fig. \ref{fig:nucl-sites}).

The equipment used in the initial phase of the high-energy programme has over the 
years developed into a broad generic programme for physics at the driplines. 
Based on these experiences a dedicated setup for studies of reactions with relativistic 
radioactive beams of the FAIR brand is under development. The new equipment to 
perform kinematically complete measurements in inverse kinematics will be built up at the 
High Energy Branch. Priority is given towards high efficiency, acceptance and resolution. 
Important ingredients in the experimental setup are a new powerful dipole magnet and a high 
efficiency array of neutron detectors. A very interesting novel detector, based on scintillation 
crystals, will surround the reaction target. It will act as a total absorption gamma-calorimeter 
and spectrometer, as well as a spectrometer for charged particles from the target residues. Around these
backbone instruments there will be all necessary detectors for particle identifications. The entire
setup will be a major ingredient in the future experimental programme at FAIR and will become, 
in its final design, a facility within the facility,

At the low-energy branch from the Super-FRS beams in the energy range 100-300 
MeV/u will be available. There will be three different approaches to study exotic nuclei in this 
experimental hall. The first has a large array of Ge detectors for in-flight gamma spectroscopy,
charged particle detection  and lifetime measurements with plunger techniques. The second option is
based on stopping the RIBs for decay studies in a stack of a highly segmented silicon-based detector array, 
surrounded by a compact Ge array, neutron detectors, fast BaF$_2$ detectors and/or a total absorption 
gamma ray spectrometer. Finally, measurements of nuclear ground-state properties like nuclear masses, spins, moments and 
isotope shifts will be performed in regions of the nuclear chart where the ISOL technique have limited
production. This is for example the case for neutron-rich refractory elements. The mass measurements will use
Penning-trap and Multi-Reflection Time-of-Flight techniques. A relative mass precision of $\Delta M/M$=10$^{-9}$ 
is expected to be reached by employing highly-charged ions and a non-destructive 
Fourier-Transform-Ion-Cyclotron-Resonance (FT-ICR) detection technique on single stored ions.
Collinear  laser spectroscopy on ions, optical pumping and collinear laser spectroscopy on atoms
and $\beta-$NMR are among the atomic physics techniques foreseen to provide 
nuclear structure information about the produced exotic nuclei.

At the Ring Branch new experiments are planned with the aim to measure the masses and the lifetimes 
of exotic nuclides out to the driplines, using 
Isochronous Mass Spectrometry for investigations of short-lived nuclear species.
A future perspective is to investigate exotic nuclei in light-ion scattering at
intermediate energies using an internal target in the HESR ring.
The reactions will be performed in inverse kinematics by using
a universal detector system providing high resolution and large solid angle coverage in kinematically 
complete measurements located around the internal gas-jet target.

\subsection{Experiments during the first years of FAIR}
 
\begin{figure}[t]
\includegraphics[width=150mm, angle=0]{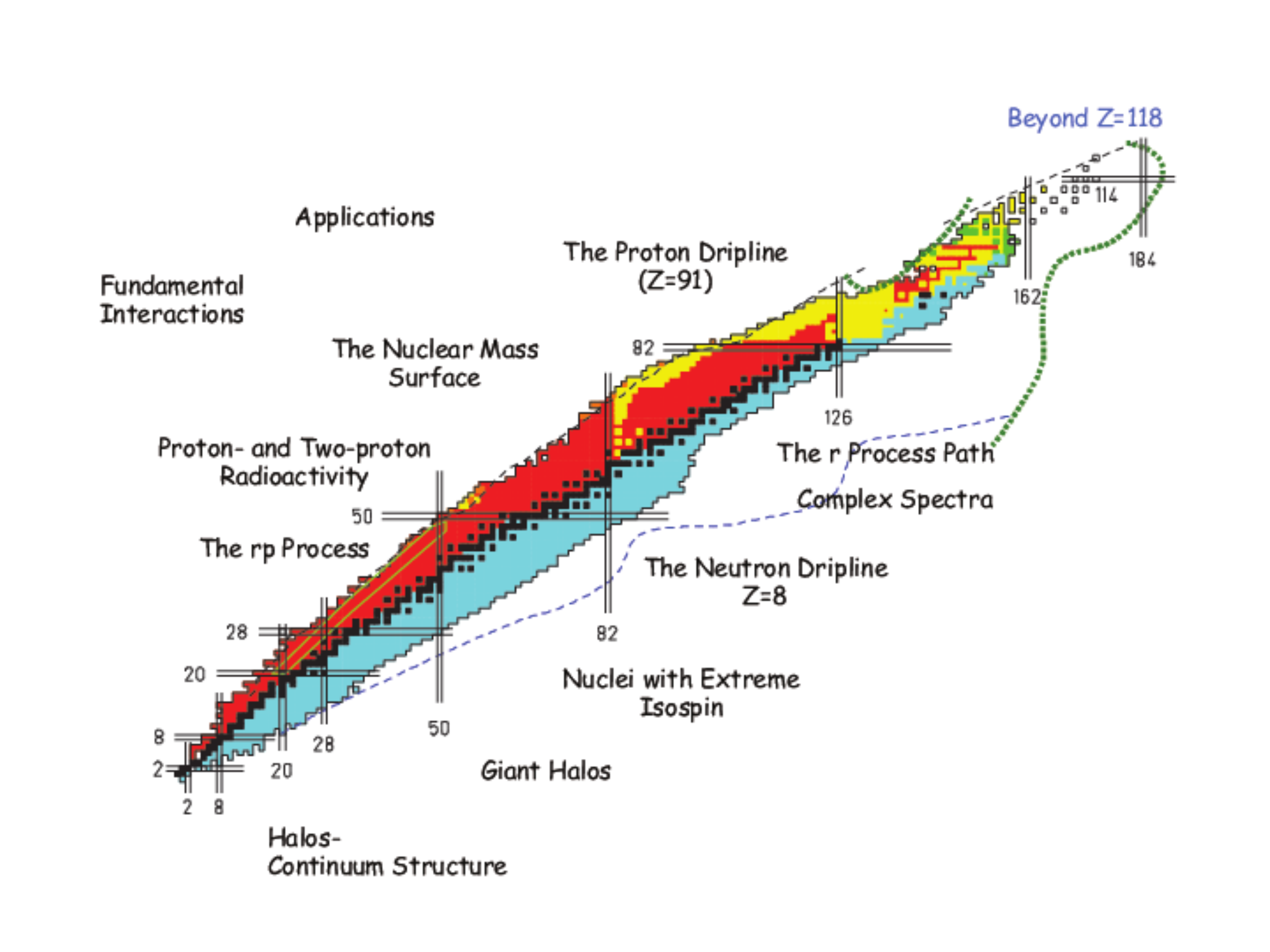}
\caption{\label{Chart} The many opportunities for nuclear physics with RIBs at  FAIR. }
\end{figure}

With the new and improved experimental equipment developed for FAIR is now up to the research groups to
select front-line experiments using the offered radioactive beams. The challenge is to
show the finger tip feeling for what can give an early scientific output.
At the same time we need to use the early phase of beam-time to test and fine-tune
the experimental equipment. The marriage between the two will lead to success. We show
in Fig.~\ref{Chart} some examples of hot subjects in different regions of the nuclear chart
and below are some examples on what one might expect.
\begin{itemize}
\item
The higher intensity at FAIR 
will advance the understanding of the structure at
the neutron dripline by giving access to heavier halo nuclei and to
unbound nuclei, maybe more than three or more mass numbers outside 
the stability line. Only Nature knows what exotic combinations of
protons and neutrons have interactions strong enough to
be classified as unbound resonances.
\item
The 3$^{rd}$ $r$-process peak by means of comprehensive measurements of 
masses, lifetimes, betas-delayed neutrons, dipole strength, and level structure 
along the $N$=126 isotones.

\item
As specific case the possible tetraneutron resonance is one of the challenges where
one finally might get an answer.  A high  $^{7}$H yield in combination with
the very efficient four-neutron efficiency expected with the new neutron
detector array will maybe provide these long-awaited experimental data that could shed
light on this problem.
\item
It has been shown experimentally that the last bound oxygen isotope is $^{24}$O, 
an isotope which also has been shown to be doubly-magic. The expected
doubly-magic nucleus $^{28}$O ($Z$=8, $N$=20) is unbound, but since shell model 
calculations, derived from microscopic $NN$ forces, would predict it to be bound it
would be highly interesting to get data for this unbound nuclear resonance. 
A broad study of the chain of the unbound oxygen isotopes $^{25-28}$O will be 
very interesting, in particular since there is a claim of the occurrence of two-neutron 
radioactivity for the unbound isotope $^{26}$O.
\item
With the high energy the full dipole response for halo nuclei will
be possible. As an example the data for $^{6}$He, limited to energies
up to about 10 MeV, would go into the region above the minimum
at 20 MeV predicted in exact six-body calculations~\cite{Bacca02}.
\item
Quasi-free one-proton knockout reactions (QFS), ($p,2p$), 
~\cite{Atar18} and ($p,pN$)~\cite{Diaz18} in inverse and
complete kinematics have recently been demonstrated as a powerful tool for systematic
studies. QFS studies will certainly be extended over large regions of the nuclear
chart at the future FAIR, from the stable nuclei out to the driplines.
\item
EoS of asymmetric matter by means of measuring the dipole 
polarisability and neutron skin thicknesses of tin isotopes with $N$ larger than 82.

\end{itemize}

\subsection{Outlook into the exotics}

With 50 years since the start of ISOLDE at CERN and 25 years since the
start of the high-energy programme at GSI it is tempting to just state that in
25 years we are just preparing for the next giant step of new developments.
But, as we have seen, the roads of science are definitively not easy to
predict, as the science of exotic nuclei have given examples on.

So, let us instead dream; in 25 years the experimental equipment for studies of the
most exotic nuclei has been completed at FAIR by finally taking the ELISe (Electron Ion scattering 
in a Storage ring, eA-collider)~\cite{Antonov2011}
experiment into operation, which will make use of a 
clean, purely electromagnetic probe, the electron, in conjunction with exotic 
radioactive beams. This new device was strongly supported by major grants from
several funding agencies after the memorable event in Stockholm  on December 10, 
in the year 20XX, where three of our present students were addressed by ``For their 
discovery of... made possible by the ingenious use of radioactive nuclear beams...''.

\section{FAIR - a new era in nuclear astrophysics}

Nuclear astrophysics aims at understanding the origin of the elements in the Universe.
This goal is intimately related to the quest to apprehend the structure and the dynamics
of the various astrophysical objects and processes which produce them. It is not surprising
that nuclear astrophysics is truely interdisciplinary. Hence progress in deeper understanding
the secrets of the universe combines astronomical observations and astrophysical modelling
with advances in nuclear physics. However, it have been the insufficient knowledge and the uncertainties of
the nuclear ingredients which often limited the desired progress. This has mainly two reasons:
1) It is often impossible to determine the nuclear ingredients directly in the laboratory
under the conditions present in the astrophysical environment. Thus nuclear modelling is required
which should, however, be based on reliable models and be constrained and guided by
experimental data as much as possible. 2) The nuclei involved in the astrophysical processes,
and often crucial for the dynamical evolution, are often short-lived isotopes with significant neutron deficiency or
excess compared to their stable counterparts. These nuclei have to be artificially produced in
the laboratory, called RIB facility, and frequently novel techniques and
instrumentation have to be developed and employed to study their properties. We will show
below that the unique combination at FAIR - high-energy primary beams from the SIS100, identification
of pure secondary beams by the SuperFragment Separator and the suite of storage rings and
instrumentation - will allow access to many of these short-lived nuclei, often for the first
time, and in this way pushing the frontier of knowledge deeper into the yet unexplored regions
of the nuclear chart (see also chapter 4). Despite these exciting possibilities, FAIR is more than a RIB facility,
as also the APPA, CBM and PANDA experiments offer unique perspectives to advance our knowledge
of astrophysical objects like core-collapse supernova and neutron stars (see chapter 2-4). {\it Hence FAIR is the
Universe in the Laboratory}.

The last two decades have witnessed that the mantra - progress in nuclear physics implies
deeper understanding of the Universe - is indeed true. Much of this progress is related
to stable beam facilities and improved instrumentation. A prominent highlight is
the solution to the 'solar neutrino problem',
i.e. the deficiency of observed neutrinos from the
Sun compared to those expected on the basis of the solar models \cite{McDonald2016}.
Here advances in solar modeling (e.g. \cite{Bahcall2001}), but also decisive experimental and theoretical progress in
determining the solar cross sections of the relevant nuclear reactions \cite{Adelberger2011}
has turned the Sun into a calibrated neutrino source for studies of neutrino physics beyond the standrard model. 
There has been also significant progress in reducing the uncertainties
related to the operation of the two key reactions in hydrostatic helium burning (triple alpha-reaction,
${}^{12}$C($\alpha,\gamma$)${}^{16}$O) in red giant stars and hence in determining the abundance
ratio of carbon and oxygen, i.e. the building blocks of life, in the Universe \cite{deBoer2017}. There are, of course, still
challenges left for stable-beam facilities as improved cross section measurements of the key reactions
of advanced hydrostatic burning stages (carbon, oxygen, silicon burning) and of the neutron sources
for the s-process, which produces half of the elements heavier than iron, are eagerly waited for.
However, with the emergence and prospect of RIB facilities or multi-purpose facilities like FAIR, it becomes possible
to experimentally study or constrain relevant nuclear physics required for a deeper understanding of the structure
and dynamics of exotic astrophysical objects like supernovae, individual and merging
neutron stars and x-ray bursters. In the following we will concentrate on how FAIR will contribute
to start a new era in our understanding of these exotic astrophysical events. This is extremely timely
as these events are all in the focus of intensive current and future astronomical and space
observations, with the discovery of gravitational waves and the accompanying electromagnetic signal
from merging neutron stars as a very recent and prime example.

\subsection{Supernovae}
Supernovae are the final fate of massive stars triggered by the collapse of its inner core
against its own gravity after this core - initially composed of the nuclei in the iron-nickel
mass region with the highest nuclear binding energy per nucleon - has run out of nuclear fusion
as its energy source \cite{Bethe1990,Janka2007}. The supernova explosion expells the outer shells of the star, including
the nuclei produced during its long life of hydrostatic burning, into the Interstellar Medium
where this matter can become birth material of future star generations, planets, ... and life.
In the very hot environment of the explosion, nucleosynthesis can occur on short timescales.
In the center of the supernova a neutron star with about 1.5-2 $M_\odot$ is born.

Obviously reliable simulations of supernova explosions and nucleosynthesis are a prerequisite
for any quantitative and consistent understanding of the origin of elements in the Universe.
This dream is likely to come true now. On the one hand, due to advances in computer hard- and software
we are at the eve of the realistic 3-dimensional simulations, reliably accounting for
convective matter flow, plasma instabilities and neutrino transport which have been identified
as crucial drivers of the explosion mechanism \cite{Just2015,Sukhbold2016}. On the other hand, RIB facilities
and in particular FAIR will decisively improve the nuclear physics input required
in the simulations. These are:

a) Electron captures on protons in nuclei are the dominant process during the collapse \cite{Langanke2003}.
Data from charge-exchange reactions and large-scale shell model calculations have impressively
reduce the uncertainty of the capture rates on the nuclei in the iron-nickel mass range (which dominate
the early core composition) from more than an order to magnitude to better than a factor 2 \cite{Langanke2000,Juodagalvis2010,Cole2012}.
At later
collapse stages heavier and more neutron-rich nuclei dominate. It has been shown that, in contrast
to earlier belief, sizable electron capture also occurs on these nuclei by allowed Gamow-Teller
transitions made possible by multi-nucleon correlations across shell gaps \cite{Langanke2003,Hix2003}. The quantitative strength
of this unblocking is yet insufficiently known, in particular for the neutron-rich nuclei
with magic neutron numbers N=50 (and 82). Ultimatively the GT strength can be directly measured
for nuclei like $^{82}$Ge in charge-exchange reactions in inverse kinematics and, at FAIR, exploiting
the storage rings ESR and HESR. Indirect information helps as well. This can be obtained by
determining the occupation numbers of the various shell-model orbitals by transfer reactions,
performed in the storage rings, by the R$^3$B experiment or by nuclear spectroscopy studies.

b) The collapse comes to a halt when the inner core (of roughly 0.5 $M_\odot$) reaches densities at or above
$2 \times 10^{14}$ g/cm$^3$, the equilibrium value of nuclear matter. At the edge of this inner core
a shock wave is created which travels outwards and initiates, supported by neutrino energy
transport, plasma instabilities and convective flow, the explosion. The energy, which is transfered
to the shock wave depends strongly on the nuclear Equation of State (EoS). 

At FAIR, the EoS will be
explored in ultrarelativistic collisions in which
a high-intensity, high-energy beam of heavy ions is impinged on a fixed target and creates, for
a short time, a hot fireball of nuclear matter, where energy is transferred into the generation of pairs
of short-lived particles. Due to their relatively long mean free paths, di-lepton pairs are well-suited
probes to study the properties of the fireball matter. This is achieved by the FAIR detectors HADES and
CBM, where CBM due to its unprecedented event rate capabilities will also observe rare probes
like particles with charm quarks created in the fireball \cite{Ablyazimov2017}. HADES has already been successfully
exploited in heavy-ion runs at GSI/SIS18, delivering primary results on potential modifications
of resonances embedded in the nuclear medium \cite{Seck2017}. 

\begin{figure}[t]
\includegraphics[width=150mm, angle=0]{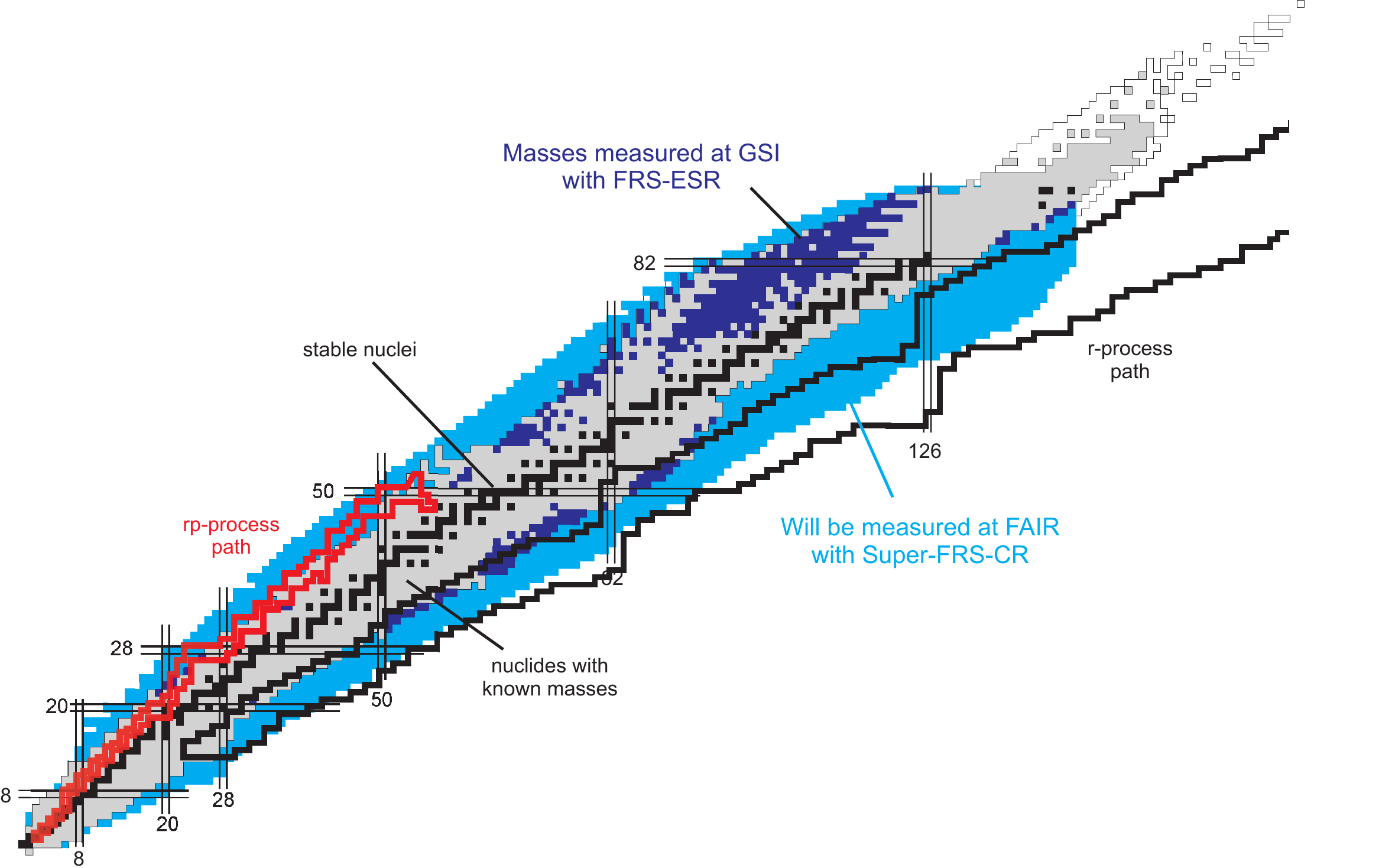}
	\caption{\label{fig:RIB-masses} The FAIR reach in nuclear masses (light blue). Masses measured at GSI are shown in dark blue }
\end{figure}

The other approach, by which FAIR will contribute
to the understanding of the EoS at supernova conditions is based on the observation of the dipole
strength distribution in neutron-rich nuclei \cite{Adrich2005}.
The symmetry energy is a crucial ingredient of EoS for asymmetric nuclear matter and hence for the neutron-rich
conditions encountered in supernova explosions. It has been shown that the symmetry energy
scales with the dipole polarizability and the difference between nuclear mass and charge radii \cite{Brown2000}.
These radii for short-lived nuclei can be measured indirectly in collision experiments and,
at a later stage of FAIR, by electron scattering, respectively. The dipole strength distribution is a
key observable of the R$^3$B experiments by means of Coulomb excitation. In nuclei with
neutron excess noticeable dipole strength is shifted from the giant dipole resonance to lower
energies. This low-lying strength (often called pygmy strength) is also sensitive to
the symmetry energy. At R$^3$B this route will be followed, studying the pygmy dipole strengths.

c) Matter is ejected from the hot surface of the freshly born neutron star as free protons and neutrons,
which, upon reaching cooler regions, are assembled to nuclei. The resulting abundance distributions
depend strongly on the ratio of protons and neutrons, which is set by interactions with neutrinos and
antineutrinos streaming out of the neutron star and depend sensitively on the neutrino energy distributions
and luminosities. Due to current understanding, the ejected matter is proton-rich at early times and
slightly neutron-rich at later times giving rize to the $\nu$p process \cite{Froehlich2005,Pruet2005} and a weak r-process, which produces
r-process elements only up to the second abundance peak at $A \sim 130$, respectively. The short-lived nuclei
involved in these processes can be generated at FAIR and other RIB facilities and
their properties (i.e. masses (see Fig.~\ref{fig:RIB-masses}), half-lives, capture cross sections) will be determined. Here the storage
rings at FAIR are a special asset allowing reaction studies in inverse kinematics.

\subsection{Neutron stars}

Neutron stars are genuine 'exotic' objects with masses upto 2 $M_\odot$ and radii of about 12-15 km.
Binaries of neutron stars rotate around each other on spiral orbits \cite{Hebeler2013}. Their final merging is a powerful
source of gravitational waves and also a source of the heaviest elements in the Universe
via the r-process (see below).

All four experimental FAIR collaborations will advance our knowledge about neutron stars and
neutron star mergers. The surface of the neutron star can be the source of tremendous
electromagnetic fields, with exciting impact on the structural forms of matter
(e.g. \cite{Neuhauser1986,Negreiros2018}). In the FAIR storage rings it is
possible to generate such extreme electric field strengths between ions passing each other at velocities
close to the speed of light and to study electron dynamics and correlations
(SPARC/APPA collaboration, see chapter 6).
The crust of the neutron star is composed of individual nuclei. With increasing density it is energetically
favorable to capture electrons by nuclei which in this beta equilibrium become hence increasingly neutron-rich. The crust
composition is then determined by the masses of neutron-rich nuclei which will be produced and measured at RIB facilities,
including the NuSTAR collaboration at FAIR (see chapter 4). Going deeper into the neutron star, electrons will be replaced
by muons and also hyperons are mixed into the composition. The PANDA collaboration has the exciting
perspective to produced $\Lambda$ and $\Lambda \Lambda$ hypernuclei with unprecedented rates and to
advance our knowledge of the $\Lambda N$, $\Lambda \Lambda$ and $\Lambda NN$ interactions (e.g. chapter 2). In the inner
neutron star core, where densities of a few times the equilibrium value of nuclear matter are expected,
the form of matter and its composition is largely unknown. Nuclear matter at such extreme densities,
although at somewhat higher temperatures than in neutron stars, can be probed by the CBM experiment
using ultrarelativistic heavy-ion collisions (see also chapter 3). The data obtained by CBM and other relativistic
heavy-ion collision experiments constrain also the nuclear EoS at the high temperature/high density conditions needed for neutron star
merger simulations.

\subsection{R-process nucleosynthesis}

About half of the elements heavier than iron and all transactinides in the Universe are produced
by the astrophysical r-process by a sequence of rapid neutron captures and beta decays. The process
occurs in an environment with extreme neutron densities, as they are encountered in
merging neutron stars and core-collapse supernovae, where the latter will likely only contribute to the
synthesis of the lighter r-process nuclei up to the mass range $A \sim 130$ (second r-process peak).
The nuclei on the r-process path
are all short-lived and have such large neutron excess that
most of them have not yet been produced in the laboratory. Hence their properties have to be modelled
introducing significant uncertainties into r-process simulations. This situation will drastically improve
once the next generation of RIB facilities is available.

The most important nuclear properties required for r-process simulations are
neutron separation energies (masses) which, for given environment conditions,
determine the nuclei on the r-process path and  beta half-lives which influence the dynamics of the process.
Neutron captures become relevant once the process stops to operate in $(n,\gamma) \leftrightarrow
(\gamma,n)$ equilibrium. For the r-process occuring in neutron-star mergers the matter flow carries
significant amout of material into the range of superheavy neutron-rich nuclei which decay by fission
(and alpha decay) requiring knowledge of fission rates and yields including the number of neutrons set free
during the fission process.


\begin{figure}[t]
\begin{center}
\includegraphics[width=0.6\textwidth, angle=0]{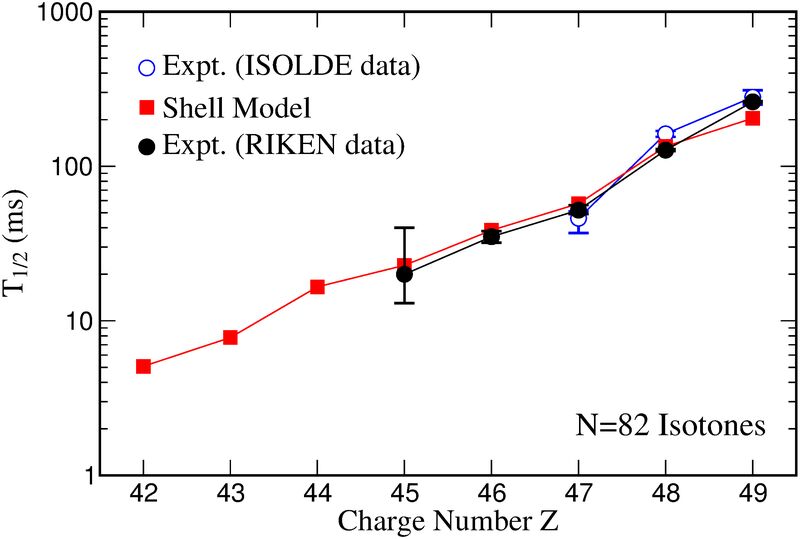}
	\caption{\label{fig:n82} 
	Halflives of r-process nuclei with neutron number $N=82$ measured at RIKEN \cite{Lorusso2015}
		compared to earlier data from CERN/ISOLDE and to predictions from large-scale shell model calculations \cite{Zhi2013}.}
        \end{center}
\end{figure}

\begin{figure}[t]
\begin{center}
\includegraphics[width=0.4\textwidth, angle=0]{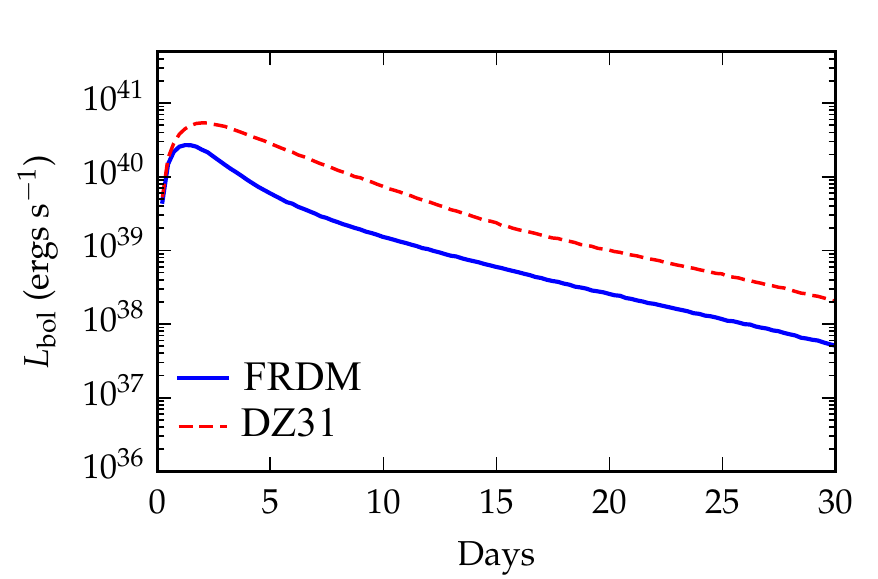}
\includegraphics[width=0.4\textwidth]{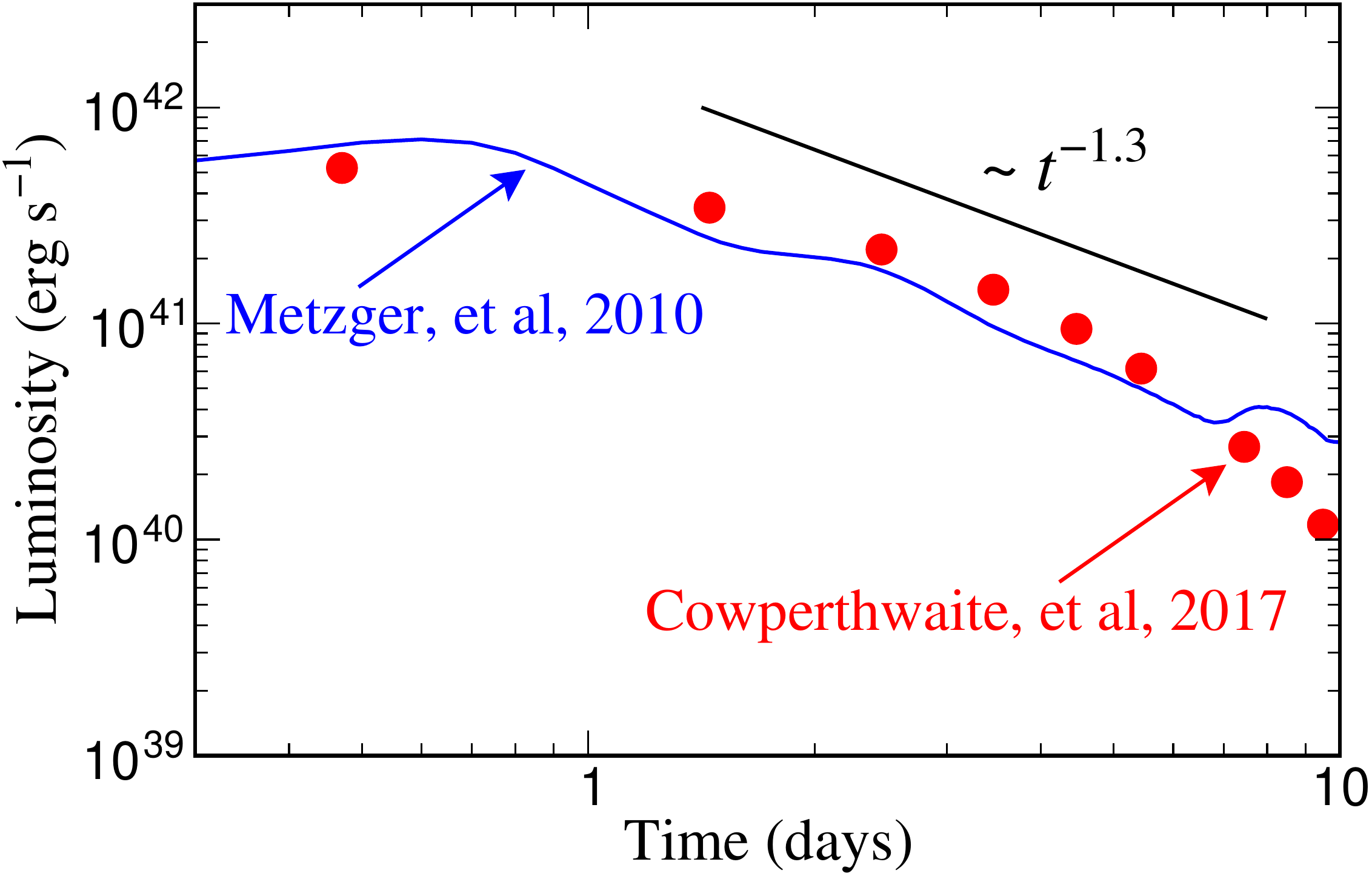}
	\caption{\label{fig:lightcurve} 
	(left) Kilonova lightcurve predictions from r-process simulations in a neutron-star merger scenario
		using two different nuclear mass sets.
        (right) Comparison of he observed lightcurve of the NS merger GW170817 (red dots, \cite{Cowperthwaite2017}) compared with the prediction (blue line, \cite{Metzger2010}). As the lightcurve is produced by an ensemble of decaying nuclei, it follows a power law rather than an exponential as is typical for supernovae lightcurves, which are dominated by the decay of an individual nucleus at a time.} 
        \end{center}
\end{figure}

The importance of experimental data for r-process simulations has recently been impressively demonstrated
by a campaign at RIBF in Riken which determined
the half-lives of many lighter r-process nuclei, including a few for the waiting-point nuclei
with magic neutron numbers $N=82$ (Fig. \ref{fig:n82}), with significant impact
for the dynamics of r-process simulations and abundance productions \cite{Lorusso2015}. For the waiting point nuclei
in the third r-process peak at $A \sim 195$, associated to the magic neutron number $N=126$,
no data yet exist. This regime will be accessible at FAIR, made possible by its high-energy, high-intensity
primary beams and the Superconducting Fragment Separator. The measurements will additionally
profit from the storage rings which, for example, allow precise mass determinations by Schottky frequency
measurements. Half-life measurements for r-process nuclei at and close to the $N=126$ magic neutron number
are crucial, as they have not only impact on the abundances in the third r-process peak,
but also regulate the mass-flow to heavier nuclei which either fission or decay by $\alpha$ emission \cite{Mendoza2015,Metzger2010}.
In the neutron-star merger scenario
fission of these heavy nuclei produces the second r-process peak around $A \sim 130$. Furthermore,
$\alpha$ decay is the potential source of the lightcurve (see Fig. \ref{fig:lightcurve}) which, for so-called kilonovae,
has served as the direct observable for active r-process nucleosynthesis in neutron-star mergers \cite{Metzger2010}. Theoretical studies
have recently argued that forbidden transitions will contribute significantly to the half-lives and will
also have considerable impact on the beta-delayed neutron emission rates. These and
half-life measurements of $N=126$ r-process nuclei will be possible exploiting storage rings,
but also by targeted gamma-spectroscopy campaigns. First data are expected already from experiments
planned for the FAIR Phase-0 program.

\subsection{Outlook}

25 years ago, the 'solar neutrino problem' constituted one of the major challenges
in nuclear astrophysics, also caused by  uncertainties in the cross sections of the solar
hydrogen reactions. The problem has been solved after
SNO and other solar neutrino detectors have confirmed neutrino oscillations and simultaneously
the validity of the Standard Solar Model.
Furthermore, cross section measurements of the reactions involved
in solar hydrogen burning performed at the underground facility LUNA or with improved
instrumentation and shielding have reduced the nuclear uncertainties to a level to turn the
Sun into a calibrated neutrino source which can be used to explore particle physics
beyond its standard model.

What can we expect in 25 years once the scientific harvest at FAIR and the next-generation RIB facilities
comes to fruit?

\begin{itemize}
\item
The direct observation of gravitational waves from compact astrophysical objects
		belongs to the major recent breakthroughs in science.
FAIR will advance our understanding of the Equation of State
of dense and hot matter and, together with progress in multidimensional simulations,
holds the prospect to exploit gravitational wave signals from such events
for searches of new physics.

\item
The RIB facilities will replace theoretical model data by experimental facts for most
of the nuclear physics input required for  simulations of the astrophysical r-process.
This will reduce the nuclear uncertainties to a level that simulations become
such a stringent constraint to allow for detailed analysis of merging neutron stars,
which have been shown to be the site
of the r-process, solving one of the millenium questions (i.e. how are the elements from
iron to uranium being made) as put forward by the US National Research Council (see Discover Magazine, February 2002).

\item
Observation of time-resolved supernova neutrino spectra by earthbound detectors
will check and advance our understanding of the supernova explosion mechanism. Furthermore it
will constrain the initial proton-to-neutron ratio which is a key ingredient
for the associated explosive nucleosynthesis. Provided the nuclear input has been measured, the nucleosynthesis
studies can be turned into measures of the dynamics and properties of the environment on top of the freshly born
neutron star.

\end{itemize}

We are bound in the coming years to unravel many mysteries related to matter at its extremes
in astrophysical objects, made possible by the triad of novel astronomical observation, advanced astrophysical
modelling and experimental (and improved theoretical) excess to exotic nuclei at FAIR and elsewhere.
Obviously we are at the eve of a new era in nuclear astrophysics.

\section{FAIR - Atomic physics from {microkelvin} to {petakelvin}}

With the development  of accelerators and ion sources that happened in the last 40 years, new possibilities have opened up for atomic physics, which have allowed to explore many uncharted territories.
With the successive help of machines like Van de Graff to the most advanced storage rings, it has been possible to study few-electron ions from light elements to uranium.
In parallel the developments of Electron-Beam Ion Traps (EBIT) and Electron-cyclotron Resonance Ion Sources (ECRIS) have also lead to advances in the field.
We have now tons of measurements for transition energies, line intensity ratios, transition rates, hyperfine structure and Landé g-factors, for all ions up to hydrogenlike uranium.
This quest to measure all these atomic parameters, in relatively simple systems, has been fed by one very important and fundamental physics goal: test bound state QED (BSQED) in a variety of conditions.
Even with such a long history, which started back in 1947 with the seminal experiment of Lamb and Retherford \cite{lar1947}, there are many issues still to address in bound state QED.

One can think, e.g. of the  unsolved, 8 years old, \emph{proton size puzzle} \cite{pana2010,ansa2013}, with a $7 \sigma$ discrepancy between the proton charge radius derived from the Lamb shift in muonic hydrogen and the average of all the measurements in normal hydrogen (for a different view, see e.g. Ref.~\cite{lorenz2012}). This problem, which now extends to the deuteron \cite{pnfa2016}, is a reminder that BSQED is not fully understood. The situation has not become clearer, as a new measurement of the $2S-4P$ transition in hydrogen \cite{bmmp2017}, provides a proton size in agreement with the muonic hydrogen value, while an improved measurement of the $1S-3S$ transitions is in agreement with the hydrogen world-average value\cite{fgtb2018}. 

More recently the simultaneous measurement of the hyperfine structure of the ground state of hydrogen-like and lithium-like bismuth at GSI \cite{uabd2017} has shown a $3\sigma$ deviation from the expected value.

\begin{figure}
\centering
	\includegraphics[clip=true,width=6cm]{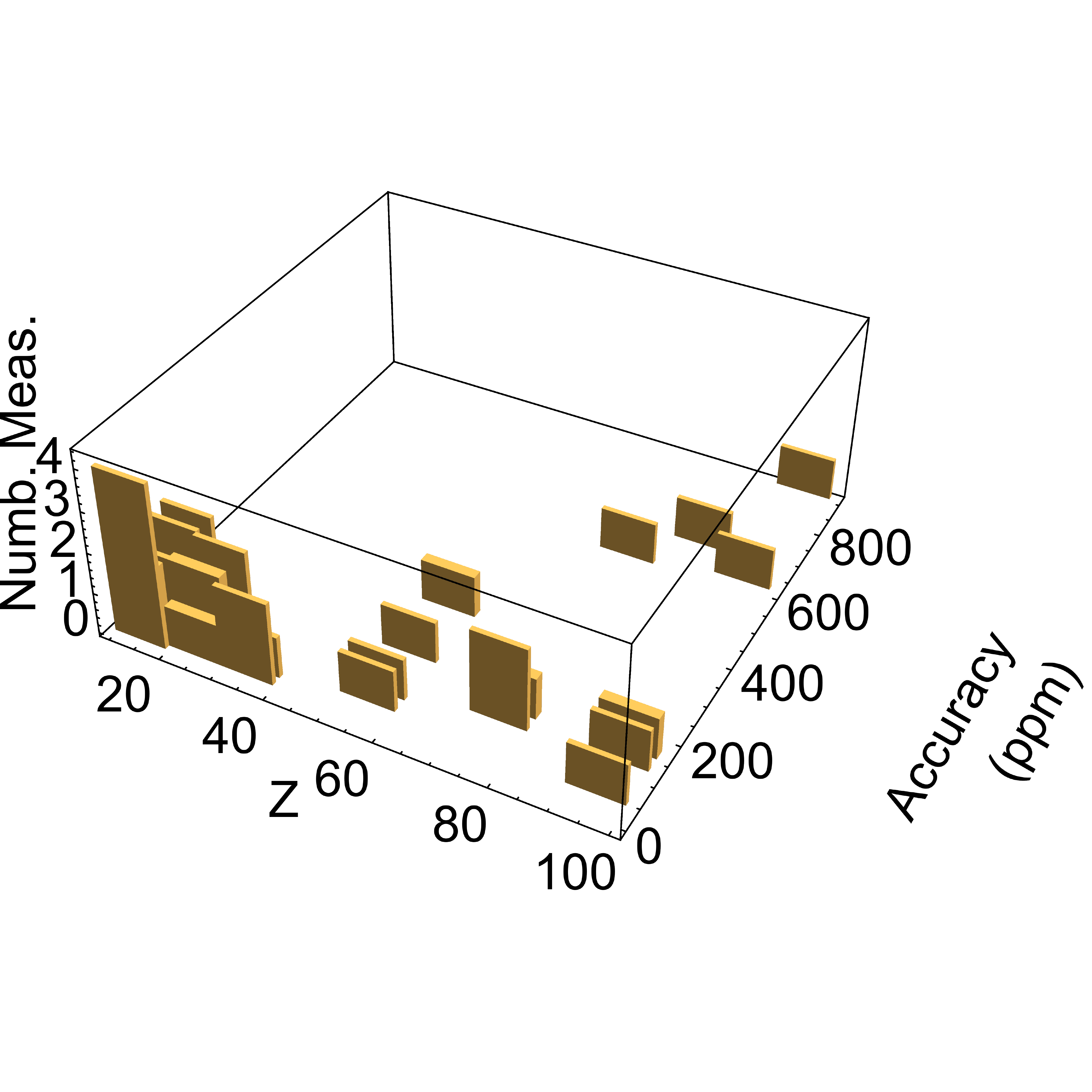}%
	\includegraphics[clip=true,width=6cm]{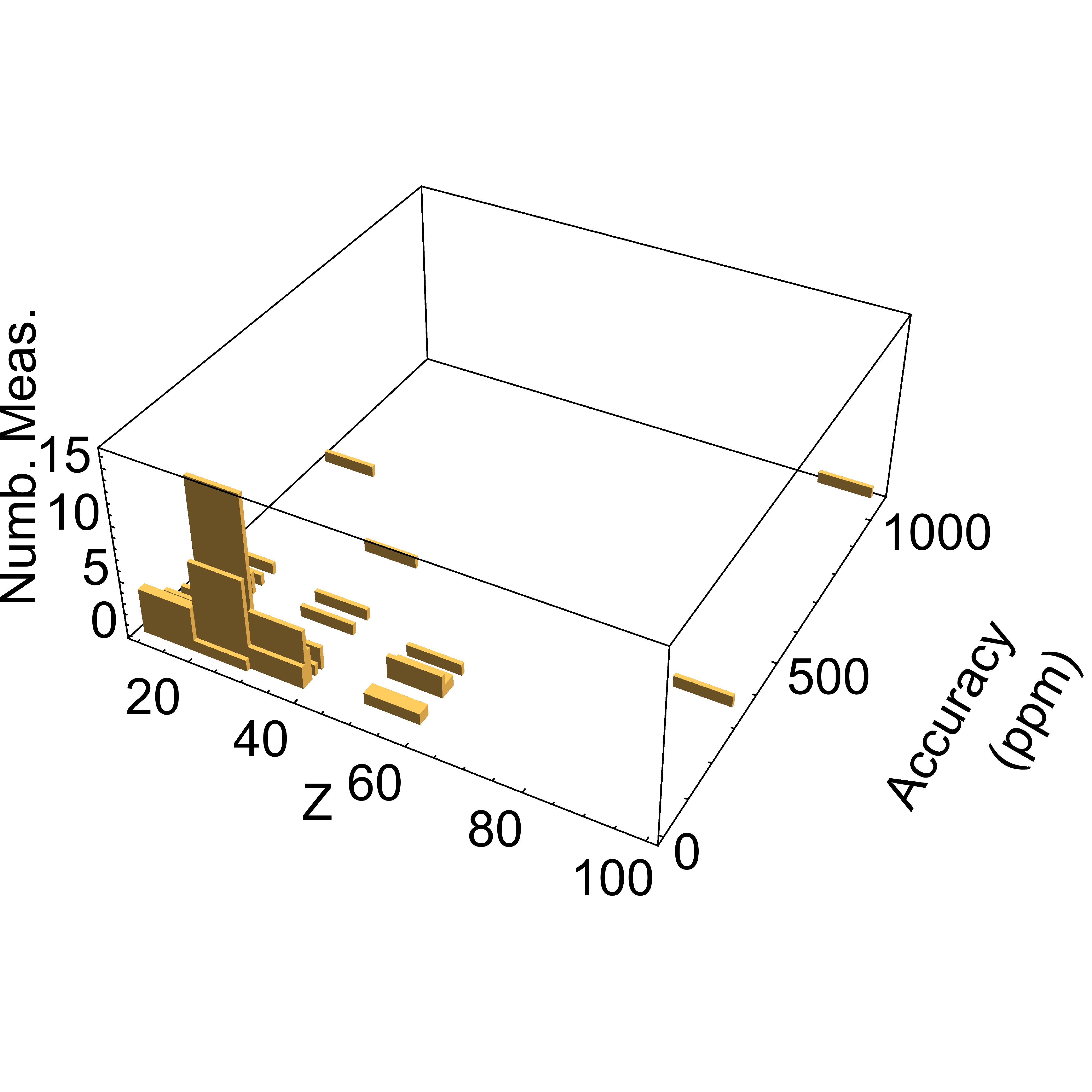}
\caption{
Histogram of all the measurements of the $n=2 \to n=1$ transition energies in hydrogenlike (left) and heliumlike ions (right) since 1970 as a function of accuracy and atomic number.
\label{fig:accuracy-h-he}
}
\end{figure}

Yet, even after  so many years of measurements of transition energies  in the medium and high-$Z$ region, there is still a strong need of more accurate measurements. If one looks at $n=2 \to n=1$ transition energies in one and two-electron ions, 
for example, there are only 4 and 5 measurements respectively, with accuracy below {10} {ppm} in the $Z \ge 10 $ region  (Figs. \ref{fig:accuracy-h-he}). Moreover, one can see that  high-precision   measurements are available only for  medium-$Z$ ions. The most precise high-$Z$ measurement is for uranium with a {41} {ppm} accuracy \cite{gsbb2005}. The situation is comparable for the $2p-2s$ transitions in lithium-like ions.

The need to have more accurate measurements at high-$Z$ originates in the fact that relativistic and QED effects scale as high powers of $Z$. 
For example the self-energy scales as $Z^4$ while the non-relativistic energy scales only as $Z^2$. 
The speed of the $1s$ electron in an atom is $\alpha Z$, where $\alpha\approx 1/137.036$ is the fine structure constant. 
Going from hydrogen to uranium thus corresponds to a speed varying from {0.7}$\%$ of the speed of light to {66} $\%$, which explains the strong enhancement of relativistic effects. 
At the same time, the nuclear electric field close to the nucleus becomes comparable to the Schwinger pair creation limit, which implies a possible breakdown of perturbation theory. In order to find possible systematic $Z$-dependent deviations with theory, there is  a need for much better measurement accuracy at very high-$Z$.

The other reason to study QED at high-$Z$  is connected to the fact that for $Z\to 1/\alpha \approx 137$, the point nucleus Dirac equation solution energy for the $1s$ level, $E_{1s}=mc^2 \sqrt{1-(Z\alpha)^2}$  becomes singular.
When taking into account the volume of the nucleus, what actually happens is that the $1s$ level energy becomes lower than $-mc^2$, and dives into the negative energy continuum for $Z\approx 173$. This should lead to the creation of electron-positron pairs (sometimes called decay of the vacuum) \cite{zap1972,mrg1972}.

The study of such qualitatively novel effects can be extended to
quantities other than energy, e.g. hyperfine structure, to test QED
in ultra-high magnetic fields, or Landé g-factors. Combining
measurements of these different operators is also important to acquire
a better understanding of the nucleus-electron cloud interaction (see,
e.g. \cite{dkmm2018}). Moreover accurate measurements in a wide range of $Z$ can lead to new methods for the determination of the fine structure constant \cite{sgor2006,ybht2016} or electron to proton mass ratio \cite{skzw2014,zsks2017}.

The physics that can be potentially impacted by precision experiments on atoms in general, on exotic atoms, or on highly-charged ions is not limited to tests of QED.
The most obvious case is astrophysics. For example experiments on highly charged ions were necessary to solve the puzzle of the intensity ratios in Ne-like iron ions observed in astrophysical plasmas \cite{bbrs2012}. In the same way,  laboratory measurements \cite{sdbs2016} provided a simple explanation of the \SI{3.5}{\kilo\electronvolt} x-ray lines observed with the XMM-Newton space x-ray telescope \cite{bmfs2014,brif2014}, proposed as possibly due to the decay of dark matter particle candidates.

There are nowadays many issues, which require new directions of research to find new ways of testing the standard model and its extensions. For example, after several years of LHC research, and the discovery of the Higgs boson, there are still no traces of supersymmetric particles nor candidates for dark matter. Direct search for dark matter particles has not been more successful either.  The recent simultaneous observation of a two neutron-stars merger  with gravitational waves, gamma rays and visible and infrared  light has put severe constraints on dark energy models. A better understanding of QED, of the structure of the vacuum or deviations from theory that could be connected to new interactions, would help sorting out some of the issues involved.
There is thus a need, stronger than ever, to explore new sectors of physics, at lower energy, providing high-precision measurements where new physics could be observed.

There is at the moment only one facility in construction able to achieve some of these physical goals - FAIR. Super EBIT facilities have produced bare or few-electron uranium ions. Yet they did not  allow making accurate measurements of $n=2 \to n=1$ transitions \cite{mek1994} or even of $2s-2p$ transitions for two-electron ions \cite{beos1996}. Only high-energy storage rings have the capacity to produce few electron heavy elements like uranium ions in large enough quantities for precision spectroscopy. For more than two decades, the GSI ESR (experimental storage ring) has provided such very heavy element beams (gold, lead, bismuth, uranium), from bare to Li-like ions. FAIR \cite{slbl2014,gsl2015,sbbb2015} will extend the possibilities further. The facility will provide higher energy beams that can be used directly or  stored, cooled and decelerated. While the ESR ring will still be available, it will be complemented by two rings, the HESR (high-energy SR) and CRYING, a low-energy ring, previously in Stockholm, that can be fed highly charged very heavy ions  \cite{laab2016}. At FAIR, using the  HITRAP beam line \cite{hbek2006,habc2015}, it will also be possible  to decelerate ions even further,  and store them in cryogenic Penning traps, allowing for high precision measurements with ions cooled at cryogenic temperatures. In the future, such ions could even be cooled to \si{\micro\kelvin} temperature by sympathetic cooling. This opens up new possibilities for high-precision measurements, including heavy ion-based atomic clocks, very insensitive to external perturbations because of the high binding energy of the electrons concerned \cite{sch2006,ddf2012}. Such frequency standards would allow for very fundamental tests on, e.g., fundamental constants drifts \cite{dssf2015}.

Some parts of the FAIR facility, like CRYRING or HITRAP are already installed and in the commissioning stage and new experiments will start in 2018.
One of the most difficult aspects of doing precision measurements on high-energy storage rings is connected to the Doppler effect. The new facility will benefit from new possibilities to measure the speed of the ions by using PTB-callibrated high-voltage dividers, which will allow to measure the cooling electron speed, and thus the ion speed to \SI{10}{ppm} accuracy  \cite{hbde2014,uabd2017}.

With its collection of beams, storage rings and traps, the FAIR facility will allow to perform a variety of experiments with improved accuracy and statistics. The variety of methods possible will allow for cross-check of different results in a way that was never possible before.

The first method that will be improved is the measurement of transition energies using resonant coherent excitation in crystals. The high energy beams available at FAIR will allow to extend the range of transition energy measurements to much heavier elements, with transition energies up to \SI{100}{\kilo\electronvolt}. The recent improvement of this method with 3D-RCE will allow for simpler measurements  \cite{nshn2012}.

The use of CRYRING \cite{laab2016}, which will allow to use strongly decelerated beams, will lead to improved measurements of transition energies, with smaller errors due to the Doppler effect. It will improve measurements of  transition energies using electron capture in the cooler, as performed at ESR in the past\cite{bmlg1995,gsbb2005}. The control of the Doppler effect in CRYRING for this method will also be improved, since it will be possible to align the detectors exactly at \SI{0}{\degree} and \SI{180}{\degree}. Moreover the use of microcalorimeters will allow to improve the resolution compared to Ge detectors by more than one order of magnitude \cite{pshk2012,kabe2017}.

The use of crystal spectrometers at the gas target of CRYRING will also benefit from the lower beam energy and improved energy measurement. It will be possible to use both reflection-type crystal spectrometers for low-energy x-rays \cite{tkbi2009} and Cauchois-type transmission spectrometers like FOCAL \cite{cbls2006,bgth2015} to cover the range from a few \si{\kilo\electronvolt} to \SI{100}{\kilo\electronvolt}.

Finally the use of dielectronic recombination with the electrons of the coolers will also allow for precise measurements of transition energies, and compare different isotopes very precisely\cite{bkhm2008}, including short-lived radioactive ones \cite{bklm2012}.

It should be noted that the high-precision QED tests considered here depend considerably on our understanding of a variety of nuclear effects, the shape and size of the charge distribution, the nuclear polarization or the magnetic moment distribution when hyperfine structure is concerned. This would mean that complementary studies, for example with muonic atoms must be conducted in parallel to improve the quality of nuclear properties knowledge (see, e.g. Ref. \cite{abbi2015}).

Laser spectroscopy will also be a major line of research at FAIR. Laser cooling at SIS100 will allow to provide cooled high energy beams \cite{wbbd2015} and provide high-accuracy spectroscopy of lithium like ions.
The spectroscopy of hyperfine structure transitions in one and three-electron ions, of the kind mentioned in the introduction is going to remain a major endeavor for the new facility, in order to understand the discrepancy reported in Ref. \cite{uabd2017}. There are also transitions between metastable states of few electron-ions that can be studied with laser spectroscopy (see, e.g. Ref. \cite{wksi2011}).

With the HITRAP facility a whole new way to perform QED tests will be possible. The HITRAP facility \cite{habc2015} will provide the low-energy beams that can be used in a variety of experiments with trapped ions. The ARTEMIS experiment will allow to perform laser and microwave double-resonance spectroscopy, yielding g-facors and nuclear magnetic moments. The SpecTrap will allow to measure the hyperfine structure of  highly-charged ions by laser spectroscopy. The ions will be cooled to \si{\micro\kelvin} by sympathetic cooling with Mg$^+$, allowing for an unprecedented accuracy. In the future, the implementation on HITRAP of the latest advances in mass measurement techniques in Penning traps, that have allowed to measure the mass of the proton to a relative accuracy of \num{2.89E-11},  will allow  measurements of the total mass of highly charged ions with a few eV accuracy. Comparing the mass of bare, hydrogen-like or helium-like heavy ions would directly yield binding energies. There are many other tests of fundamental physics that could be performed, like, e.g. parity-non conservation experiments in helium-like ions \cite{ssim1989,msgi1996}.

FAIR  will help to achieve significant advances in our understanding of the vacuum decay and on the structure of super-heavy atoms. It was found many years ago that to derive the binding energy of a compound nucleus through quasi-molecular x rays in a collision, one had to have a projectile with at least one K hole (see, e.g. Ref. \cite{anh1985}). This was demonstrated for the first time on Cl$^{16+}$ on Ar collisions \cite{sstj1985}. It requires first accelerating low-charge ions to high energy to produce the bare or hydrogen-like ion, and then to decelerate them down to energies around the Coulomb barrier to observe the quasi-molecular x-rays.  The detection in coincidence of the x rays emitted by the collision of the highly-charged ion with an atom of a target in a storage ring, and of the recoil ion to obtain the impact parameter will allow to obtain the $1s$ binding energy of any compound nucleus in such a collision. Preliminary tests on Xe$^{54+}$ on Xe have been performed at the ESR \cite{gdbg2010}. One of the main difficulties will be to design targets compatible with insertion in a storage ring to allow for cases like U$^{91+}$ on U collisions.

In conclusion, the start of the FAIR facility will provide many new opportunities to develop highly-charged ion physics, for both tests of fundamental theories, and applications to other fields of physics like astrophysics. Many more opportunities that could be accounted for here will become possible as and when the different facilities will become available.

\section{Low-energy antiprotons at FAIR}

From early on in the development of FAIR it was recognized that with the existing FAIR storage rings also a facility for low-energy antiprotons can be created \cite{Widmann:2005vz} that would supersede the at that time existing facility, the Antiproton Decelerator at CERN \cite{Baird:1997qd}, in intensity and lowest energy. Thus the idea of FLAIR, the Facility for Low-energy Antiproton and Ion Research, was created in 2005 \cite{FLAIR:2005}. 
When FAIR was started, however, FLAIR was not included into  its phase 1  that is currently under construction.

Originally FLAIR was designed to use the new NESR storage ring to decelerate antiprotons and consisted of two dedicated storage rings LSR and USR and the HITRAP facility. For the Low-energy Storage Ring LSR, CRYRING was chosen, while the electrostatic Ultra-low energy Storage Ring USR is a new development similar to the Cryogenic storage Ring CSR \cite{vonHahn:2016wjl} which in the mean time was realised at MPI-K Heidelberg. The core features of FLAIR are the simultaneous use of the facility for low-energy antiprotons and highly charged ions, and the availability of both fast and slow extracted antiprotons at energies down to 20 keV as well as at rest in HITRAP.

In the mean time, CRYRING was moved to GSI/FAIR and installed behind the ESR \cite{laab2016} where also HITRAP \cite{hbek2006,habc2015} is now coming into operation, as described in the previous section. This way, two of the three main components of FLAIR have already been realized and will start the physics program with highly charged ions. A study was launched to evaluate the feasibility of bringing antiprotons back to the ESR and to create the full FLAIR facility at this location. It concluded that it is in principle indeed feasible to transfer antiprotons through the HESR into the ESR at a rate similar to what is now available at the CERN-AD (in average $\sim 2\times 10^5 \, \overline{\mathrm{p}}$/s) and what was estimated initially for FLAIR, needing only a transfer line from HESR back to the ESR \cite{Widmann:2015lna,Katayama:2015wsm,Stoehlker:2015}. In this way, also the antiproton part of FLAIR could be created at much lower cost than originally foreseen.

Experiments using CRYRING and HITRAP with highly charged ions have already been described in the previous section. The physics program of FLAIR with antiprotons consists of several topics: \textit{i)} tests of CPT symmetry and the weak equivalence principle using antihydrogen and atoms containing antiprotons, \textit{ii)} atomic collisions with antiprotons, and \textit{iii)} the usage of antiprotons as nuclear probes. Antihydrogen studies are the main topics at CERN, where the AD is now being extended by ELENA \cite{Oelert:2017mud}, a storage ring with similar parameters as CRYRING albeit with a slightly lower energy of 100 keV. ELENA is currently being commissioned and is expected to go into operation in 2018 for the GBAR experiment and after the CERN long shutdown, in 2021 for all experiments, providing pulsed antiproton beams at fixed energy of 100 keV. Recently the ALPHA collaboration has published the first spectroscopy results of the 1s--2s laser spectroscopy \cite{Ahmadi:2016fir} as well as the ground-state hyperfine splitting \cite{Ahmadi:2017gwe}, still with moderate precision but improvements are expected in the near future. Also three experiments will from this year on work on gravity measurements with antihydrogen, so that the topic \textit{i)} will see a rapid progress at CERN in the years to come, in spite of the frequent long shutdowns at CERN. With the current number of experiments the available space at AD/ELENA is now fully occupied, leaving little opportunities for new ideas.

\begin{figure}
  \centering
  \includegraphics[width=12cm]{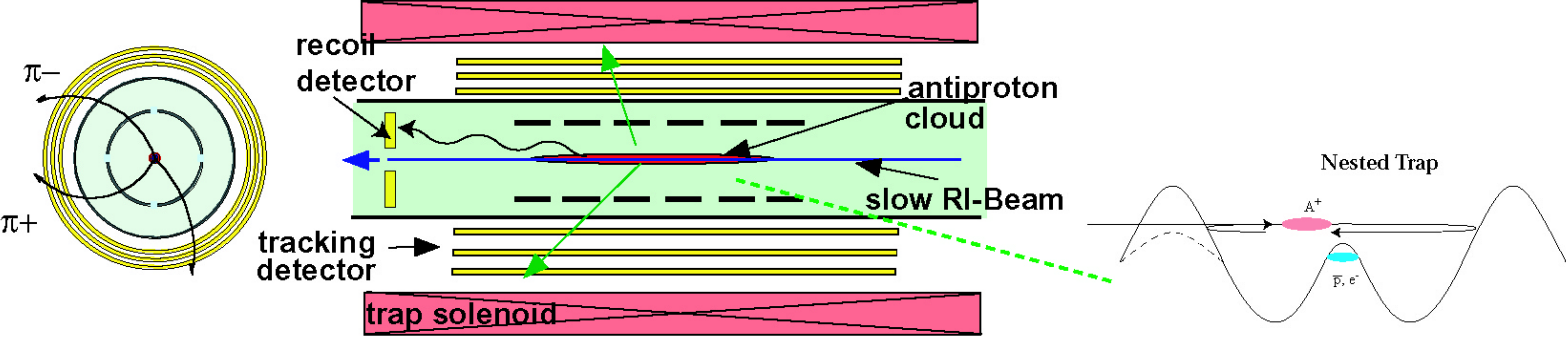}
  \caption{\label{fig:EW-exo+pbar} Schematic setup for the experiment Exo+pbar as proposed for FLAIR \cite{Wada:2004ee}. Antiprotonic atoms are formed in a solenoid magnet (center) containing a nested Penning trap (right). The proposed detection method is to track charged particles inside the solenoid (left) and to extract halo factors from the number of charged pions produced in the annihilation of the antiproton with a nucleon at the nuclear surface. From \cite{Widmann:2015lna}.}
\end{figure}

The remaining physics topics of the FLAIR program cannot be done at ELENA due to the lack of an internal target and of slow extracted beams. They would therefore be uniquely possible at FLAIR. Using an internal target and a so-called reaction microscope, as is planned to be installed in CRYRING in the near future \cite{laab2016}, kinematically complete atomic collision processes (topic \textit{ii)}), {\em e.g.} ionization processes, could be studied \cite{Welsch:2008zzd}. Using antiprotons as probes constitutes the cleanest system for theoretical studies due to lack of screening effects of the projectile, therefore these measurements would provide benchmark data for theoretical models. Topic \textit{iii)}, using antiprotons as nuclear probes, needs slowly extracted beams as will be available at FLAIR. A very attractive scheme proposes to study nuclear halo effects using antiprotonic atoms as was already done at LEAR \cite{Trzcinska:2001sy}, but extending this method to unstable isotopes, and to determine halo factors by distinguishing $\overline{\mathrm{p}}$p from $\overline{\mathrm{p}}$n annihilations by the number of charged pions created in each event \cite{Bugg:1973zz}. The proposal for FLAIR \cite{Wada:2004ee} is based on nested Penning traps like they are used in the formation of antihydrogen to generate these antiprotonic atoms (cf. Fig.~\ref{fig:EW-exo+pbar}), thus requiring a connection to both a source of antiprotons and unstable nuclei. Due to the opposite electric charge, the same storage rings cannot be used for the deceleration of two species and the isotopes have to be obtained from a stopping cell located in close proximity. A different approach towards such a measurement has been proposed for CERN \cite{Obertelli:2017}, where antiprotons are supposed to be brought in a movable Penning trap from the AD to ISOLDE and merged with unstable ions there. The attempt of performing such a technically extremely challenging experiment shows the big interest in such measurements.

\begin{figure}
  \centering
  \includegraphics[width=10cm]{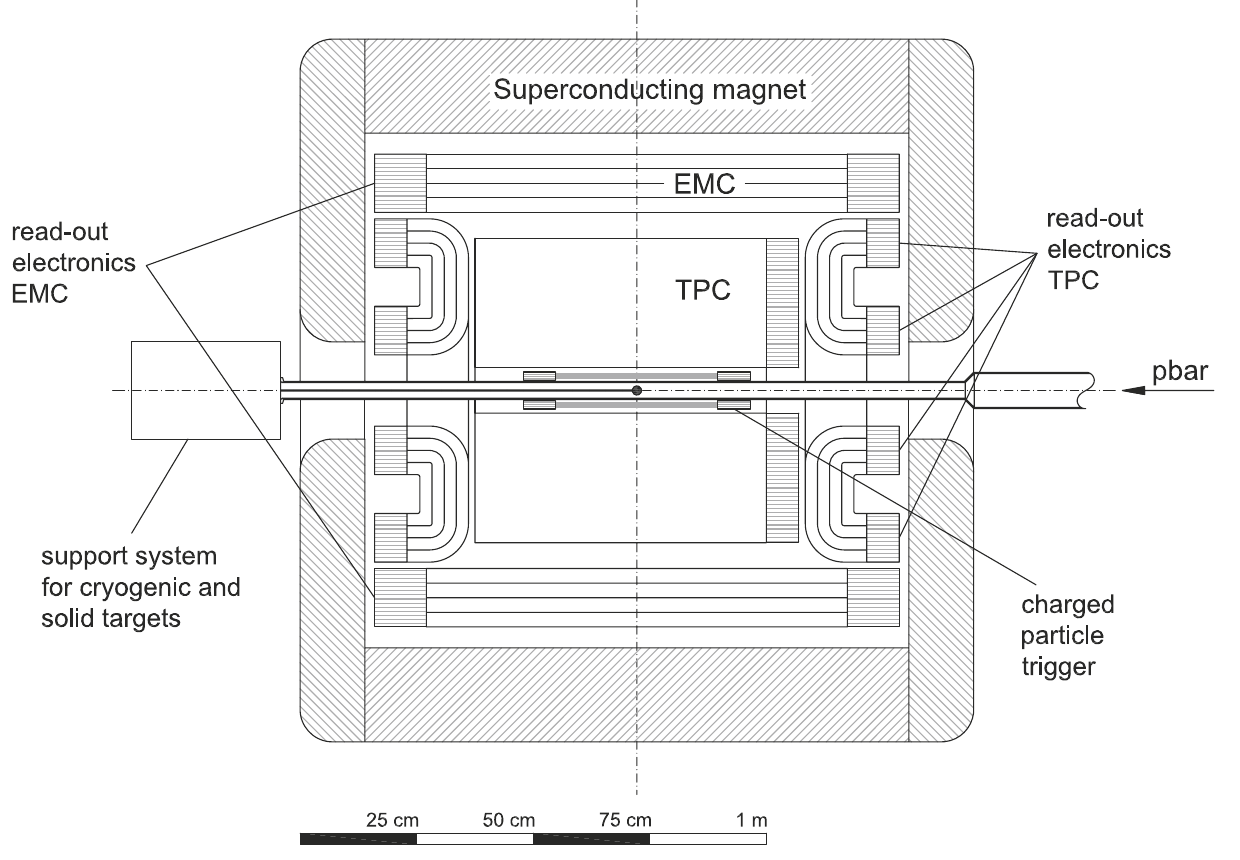}
  \caption{\label{fig:EW-strangeness-detector} Schematic layout of a possible 4$\pi$ detector for use with stopped antiprotons. From \cite{Zmeskal:2009jep}.}
\end{figure}

A further topic of interest that has been discussed for some time is the double-strangeness production with antiprotons producing double-kaonic nuclear bound states \cite{Kienle:2005gn,Kienle:2007zza}. Such states where  predicted to exist with extremely high binding energy and density \cite{Akaishi:2002bg} and could be produced from stopped or low-energy antiprotons.  Following analysis of LEAR data which showed evidence for events in which two $K^+$ were produced \cite{Bendiscioli:2007zzb}, studies have been preformed on the feasibility of such an experiment using a 4$\pi$ detector (see Fig.~\ref{fig:EW-strangeness-detector}) capable of detecting both charged and neutral reaction products for FLAIR \cite{Zmeskal:2009jep}, and for J-PARC \cite{Sakuma:2012wva} using a secondary beam. Due to the necessary degradation of antiprotons produced at high energy in a secondary beam, the intensity available from a storage ring is orders of magnitude higher, making FLAIR an ideal place for such an experiment. Furthermore, such a detector can be used for the study of other hadrons created in the annihilation process.

In summary, low-energy antiprotons at FAIR offer unique experiments using internal targets as well as slowly extracted antiproton beams that are currently not available  anywhere. With the availability of the core components CRYRING and HITRAP the major infrastructure is already available and the cost of adding low-energy antiprotons has become much lower as initially anticipated in the FLAIR technical proposal.


\section{FAIR - Exploring stellar and planetory plasmas }
\begin{figure}[h!]
  \centering
  \includegraphics[width=10cm]{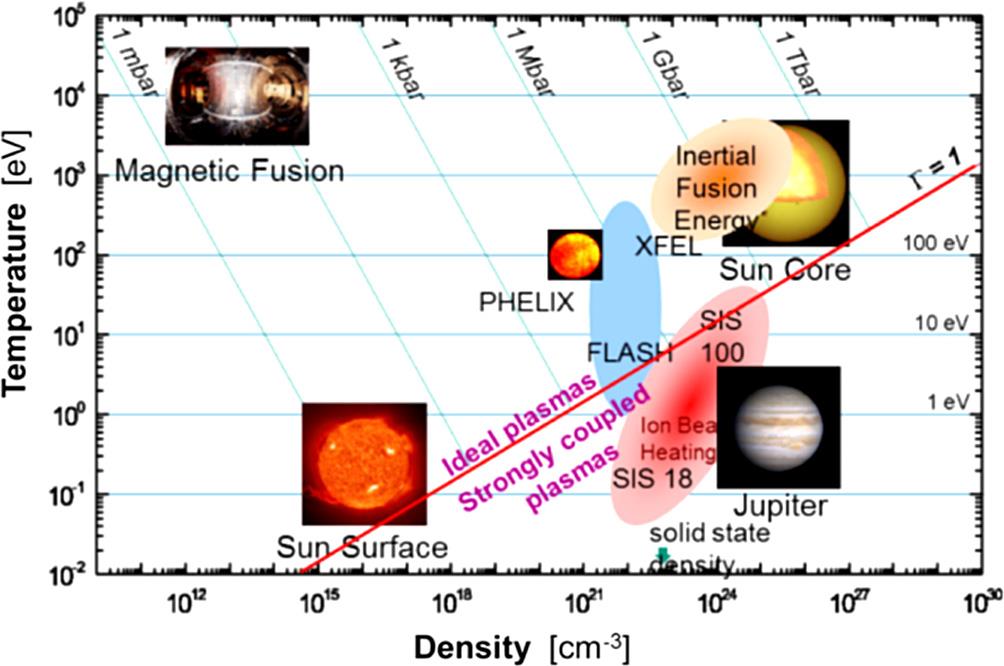}
  \caption{\label{fig:plasma_figure1} Schematic phase diagram showing the areas accessible with FAIR in comparison to other facilities. The conditions in some existing astrophysical objects are displayed in addition \cite{Stoehlker2015}.}
\end{figure}

In case of the interdisciplinary research of APPA (Atomic, Plasma, and Applied Research) [STO2015], the particular focus is the study of matter under extreme conditions. It comprises the interaction of matter at highest electromagnetic fields (SPARC, atomic physics see previous chapter) as well as properties of plasmas and of solid matter under extreme pressure, density, and temperature conditions such as they prevail in stellar objects.  The latter is the research focus of the High Energy Density at FAIR collaboration (HED@FAIR), exploiting the unique properties of intense heavy ion pulses to heat matter volumetrically. More specifically, the planned HED@FAIR program aims on the equation of state as well as on the transport properties of different materials in the so far widely unexplored regions of the phase diagram related to warm dense matter and high-energy density (see Fig.~\ref{fig:plasma_figure1}) \cite{Stoehlker2015,SHARKOV2013,VARENTSOV2013,HOFFMANN2002,TAHIR2000,TAUSCHWITZ2007}.

At FAIR, the ion pulses deposit energies in excess of 10 kJ/g, enabling to drive samples into the regime of high-energy density (HED) matter with pressures in the range of several Mbar. In addition, the volume of matter that can be uniformly heated is in the range of cubic millimeters and the characteristic times are sub-microseconds, such that samples are also created in local thermodynamical equilibrium. Matter at such extreme conditions is ubiquitous in the universe e.g. inside compact astrophysical objects such as planets, brown dwarfs and stars. FAIR will enable the exploration of matter under such extremes in the laboratory. These challenging experiments will allow, by benchmarking theoretical models, to test our understanding of such objects where accurate knowledge of the material properties over an extensive parameter range is required. In addition, the realm of warm dense matter remains a particular challenge for theoretical modeling due to the strongly coupled ion component and partial electron degeneracy. Heating with heavy ions offers an entirely new approach for generating HED samples, complementary to the highly non-equilibrium states generated in the femtosecond interaction with X-ray free-electron lasers, or ablation-pressure-driven shockwaves generated by high-power laser facilities. Naturally, the field has overlap with materials science and atomic physics. For materials research, this concerns the area of strongly excited solids, thermally induced stress waves and radiation damage. In the field of atomic physics, the influence of the dense medium on atomic structure and continuum lowering will be studied. For the success of this research program, theoretical support by sophisticated computational tools and techniques is of utmost importance in order to provide the required simulation for matter under such extreme conditions. In addition, only by simulations the design of special target configurations will provide the basis for precise diagnostics of the target state.

The experimental station to be used for plasma physics at FAIR will be the APPA cave. The APPA Cave at SIS100 is a multipurpose user facility, designed for ions with a magnetic rigidity of 100 Tm for the research fields of atomic physics, materials research and plasma physics. For plasma physics, this cave enables experiments with beams at highest intensity and low charge state e.g. $5 \times 10^{11}$ U$^{28+}$ ions at 2.7 GeV/u. More specifically, the HED@FAIR collaboration develops and exploits the following experimental scenarios at the dedicated plasma physics beamline within the APPA cave:

\begin{figure}[h!] 
  \centering
  \includegraphics[width=10cm]{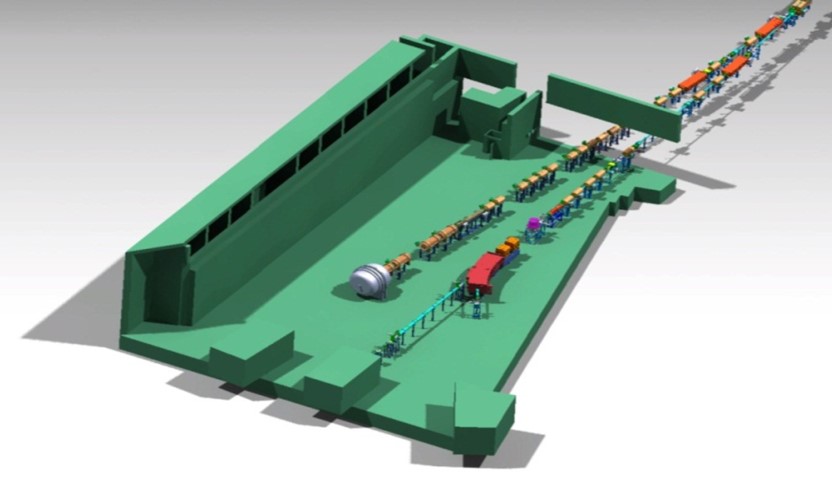}
  \caption{The APPA cave with two beamlines and several experimental stations, serving the atomic physics, plasma physics and materials research communities.}
\end{figure}

More specifically, the FAIR plasma physics collaboration develops and exploit the following experimental scenarii at the dedicated plasma physics beamline within the APPA cave:

In \textbf{HIHEX} (Heavy Ion Heating and Expansion) experiments, targets are quickly and uniformly heated by an intense heavy-ions pulse to reach high energy density states. After heating, the sample will isentropically expand and pass through various regions of the phase diagram. Such samples will be used to study the Equation of State and transport properties of high-entropy phases of materials of interest \cite{HOFFMANN2002}.

\textbf{PRIOR} is a high-energy proton microscope facility which will be integrated into the plasma physics beam line at the APPA cave behind SIS100. The high-energy proton beams will enable worldwide unique research opportunities by providing both high resolution and large field of view options for proton radiography \cite{SHARKOV2013}. Research with PRIOR will focus on the fundamental properties of materials in extreme dynamic environments which are generated by external drivers (e.g. laser or particle beams). It will be open for multidisciplinary research projects comprising research fields such as plasma physics, materials research under extreme conditions and medical physics (see PaNTERA project below) \cite{DURANTE2014,DURANTE2011}.

\textbf{LAPLAS} ({\bf LA}boratory {\bf PLA}netary Sciences) experiments at FAIR aim at compressing matter, at moderate temperatures, to states relevant to planetary interiors. LAPLAS targets are made of a high-Z cylindrical tamper that is heated by the ion beam and compresses its payload. For low-Z elements like hydrogen, LAPLAS will make use of cryogenic targets, confined by an outer cylinder of a heavier material \cite{TAHIR2000}. For high-Z elements like iron, the ion beam needs to exhibit an annular focal spot, obtained by applying a radio frequency beam rotator. This leads to a low-entropy compression of the cold inner material to pressures of several Mbar \cite{TAHIR2017}.

In summary, plasma physics research will exploit the unique capabilities of the high-intensity ion beams of FAIR for research of matter under extreme conditions. These studies will allow to access the properties of plasmas and of solid matter under extreme pressure, density, and temperature at conditions as they prevail in stellar environments such as stars and giant planets.  

\section{Serving Society at FAIR - Biomedical application}

Nuclear physics finds many successful applications in biology and medicine. Charged particle therapy, medical imaging, and radioisotope
production are the best examples of the medical benefits stemming from nuclear physics \cite{NuPECC}. Most of these applications are based
at ion accelerators. Particle therapy, in particular, is nowadays largely spread all over the world and exploits cyclotrons or
synchrotrons accelerating protons or carbon ion beams at energies below 400 MeV/n, generating particles with a range in tissue of
approximately 25 cm, corresponding to deep-seated tumors \cite{Durante2016}.  Pre-clinical research in medical physics and radiobiology is often
performed at the same facilities where patients are treated.

Research in space radiation protection also needs high-energy beams of protons and heavy ions \cite{Durante2011}. Cosmic rays can reach energies
above 10 GeV/n, and therefore only part of the spectrum can be reproduced at therapy facilities. A few facilities, such as the
NASA Space Radiation Laboratory (NSRL) at the Brookhaven National Laboratory (BNL) in Upton, NY (USA), and the SIS18 of the
GSI Helmholtz Center for Heavy Ion Research in Darmstadt, Germany, can provide heavy ion beams up to 1 GeV/n and run research
programs in space radiation protection sponsored by NASA \cite{LaTessa2016} and ESA \cite{Durante2010}, respectively.

\begin{figure}[ht!]
	\begin{center}
		\includegraphics[width=0.50\textwidth]{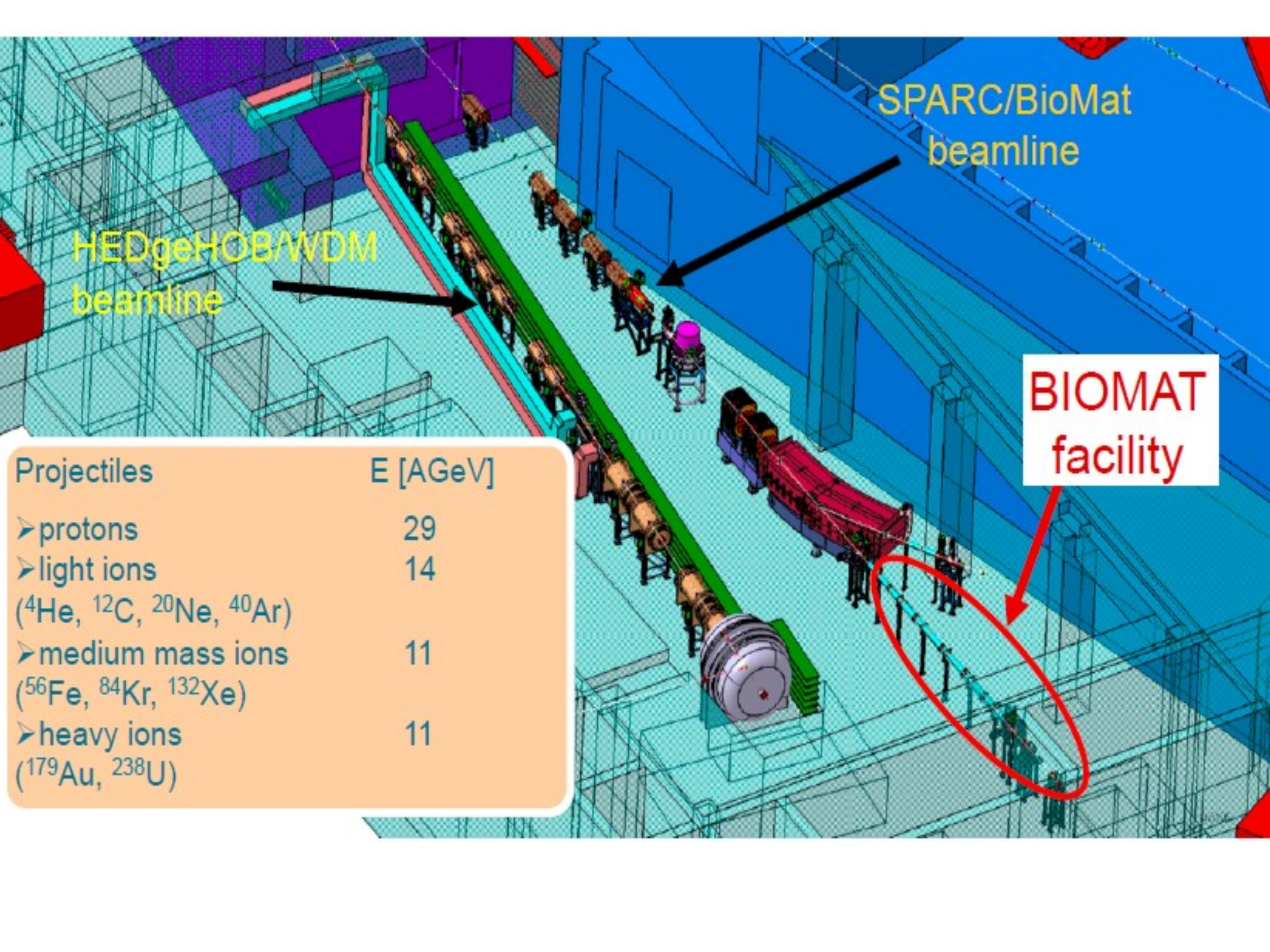}
		\caption{Layout of the future BIOMAT cave that will be used at FAIR for the biomedical applications within the APPA experiment. The vault includes the beamline for plasma physics and the SPARC experiment. Maximum energies of the beam that will be transported in BIOMAT are indicated in the insert.}
		\label{fig:bio1}
	\end{center}
\vspace{-3mm}
\end{figure}

The new high-energy facilities
under construction in Europe, America, and Asia have plans for applied nuclear physics, especially toward atomic physics,
plasma physics, material research, and medicine. APPA \cite{Stoehlker2015} is one of the four pillars of FAIR and its program covers several
applications of high-energy heavy ion beams, including biomedical applications. Within the APPA cave (Figure~\ref{fig:bio1}), a beamline is
dedicated to biophysics and material research (BIOMAT collaboration). Cutting-edge research topics currently under study at GSI will
of course be included in the FAIR program: new ions \cite{Tommasino2015}, radioimmunology \cite{Durante2013},
and treatment of noncancer diseases,
especially heart arrhythmia \cite{Lehmann2016}.
What can be done at FAIR that is not possible at the current medical accelerators or at GSI today?
In fact, many therapy centers have intense research programs – e.g. HIT in Heidelberg (Germany), CNAO in Pavia (Italy) or
HIMAC in Chiba (Japan). NSRL and GSI are reference facilities for space radiation research in USA and Europe.
Therefore, interest in biophysics at FAIR will be based on opportunities that cannot be offered by the current
(clinical or research) centers. This essentially means biomedical applications that can benefit from higher energies,
production of radioactive isotope beams (RIB), and higher intensities.

\subsection{High Energy}

\subsubsection{Space radiation protection}

\begin{figure}[ht!]
	\begin{center}
		\includegraphics[width=0.50\textwidth]{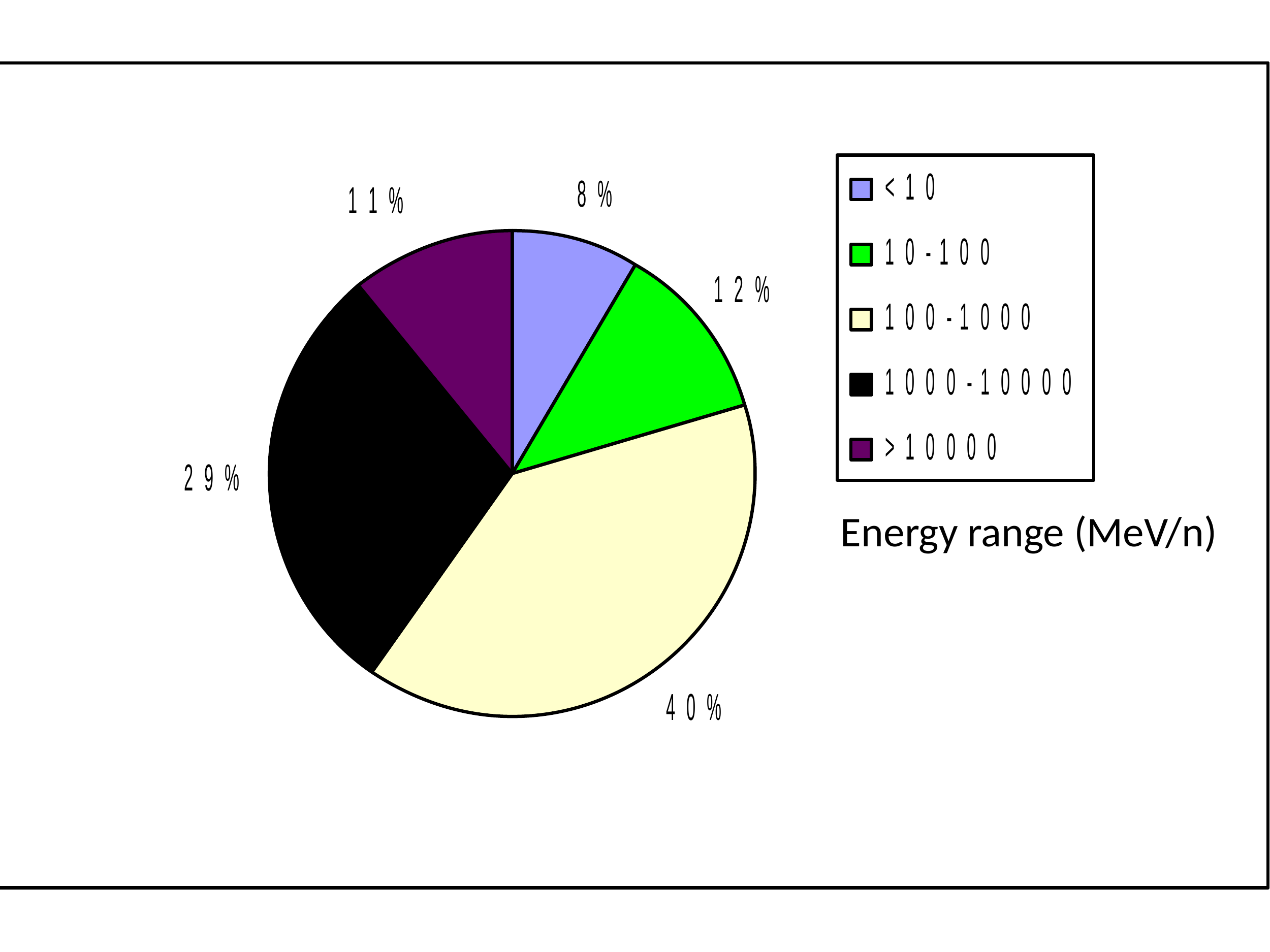}
	\caption{Galactic cosmic ray energy contributions to blood forming organ dose equivalent in deep space behind a shielding
		of 5 g/cm2 Al (averaged over 1 year at solar minimum). Simulations by the HZETRN code by NASA, adapted from
		\cite{Wilson1997}.}
          \label{fig:bio2}
	\end{center}
\vspace{-3mm}
\end{figure}

Research in space radiation protection can greatly benefit from facilities able to accelerate heavy ions beyond 1 GeV/n.
Space programs of all major space agencies worldwide are now concentrating on manned exploration of the Solar system: the moon
base and the mission to Mars \cite{ISEGG}. Exposure to space radiation is generally acknowledged as the main health risk for crews of
exploratory-class missions \cite{Chancellor2014}. Even if microgravity and isolation pose serious health concerns for long-term space travel,
radiation risk is affected by large uncertainties, and countermeasures are difficult to implement. For reducing risk uncertainty,
accelerator-based radiobiology is essential because heavy ions, main contributors to the dose equivalent in deep space, are not present
on earth, and therefore no epidemiological studies are available \cite{Durante2005}. Research at NSRL and GSI provided significant
information about the effects of heavy ions with energies of 1 GeV/n or below in cell, tissue, and animal models. Ions with energies
between 1 and 10 GeV/n represent about 50$\%$ of the total Galactic Cosmic Ray (GCR) flux and can contribute up to 30$\%$ of the total dose equivalent in deep space \cite{Wilson1997} (Figure~\ref{fig:bio2}).
Therefore, radiobiology
studies using these ions are important to understand the risk from radiation exposure in space travel. FAIR will be the main center of these studies, and its role has been endorsed by ESA with a Memorandum of Understanding signed in February 2018. It is expected that FAIR will be the most realistic ground-based simulator of GCR and will contribute to research in health and material hardness. 

\subsubsection{Particle radiography}

Range uncertainty is arguably the main physics limitation of particle therapy \cite{Paganetti2012}. The uncertainty comes from different
sources, including uncertainties in the tissue mean excitation energy and inter- or intra-fractional organ motion. One of the main
sources of uncertainty is the conversion from Hounsfield units (from the CT scans used for planning) into water-equivalent path lengths
(defining the particle range). Using the same particle beam for radiography would obviously eliminate this uncertainty. In addition,
using the therapeutic beam for imaging can allow online beam adjustment for irradiation of moving targets. For this reason, many studies
in proton \cite{Poludniowski2015} and heavy ion \cite{Parodi2014} radiography are ongoing in several centers.

\begin{figure}[ht!]
	\begin{center}
		\includegraphics[width=0.50\textwidth]{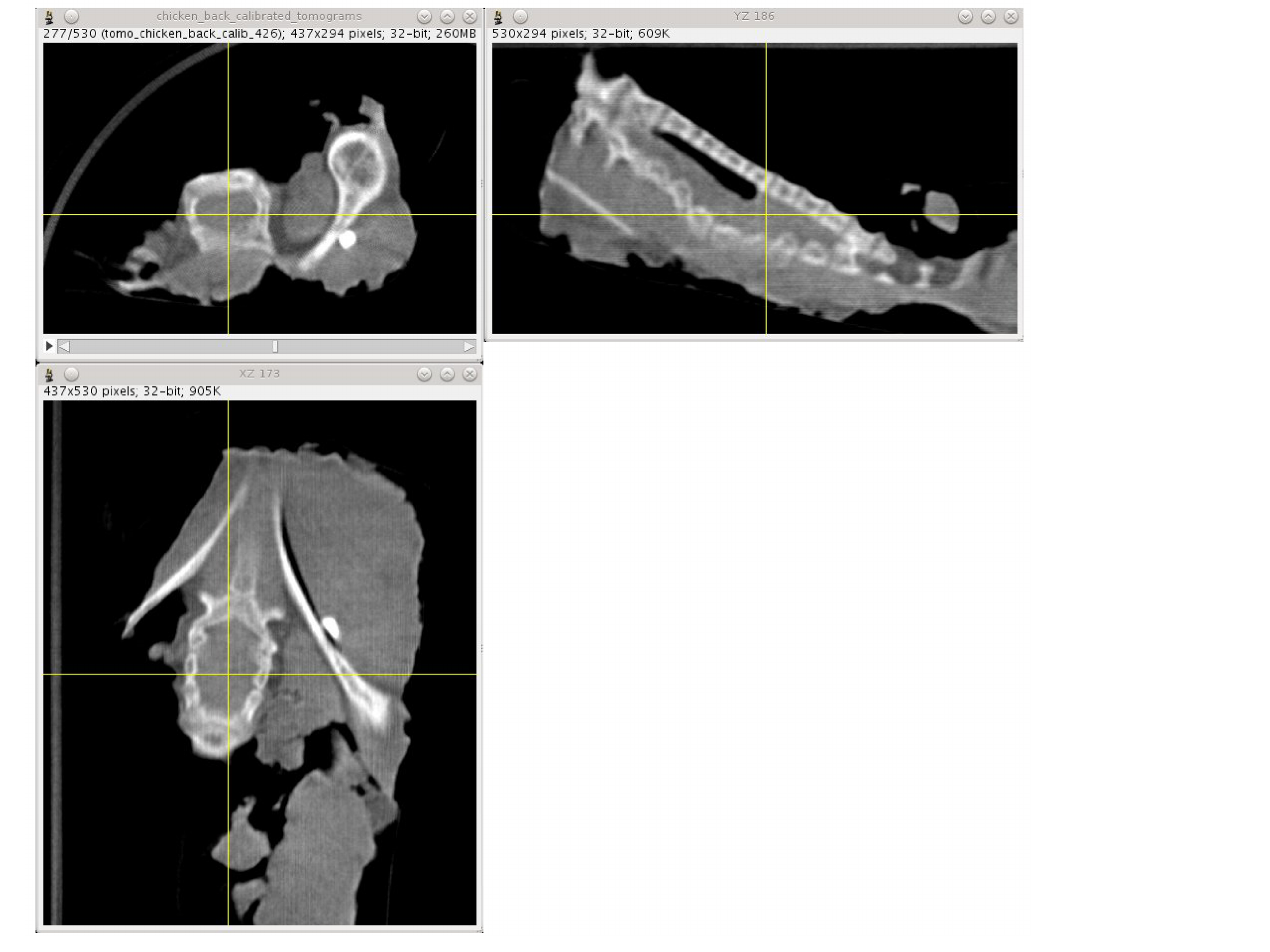}
	\caption{Proton CT of chicken samples obtained with 800 MeV at Los Alamos National Laboratory within the
		experiment PANTERA with GSI Helmholtz Center. Bones and soft tissues are clearly resolved.
		Arrows point to a tiny metal implant close to the bone. Results are discussed in \cite{Prall2016}. }
		\label{fig:bio3}
		
	\end{center}
\vspace{-3mm}
\end{figure}

The use of particles for radiography and tomography is hampered by the low energy used in clinics. For radiography, the beam must
obviously cross the patient’s body, and using the maximum therapeutic energy of the current facility this is possible only for the
head-and-neck region. Moreover, proton scattering blurs the images and the quality is hardly usable for planning and for range verification.
The quality is highly improved by using high-energy beams. First images of biological targets have been obtained with 800 MeV protons at
ITEP in Moscow, Russia \cite{Varentsov2013},  and at the Los Alamos National Laboratory in USA \cite{Prall2016}. The promising results (Figure~\ref{fig:bio3}) suggest that
high-energy particle beams can be used for biomedical imaging, and even for microscopy using ultra-relativistic beams, where the
resolution drops below 1 mm. The PRIOR project in APPA at FAIR is planning a 4.5 GeV beam for proton microscopy of thick solid matter
and biological targets \cite{Prall2015}.
Much research is needed for using this system in image-guided particle radiosurgery, because the dose has
to be significantly reduced, unless the same high-energy beams are used for simultaneous treatment and imaging \cite{Durante2012}.

\subsection{Radioactive beams}

RIB beams are one of the main motivations for the construction of new high-intensity and high-energy accelerators.
RIB can be very useful for beam monitoring and range verification in cancer therapy. The use of RIB for image-guided particle
therapy had been already
proposed at LBNL during the pilot project in the 70s \cite{Chu1993}.
Induced radioactivity is currently used in several centers for beam monitoring.
Positron-emitting, neutron-poor isotopes produced by nuclear fragmentation, can be detected by positron emission tomography (PET).
In particular, the production of $^{11}$C and $^{10}$C by fragmentation of the projectile $^{12}$C, and of $^{15}$O by target fragmentation of $^{16}$O, is used for
PET monitoring in C-ion or proton therapy \cite{Parodi2015}.
These range verification methods would be greatly enhanced by using a radioactive beam
for therapy. Such a beam could be visualized in the patient’s body, and would provide unprecedented accurate range verification.
Following early attempts at LBNL \cite{Llacer1984}, the idea has been implemented at HIMAC in Chiba (Japan) \cite{Sakurai2014},
with a prototype for an
open-PET to visualize the full path of a $^{11}$C in the body \cite{Hirano2016}. Monte Carlo simulations have also been performed at CERN, where the
radioactive beam can be produced at ISOLDE \cite{Augusto2016}.
The annihilation of antiprotons can also be exploited for an accurate visualization of
the beam position \cite{Sellner2011}. However,
all these efforts have been hampered so far by the low intensity of the radioactive beam. At HIMAC,
the $^{11}$C intensity was below $10^5$ particles per second (pps), which is insufficient to give a good signal/noise ratio
and for a therapeutic plan. At FAIR, the intensity of RIB is supposed to be increased by a factor of 10,000 compared to the current beams.
FAIR will also produce very intense antiproton beams, which can be used for tests of new detectors of antiproton beam imaging.
FAIR will be the perfect testbed for the applications of RIB in image-guided particle therapy.

\subsection{High intensity}

A typical feature of the new facilities will be a substantial increase in beam intensity. At FAIR, the intensity of the primary
heavy ion beam should be increased by a factor 100. Current synchrotrons for therapy work at $10^8$-$10^9$ pps for fast beam
scanning \cite{Weinrich2011}.
Higher intensity would be desirable for reducing treatment time and minimizing organ movement during treatment \cite{Bert2011}.
Multi-energy extraction with extended flat top provides different energies during the same cycle, and therefore can take full
advantage of higher intensities \cite{Mizushima2014}. The limitation is given by the dose control system in the beam delivery, and in
particular by the inter-spot dose, i.e. the dose delivered while moving from one spot to the next during raster scanning.

The normal dose rates achieved with the current intensities are between 0.5 and 10 Gy/min. Dose rates $>$ 2000 Gy/min are now
becoming of increasing interest in radiotherapy. In fact, experiments in a mouse model have recently measured a very large sparing
effect of the normal tissue by 4-6 MeV electrons delivered at this ultra-high dose rate \cite{Favaudon2014}. Lung tumor control was similar for
conventional dose–rate and FLASH radiotherapy, but fibrosis was much higher at conventional 1.8 Gy/min. In a second study,
the FLASH irradiation also proved to be able to preserve spatial memory after whole-brain mouse irradiation, while at the
same dose (10 Gy) conventional dose-rates totally impairs the animals’ spatial memory \cite{Montay2017}.

There is not a general agreement on the mechanism underlying the sparing effect of FLASH radiotherapy, even if the oxygen depletion
hypothesis, based on fast oxygen consumption leading to hypoxia and therefore increased resistance of well-oxygenated normal tissue,
is considered the most plausible explanation \cite{Durante2017}. Even without a theoretical justification, FLASH is attracting great interest
in the radiotherapy community, as a potential system to enlarge the therapeutic window, i.e. to reduce the toxicity while keeping
the same effectiveness in the tumor. These very high dose rates are, however, difficult to achieve with photons and with charged particles.
FAIR will be able to test the FLASH radiotherapy with ions, because the increased intensity will allow irradiations in the range
where the effect has been observed with electrons.

\subsection{Outlook}

There are several facilities worldwide providing high-energy beams for biomedical research. Medical facilities are limited to protons and carbon ions with energies up to 400 MeV/n. Higher energies are needed for space radiation protection studies, and currently these studies are limited to the NSRL accelerator in Brookhaven National Laboratory in the USA and to the SIS18 accelerator at GSI in Germany. FAIR will be the first facility able to study heavy ions at energies above 1 GeV/n, an important component of the GCR. The development of efficient countermeasures for high-energy heavy ions in the GCR is a mandatory step towards the planned manned mission to the moon and to Mars. Therefore, these studies are urgent and highly needed. In oncology, even if current facilities cover the range necessary to treat deep tumors in humans, higher energies can be beneficial for imaging. Using the same beam for imaging and treatment would bring therapy into the modern area of theranostics and could facilitate the treatment of moving targets. Theranostics is also an attractive possibility when using radioactive ion beams for therapy, currently produced at intensity too low for practical treatments. The high intensity at FAIR will make possible to test radioactive ion beams for simultaneous treatment and beam visualization by online PET. Finally, the high intensity at FAIR offers other unique opportunities. Will it be possible to treat tumors in seconds, thus making particle therapy cheaper and easier, especially when treating moving organs? Is this high dose rate beneficial in widening the therapeutic window, as suggested by FLASH experiments with electron beams? FAIR will give answers to these questions and more. Obviously the potential of the FAIR facilities cannot be translated elsewhere in the present days, but the results in this special facility will drive the technological developments in the next 20 years.

\section{Outlook and more fun at FAIR}

The Facility for Antiproton and Ion Research FAIR will be a global flagship for fundamental science. 
FAIR not only opens up unprededented research opportunities in hadron, nuclear and atomic physics
and in associated applied sciences, but will also push the current limits in accelerator and detector
technologies. Several of the exciting perspectives have been discussed in this manuscript. 
They encompass a large area of research in hadron and nuclear physics, as well as in atomic and plasma physics and in applied
sciences. As general themes, FAIR research will advance our understanding at three frontiers
of modern science: 
\begin{itemize}
\item
It will deepen our general understanding
of fundamental theories of Nature like QCD and QED and will put their applicabilities
to the test in regimes not adequately accessible at other facilities like
QCD in the transition region between perturbative and non-perturbative regimes and QED with highly charged ions where
perturbation theory is not easily applicable. 

\item It will advance our understanding how
complexity arizes from fundamental building blocks and the interaction among theme. This research is at the center
of all experimental collaborations. For example, PANDA will search for and unravel the structure of exotic hadronic states made
of quarks and gluons, NUSTAR aims at the development of a global and unified picture which describes the
many nuclei and nuclear phenomena from nucleonic degrees of freedom and their interaction, where not the least effort goes
into deriving the interaction among nucleons from the fundamental properties and  symmetries of the underlying theory
of strong interaction, QCD. The CBM experiment contributes to this quest by exploring the bulk properties
of nuclear matter and the various phases of the QCD phase diagram which, in particular in the high-density regime, i.e.
the home turf of the CBM experiment, is largely unknown. 

\item It will significantly progress modern accelerator, detector and also IT technologies
and has strong societal impact, with the development of novel accelerator-based medical treatments as a prime example.
\end{itemize}
The fundamental research in all experimental collaborations has many applications in nuclear astrophysics, as outlined throughout
the manuscript.

Obviously there will be a lot of science excitement once FAIR becomes operational. But scientists think already 
for the times beyond the first-years operations of FAIR. And indeed there exist many options how FAIR can become even more
powerful than the version currently under construction. The accelerator tunnel which will host the new SIS100 synchroton,
FAIR's heart and work horse, allows for the installation of a second ring accelerator. Of course the realization of this option
and its design will depend on the physics results and scientific surprises, which FAIR will deliver in its first years
of operation, and naturally on the advances of accelerator technologies. 

As outlined above, the fun for researchers at FAIR, with its worldwide unique combination of powerful accelerators, storage rings
covering many orders of magnitude in energies, and detectors is to explore science at all length scales
from the internal structure of the nucleons and mesons governed by Quantum Chromodynamics to the
mysteries of exotic astrophysical objects in the Universe like colliding neutron stars or stellar explosions. The fun is also to follow
the history of our Universe from fractions of seconds after the Big Bang when matter existed as a correlated plasma
of quarks and gluons to the present and future challenges of mankind by developing novel accelerator-based health treatments
and materials to support space travel. New ideas might involve polarized particle beams opening up new  tests of
the standard model of strong and electromagnetic interactions. The combination of a powerful laser with the intense heavy-ion beams
is a unique tool to generate electromagnetic plasmas and to simultaneously diagnose it. 

The greatest fun of all at FAIR will be the surprising scientific results which have not been anticipated
and which will deepen and advance our understanding of our Universe. There will be excitement for many science generations to come.

\section*{Acknowledgments}
\hspace*{\parindent}
V.K. was supported by the U.S. Department of Energy,
Office of Science, Office of Nuclear Physics, under contract number
DE-AC02-05CH11231. The work of UGM was supported in part by the DFG 
and NSFC through funds provided to the Sino-German CRC 110 (DFG Grant No.
TRR 110), by the VolkswagenStiftung (Grant No. 93562) and by the CAS President's
International Fellowship Initiative (PIFI) (Grant No. 2018DM0034).

\newcommand{\newblock}{}


\begin{thebibliography}{999}
\bibitem{FAIR}
		W.F. Henning {\it et al.},
		The International Accelerator Facility for Beams of Ions
		Antiprotons (GSI publication, November 2001)

	\bibitem{NuPECC2017}
          NuPECC, Long Range Plan {\it Perspectives in Nuclear Physics}
		(November 2017)



\bibitem{Fritzsch:1973pi}
  H.~Fritzsch, M.~Gell-Mann and H.~Leutwyler,
  Phys.\ Lett.\  {\bf 47B} (1973) 365.

\bibitem{Gross:1973id}
  D.~J.~Gross and F.~Wilczek,
  Phys.\ Rev.\ Lett.\  {\bf 30}, 1343 (1973).

\bibitem{Politzer:1973fx}
  H.~D.~Politzer,
  Phys.\ Rev.\ Lett.\  {\bf 30}, 1346 (1973).

\bibitem{GellMann:1964nj}
  M.~Gell-Mann,
  Phys.\ Lett.\  {\bf 8} (1964) 214.

\bibitem{Dalitz:1959dn}
  R.~H.~Dalitz and S.~F.~Tuan,
  Phys.\ Rev.\ Lett.\  {\bf 2} (1959) 425.

\bibitem{Aubert:2003fg}
  B.~Aubert {\it et al.} [BaBar Collaboration],
  Phys.\ Rev.\ Lett.\  {\bf 90} (2003) 242001

\bibitem{Besson:2003cp}
  D.~Besson {\it et al.} [CLEO Collaboration],
  Phys.\ Rev.\ D {\bf 68}, 032002 (2003)
  Erratum: [Phys.\ Rev.\ D {\bf 75}, 119908 (2007)].

\bibitem{Godfrey:1985xj} 
  S.~Godfrey and N.~Isgur,
  Phys.\ Rev.\ D {\bf 32}, 189 (1985).

\bibitem{Choi:2003ue}
  S.~K.~Choi {\it et al.} [Belle Collaboration],
  Phys.\ Rev.\ Lett.\  {\bf 91} (2003) 262001

\bibitem{Aaij:2015tga}
  R.~Aaij {\it et al.} [LHCb Collaboration],
  Phys.\ Rev.\ Lett.\  {\bf 115}, 072001 (2015)


\bibitem{Hicks:2012zz}
  K.~H.~Hicks,
  Eur.\ Phys.\ J.\ H {\bf 37}, 1 (2012).


\bibitem{Chen:2016qju}
  H.~X.~Chen, W.~Chen, X.~Liu and S.~L.~Zhu,
  Phys.\ Rept.\  {\bf 639}, 1 (2016)

\bibitem{Lebed:2016hpi}
  R.~F.~Lebed, R.~E.~Mitchell and E.~S.~Swanson,
  Prog.\ Part.\ Nucl.\ Phys.\  {\bf 93}, 143 (2017)

\bibitem{Guo:2017jvc}
  F.~K.~Guo, C.~Hanhart, U.-G.~Mei{\ss}ner, Q.~Wang, Q.~Zhao and B.~S.~Zou,
  Rev.\ Mod.\ Phys. {\bf 90} (2018) 015004

\bibitem{Durr:2008zz}
  S.~D\"urr {\it et al.},
  Science {\bf 322}, 1224 (2008)

\bibitem{Luscher:1986pf}
  M.~L\"uscher,
  Commun.\ Math.\ Phys.\  {\bf 105}, 153 (1986).


\bibitem{Feng:2010es}
  X.~Feng, K.~Jansen and D.~B.~Renner,
  Phys.\ Rev.\ D {\bf 83}, 094505 (2011)

\bibitem{Mohler:2013rwa}
  D.~Mohler, C.~B.~Lang, L.~Leskovec, S.~Prelovsek and R.~M.~Woloshyn,
  Phys.\ Rev.\ Lett.\  {\bf 111}, no. 22, 222001 (2013)

\bibitem{Moir:2016srx}
  G.~Moir, M.~Peardon, S.~M.~Ryan, C.~E.~Thomas and D.~J.~Wilson,
  JHEP {\bf 1610}, 011 (2016)

\bibitem{Abbott:2016blz}
  B.~P.~Abbott {\it et al.} [LIGO Scientific and Virgo Collaborations],
  Phys.\ Rev.\ Lett.\  {\bf 116}, no. 6, 061102 (2016)

\bibitem{Bauswein:2017vtn}
  A.~Bauswein, O.~Just, H.~T.~Janka and N.~Stergioulas,
  Astrophys.\ J.\  {\bf 850}, no. 2, L34 (2017)

\bibitem{Baym:2017whm}
  G.~Baym, T.~Hatsuda, T.~Kojo, P.~D.~Powell, Y.~Song and T.~Takatsuka,
   Rept.\ Prog.\ Phys.\  {\bf 81}, no. 5, 056902 (2018).


\bibitem{Haidenbauer:2013oca}
  J.~Haidenbauer, S.~Petschauer, N.~Kaiser, U.-G.~Mei{\ss}ner, A.~Nogga and W.~Weise,
  Nucl.\ Phys.\ A {\bf 915}, 24 (2013)

\bibitem{Beane:2012ey}
  S.~R.~Beane {\it et al.},
  Phys.\ Rev.\ Lett.\  {\bf 109}, 172001 (2012)

\bibitem{Inoue:2013nfe}
  T.~Inoue {\it et al.} [HAL QCD Collaboration],
  Phys.\ Rev.\ Lett.\  {\bf 111}, no. 11, 112503 (2013)


\bibitem{Jaffe:1976yi}
  R.~L.~Jaffe,
  Phys.\ Rev.\ Lett.\  {\bf 38}, 195 (1977)
  Erratum: [Phys.\ Rev.\ Lett.\  {\bf 38}, 617 (1977)].

\bibitem{Doi:2015oha}
  T.~Doi {\it et al.},
  PoS LATTICE {\bf 2015}, 086 (2016)

\bibitem{Haidenbauer:2011za}
  J.~Haidenbauer and U.-G.~Mei{\ss}ner,
  Nucl.\ Phys.\ A {\bf 881}, 44 (2012)

\bibitem{Lutz:2009ff}
  M.~F.~M.~Lutz {\it et al.} [PANDA Collaboration],
  arXiv:0903.3905 [hep-ex].


\bibitem{Morningstar:1999rf}
  C.~J.~Morningstar and M.~J.~Peardon,
  Phys.\ Rev.\ D {\bf 60}, 034509 (1999)

\bibitem{Dudek:2010wm}
  J.~J.~Dudek, R.~G.~Edwards, M.~J.~Peardon, D.~G.~Richards and C.~E.~Thomas,
  Phys.\ Rev.\ D {\bf 82}, 034508 (2010)

\bibitem{Berwein:2015vca}
  M.~Berwein, N.~Brambilla, J.~Tarrus Castella and A.~Vairo,
  Phys.\ Rev.\ D {\bf 92}, no. 11, 114019 (2015)


\bibitem{Godfrey:2003kg}
  S.~Godfrey,
  Phys.\ Lett.\ B {\bf 568}, 254 (2003)

\bibitem{Liu:2012zya}
  L.~Liu, K.~Orginos, F.~K.~Guo, C.~Hanhart and U.-G.~Mei{\ss}ner,
  Phys.\ Rev.\ D {\bf 87}, no. 1, 014508 (2013)


\bibitem{Patrignani:2016xqp}
  C.~Patrignani {\it et al.} [Particle Data Group],
  Chin.\ Phys.\ C {\bf 40}, no. 10, 100001 (2016).

\bibitem{Mertens:2012kpa}
  M.~C.~Mertens [PANDA Collaboration],
  Hyperfine Interact.\  {\bf 209} (2012) no.1-3,  111.

\bibitem{Du:2017zvv}
  M.~L.~Du, M.~Albaladejo, P.~Fern\'andez-Soler, F.~K.~Guo, C.~Hanhart, U.-G.~Mei{\ss}ner, J.~Nieves and D.~L.~Yao,
Phys.\ Rev.\ D {\bf 98}  no.9,  094018 (2018).
  

\bibitem{Adamuscin:2006bk}
  C.~Adamuscin, E.~A.~Kuraev, E.~Tomasi-Gustafsson and F.~E.~Maas,
  Phys.\ Rev.\ C {\bf 75}, 045205 (2007)

\bibitem{Haidenbauer:2014kja}
  J.~Haidenbauer, X.-W.~Kang and U.-G.~Mei{\ss}ner,
  Nucl.\ Phys.\ A {\bf 929} (2014) 102

\bibitem{Braun-Munzinger:2015hba}
P.~Braun-Munzinger, V.~Koch, T.~Sch{\"a}fer, and J.~Stachel,
\newblock Phys. Rept. {\bf 621}  (2016) 76

\bibitem{Wang:2016opj} Quark Gluon Plasma 5, X.N. Wang editor, World
  Scientific, 2016

\bibitem{Abelev:2010rv}
 B.~I. Abelev {\em et~al.} [STAR collaboration],
\newblock Science {\bf 328}, (2010) 58

\bibitem{Adam:2015yta}
J.~Adam  {\em et~al.}  [ALICE collaboration],
\newblock Phys. Lett. {\bf B754}, (2016) 360

\bibitem{Adam:2015vda}
J.~Adam {\em et~al.} [ALICE collaboration],
\newblock Phys. Rev. {\bf C93}, (2016) 024917

\bibitem{Randrup:2006nr}
J.~Randrup and J.~Cleymans,
\newblock Phys. Rev. {\bf C74}, (2006) 047901

\bibitem{Friman:2011zz}
B.~Friman {\em et~al.},
\newblock Lect. Notes Phys. {\bf 814}, (2011), Springer.

\bibitem{Aoki:2006we}
Y.~Aoki, G.~Endrodi, Z.~Fodor, S.~D. Katz, and K.~K. Szabo,
\newblock Nature {\bf 443}, (2006) 675

\bibitem{Stephanov:2004wx}
M.~A. Stephanov,
\newblock Prog. Theor. Phys. Suppl. {\bf 153}, (2004) 139

\bibitem{Stephanov:2008qz}
M.~Stephanov,
\newblock Phys.Rev.Lett. {\bf 102},  (2009)032301

\bibitem{Stephanov:2011pb}
M.~Stephanov,
\newblock Phys. Rev. Lett. {\bf 107}, (2011) 052301

\bibitem{Luo:2015doi}
X.~Luo, Nucl. Phys. {\bf A956}, (2016) 75

\bibitem{Luo:2017faz}
X.~Luo and N.~Xu,
\newblock Nucl. Sci. Tech. {\bf 28},  (2017) 112

\bibitem{He:2017zpg}
S.~He and X.~Luo,
\newblock Phys. Lett. {\bf B774}, (2017) 623

\bibitem{Bzdak:2016jxo}
A.~Bzdak, V.~Koch, and V.~Skokov,
\newblock Eur. Phys. J. {\bf C77}, (2017) 288

\bibitem{McLerran:2007qj}
L.~McLerran and R.~D. Pisarski,
\newblock Nucl. Phys. {\bf A796}, (2007) 83

\bibitem{Glendenning:1984jr}
  N.~K.~Glendenning,
  Astrophys.\ J.\  {\bf 293} (1985) 470.

\bibitem{Gongyo:2017fjb}
S.~Gongyo {\em et~al.},
\newblock (2017), arXiv:1709.00654.

\bibitem{Steinheimer:2008hr}
  J.~Steinheimer, M.~Mitrovski, T.~Schuster, H.~Petersen, M.~Bleicher and H.~Stoecker,
  Phys.\ Lett.\ B {\bf 676} (2009) 126

\bibitem{Bauswein:2015vxa}
  A.~Bauswein, N.~Stergioulas and H.~T.~Janka,
  \newblock Eur.\ Phys.\ J.\ A {\bf 52} (2016)   56
  

\bibitem{Lippuner:2017bfm}
J.~Lippuner, R.~Fernández, L.~F.~Roberts, F.~Foucart, D.~Kasen, B.~D.~Metzger and C.~D.~Ott, ``Signatures of hypermassive neutron star lifetimes on r-process nucleosynthesis in the disc ejecta from neutron star mergers,''
\newblock  Mon.\ Not.\ Roy.\ Astron.\ Soc.\  {\bf 472} (2017)  904
 
\bibitem{Barnes:2013wka}
J.~Barnes and D.~Kasen,
\newblock Astrophys. J. {\bf 775},  (2013) 18

\bibitem{Fernandez:2016sbf}
R.~Fernandez {\em et~al.},
\newblock Class. Quant. Grav. {\bf 34}, (2017) 154001

\bibitem{Kasen:2017sxr}
D.~Kasen, B.~Metzger, J.~Barnes, E.~Quataert, and E.~Ramirez-Ruiz,
\newblock Nature {\bf 551},  (2017) 80

\bibitem{Senger:2017oqn}
  P.~Senger [CBM Collaboration],
  Nucl.\ Phys.\ A {\bf 967} (2017) 892.
  
  \bibitem{HADES:2017oqn}
J.~Friese, Nucl.\ Phys.\ A {\bf 654} (1999) 1017 c; G.~Agakishiev et al. [HADES Collaboration], Phys.\ Rev.\ C {\bf 84, 014902} (2011).
  
  \bibitem{HADESSIS18:2017oqn}
  G.~Agakishiev et al. [HADES Collaboration], Phys.\ Lett.\ B {\bf 663} (2008) 43; G.~Agakichiev et al. [HADES Colalboration], Phys.\ Rev.\ Lett.\ {\bf 98} (2007) 052302; E.~Santini, M. D.~Cozma, A.~Faessler, C. ~Fuchs, M. I.~Krivoruchenko and B.~Martemyanov, Phys.\ Rev.\ C {\bf 78} (2008) 034910; E. L.~Bratkovskaya and W.~Cassing, Nucl.\ Phys.\ A {\bf 807} (2008) 214.



\bibitem{Hansen69} P.G. Hansen {\it et al}, Physics Letters {\bf B28} (1969) 415 

\bibitem{Westfall79} G.D. Westfall {\it et al.}, Phys. Rev. Lett. {\bf 43} (1979) 1859



\bibitem{Tanihata85} I. Tanihata {\it et al.}, Phys. Rev. Lett. {\bf 55} (1985) 2676

\bibitem{Han87} P.G. Hansen and B. Jonson, Europhys. Lett. {\bf 4} (1987) 409

\bibitem{Jon04} B. Jonson, Phys. Rep. {\bf 389} (2004) 1

\bibitem{Humbert95} F. Humbert {\it et al.},  Physics Letters {\bf B347} (1995) 198

\bibitem{Aksyutina08} Y. Aksyutina {\it et al.}, Phys. Lett. {\bf B666} (2008) 430

\bibitem{Aumann99} T. Aumann {\it et al.}, Phys. Rev. {\bf C59} (1999) 1252

\bibitem{Aksyutina13} Y. Aksyutina {\it et al.}, Phys. Rev. {\bf C87} (2013) 064316

\bibitem{Wamers14} F. Wamers {\it et al.}, Phys. Rev. Lett. {\bf 112} (2014) 132502

\bibitem{Mukha07} I. Mukha {\it et al.}, Phys. Rev. Lett. {\bf 99} (2007) 182501

\bibitem{Bernas97} M. Bernas {\it et al.}, Phys. Lett. {\bf B415} (1997) 111

\bibitem{Hinke12} C.B. Hinke {\it et al.}, Nature {\bf 486} (2012) 341


\bibitem{Rossi13} D.M. Rossi {\it et al.}, Phys. Rev. Lett. {\bf 111} (2013) 242503

\bibitem{Bosch13} F. Bosch and Yuri A. Litvinov, Int. Journ. of Mass Spectr. {\bf 349-350} (2013) 151

\bibitem{Gutbrod06} 	H. H. Gutbrod {\it et al.}, 
	FAIR Baseline Technical Report (GSI) (2006)


\bibitem{Bacca02} S. Bacca {\it et al.}, Phys. Rev. Lett. {\bf 89} (2002) 052502

\bibitem{Atar18} L. Atar {\it et al.}, Phys. Rev. Lett. {\bf 120} (2018) 052501

\bibitem{Diaz18} P. Diaz Fern\'andez {\it et al.}, Phys. Rev. {\bf C97} (2018) 024311

\bibitem{Antonov2011} A.N. Antonov {\it et al.}, Nucl. Inst. Meth. {\bf A637} (2011) 60




\bibitem{McDonald2016}
	A.B. McDonald, Nobel Lecture, Rev. Mod. Phys. {\bf 88} (2016) 030502
    
\bibitem{Bahcall2001} J.N. Bahcall, M.H. Pinsonneault and S. Basu, Astr. J. {\bf 555} (2001) 990

\bibitem{Adelberger2011} E.G. Adelberger {\it et al.}, Rev. Mod. Phys. {\bf 83} (2011) 195

\bibitem{deBoer2017}
	R.J. deBoer, J. Goerres, M. Wiescher {\it et al.}, Rev. Mod. Phys. {\bf 89} (2017) 035007

\bibitem{Bethe1990}
H.A. Bethe, Rev. Mod. Phys. {\bf 62} (1990) 801


\bibitem{Janka2007}
	H.-T. Janka, K.~Langanke, A.~Marek,
  G.~Mart{\'i}nez-Pinedo, B.~M{\"u}ller, Phys. Repts. \textbf{442}
  (2007) 38.

  \bibitem{Janka2012}
	  H.-T. Janka,
  Annu. Rev. Nucl. Part. Sci. \textbf{62}, 407 (2012).



\bibitem{Just2015} O. Just, M. Obergaulinger and H.-Th. Janka, Monthly Notices of Royal Astronomical Society {\bf 453}
	(2015) 3386
	
\bibitem{Sukhbold2016} T. Sukhbold, T. Ertl, S.E. Woosley {\it et al.}, Astr. Journ. {\bf 821} (2016) 38

\bibitem{Langanke2003} K. Langanke {\it et al.}, Phys. Rev. Lett. {\bf 90} (2003) 241102

\bibitem{Langanke2000}
	K. Langanke and G. Mart\'{\i}nez-Pinedo,
  Nucl. Phys. {\bf A673} (2000) 481

\bibitem{Juodagalvis2010} A. Juodagalvis {\it et al} 2010, Nucl. Phys.
  A {\bf 848} 454


\bibitem{Cole2012}
	A.L. Cole {\it et al.}, Phys. Rev. C \textbf{86}
  (2012) 015809


\bibitem{Langanke2003a}
	K. Langanke {\it et al.}, Phys. Rev. Lett. {\bf 90} (2003) 241102


\bibitem{Hix2003}
R.W. Hix {it et al.}, Phys. Rev. Lett. {\bf 91} (2003) 210102




 \bibitem{Ablyazimov2017}
	 T. Ablyazimov, A. Abuhoza, R.P. Adak {\it et al.}, Eur. Phys. Journ. {\bf 53} (2017) 60

\bibitem{Seck2017} F. Seck, T. Galatyuk, R. Rapp and J. Stroth,
Acta Phys. Polon. Suppl. {\bf 10} (2017) 717

\bibitem{Adrich2005}
P. Adrich, A. Klimkiewicz, M. Fallot {\it et al.}, Phys. Rev. Lett. {\bf 95} (2005) 142501

 \bibitem{Brown2000}
S. Typel and B.A. Brown, Phys. Rev.
{\bf C64} (2001) 027302

 \bibitem{Froehlich2005}
	 C. Fr\"ohlich, G. Martinez-Pinedo {\it et al.},
  Phys. Rev. Lett. \textbf{96} (2006) 142502


 \bibitem{Pruet2005}
	 J. Pruet {\it et al.}, Astrophys. J. {\bf 623}
  (2005) 325


 \bibitem{Hebeler2013}
	 K. Hebeler, J.M. Lattimer, C.J. Pethick {\it et al.}, Astr. Journ. {\bf 773} (2013) 11
     
 \bibitem{Neuhauser1986} D. Neuhauser, K. Langanke and S.E. Koonin, Phys. Rev. {\bf A33} (1986) 2084
 
 \bibitem{Negreiros2018} R. Negreiros {\it et al.}, Universe {\bf 4} (2018) 43

 \bibitem{Cowan1991}
        J.J. Cowan, F.-K. Thielemann and J.W. Truran,
	Phys. Rep. {\bf 208} (1991) 267

 \bibitem{Lorusso2015}
       G. Lorusso {\it et al.}, Phys. Rev. Lett. {\bf 114} (2015) 192501

 \bibitem{Mendoza2015}
	 J. Mendoza-Temis, M.R. Wu, K. Langanke {\it et al.}, Phys. Rev. {\bf C92} (2015) 055805

 \bibitem{Metzger2010}
         B.D. Metzger, G. Martinez-Pinedo, S. Darbha {\it et al.}, Monthly Notices of Royal Astronomical Society {\bf 406} (2010) 2650
         
 \bibitem{Cowperthwaite2017} 
 P.S. Cowperthwaite, E. Berger, V.A. Villar {\it et al}, Astr. Journ. Lett. {\bf 848} (2017) L17 

 \bibitem{Zhi2013} Q. Zhi {\it et al.}, Phys. Rev. {\bf C87} (2013) 025803

 \bibitem{lar1947} W.E. Lamb, R.C. Retherford, Phys. Rev. {\bf 72} (1947) 241


 \bibitem{pana2010} R. Pohl, A. Antognini, F. Nez {\it et al.}, Nature {\bf 466} (2010) 213

\bibitem{ansa2013} A. Antognini, F. Nez, K. Schuhmann {\it et al.}. Science {\bf 339} (2013) 417

\bibitem{lorenz2012} I.~T. Lorenz, H.-W. Hammer, U.-G. Mei{\ss}ner, Eur. Phys. J. A{\bf 48}
(2012) 151

 \bibitem{pnfa2016} R. Pohl, F. Nez, L.M.P. Fernandes {\it et al.}, Science {\bf 353} (2016) 669

 \bibitem{bmmp2017} A. Beyer, L. Maisenbacher, A. Matveev {\it et al.}, Science {\bf 358} (2017) 79

\bibitem{fgtb2018}H. Fleurbaey, S. Galtier, S. Thomas, M. Bonnaud, L. Julien, F. Biraben, F. Nez, M. Abgrall, J. Guéna, 	arXiv:1801.08816 (2018)


 \bibitem{uabd2017} J. Ullmann, Z. Andelkovic, C. Brandau {\it et al.}, Nature Comm. {\bf 8} (2017) 15484

 \bibitem{gsbb2005} A. Gumberidze, T. St\"ohlker, D. Banas {\it et al.}, Phys. Rev. Lett. {\bf 94} (2005) 223001

 \bibitem{zap1972} Y.B. Zeldovich, V.S. Popov, Sov. Phys. Usp. {\bf 14} (1972) 673

 \bibitem{mrg1972} B. M\"uller, J. Rafelski and W. Greiner, Z. Phys. {\bf A257} (1972) 62

 \bibitem{dkmm2018} A.E. Dorokhov, N.I. Kochelev, A.P. Martynenko {\it et al.}, Phys. Lett. {\bf B776} (2018) 105

 \bibitem{sgor2006} V.M. Shabaev, D.A. Glazov, N.S. Oreshhkina {\it et al.}, Phys. Rev. Lett. {\bf 96} (2006) 253002

 \bibitem{ybht2016} V.A. Yerokhin, E. Berseneva, Z. Harman {\it et al.}, Phys. Rev. {\bf A94} (2016) 022502

 \bibitem{skzw2014} S. Sturm, F. Kohler, J. zatorski {\it et al.}, Nature {\bf 506} (2014) 467

 \bibitem{zsks2017} J. Zatorski, B. Sikora, S.G. Karshenboim {\it et al.}, Phys. Rev. {\bf A96} (2017) 012502	
	
 \bibitem{bbrs2012} S. Bernitt, G.V. Brown, J.K. Rudolp {\it et al.}, Nature {\bf 492} (2012) 225

 \bibitem{sdbs2016} C. Shah, S. Dobrodey, S. Bernitt {\it et al.}, Astr. Journ. {\bf 833} (2016) 52

 \bibitem{bmfs2014} E. Bulbul, M. Markevitch, A. Foster {\it et al.}, Astr. Journ. {\bf 789} (2014) 13

 \bibitem{brif2014} A. Boyarsky, O. Ruchayzkiy, D. Iakubovskyi, J. franse, Phys. Rev. Lett. {\bf 113} (2014) 251301

 \bibitem{mek1994} R. Marrs, S. Elliott and D. Knapp, Phys. rev. Lett. {\bf 72} (1994) 4082

 \bibitem{beos1996} P. Beiersdorfer, S.R. Eliott, A. Osterheld {\it et al.}, Phys. Rev. {\bf A53} (1996) 4000

 \bibitem{slbl2014} T. St\"ohlker, Y. Litvonov, A. Br\"auning-Demian {\it et al.}, Hyperfine Int. {\bf 227} (2014) 45

 \bibitem{gsl2015} A. Gumberidze, S. Th, Y.A. Litvinov {\it et al.}, Phys. Scr. {\bf 2015} (2015) 014076

 \bibitem{sbbb2015} T. St\"ohlker, V. Bagnoud, K. Blaum {\it et al.}, Nucl. Instr. Meth. {\bf B365} (2015) 680

 \bibitem{laab2016} M. Lestinsky, V. Andrianov, B. Aurand {\it et al.}, Eur. Phys. Journ. {\bf 225} (2016) 797

 \bibitem{hbek2006} F. Herfurth, K. Blaum, S. Eliseev {\it et al.}, Hyperfone Int. {\bf 173} (2006) 93

 \bibitem{habc2015} F. Herfurth, Z. Andelkovic, W. Barth {\it et al.}, Phys. Scr. {\bf 2015} (2015) 014605

 \bibitem{sch2006} S. Schiller, Phys. Rev. Lett. {\bf 98} (2007) 180801

 \bibitem{ddf2012} A. Derevianko, V.A. Dzuba, V.V. Flambaum, Phys. Rev. Lett. {\bf 109} (2012) 180801

 \bibitem{dssf2015} V.A. Dzuba, M.S. Safronova, U.I. Safronova, V.V. Flambaum, Phys. Rev. {\bf A92} (2015) 060502

 \bibitem{hbde2014} J. H\"allstr\"om, A. Bergman, S. Dedeoglu {\it et al.}, IEEE Trans. Instr. Meas. {\bf 63} (2014) 2264

 \bibitem{nshn2012} Y. Nakano, S. Suda, A. Hatakeyama {\it et al.}, Phys. Rev. {\bf A85} (2012) 020701

 \bibitem{bmlg1995} H. Beyer, G. Menzel, D. Liesen {\it et al.}, Z. Phys. {\bf D35} (1995) 169

 \bibitem{pshk2012} C. Pies, S. Sch\"afer, S. Heuser {\it et al.}, Jour, Low Temp. Phys. {\bf 167} (2012) 269

 \bibitem{kabe2017} S. Kraft-Bermuth, V. Andrianov, A. Bleile {\it et al.}, J. Phys. {\bf B50} (2017) 055603

 \bibitem{tkbi2009} M. Trassinelli, A. Kumar, H. Beyer {\it et al.}, Europhys. Lett. {\bf 87} (2009) 53001

 \bibitem{cbls2006} S. Chatterjee, H.F. Beyer, D. Liesen {\it et al.}, Nucl. Instr. Meth. {\bf B245} (2006) 67

 \bibitem{bgth2015} H.F. Beyer, T. Gassner, M. Trassinelli {\it et al.}, J. Phys. {\bf B48} (2015) 144010

 \bibitem{bkhm2008} C. Brandau, C. Kozhuharov, Z. Harman {\it et al.}, Phys. Rev. Lett. {\bf 100} (2008)

 \bibitem{bklm2012} C. Brandau, C. Kozhuharov, A.L. Yu {\it et al.}, Journ. Phys. Conf. Ser. {\bf 388} (2012) 062042

 \bibitem{abbi2015} A. Antognini, N. Berger, D. vom Bruch {\it et al.}, (2015) Paul Scherrer Institute Proposal R-16-01.1

 \bibitem{wbbd2015} D. Winters, T. Beck, G. Birkl {\it et al.}, Phys. Scr. {\bf 2015} (2015) 014408

 \bibitem{wksi2011} D.F.A. Winters, T. K\"uhl, D.H. Schneider {\it et al.}, Phys. Scr. {\bf T144} (2011) 014013

 \bibitem{ssim1989} A. Sch\"afer, G. Soff, P. Indelicato {\it et al.}, Phys. Rev. {\bf A40} (1989) 7362

 \bibitem{msgi1996} M. Maul, A. Sch\"afer, W. greiner, P. Indelicato, Phys. Rev. {\bf A53} (1996) 3915

 \bibitem{anh1985} R. Anholt, Rev. Mod. Phys. {\bf 57} (1985) 995

 \bibitem{sstj1985} R. Schuch, H. Schmidt-B\"ocking, I. Teserruya {\it et al.}, Z. Phys. {\bf A320} (1985) 185

 \bibitem{gdbg2010} A. Gumberidze, R. DuBois, F. Bosch {\it et al.}, (2010) GSI proposal E103	


\bibitem{Widmann:2005vz}
  E.~Widmann,
  Phys.\ Scripta {\bf 72} (2005) C51.
\bibitem{Baird:1997qd}
  S.~Baird {\it et al.},
  Nucl.\ Instrum.\ Meth.\ A {\bf 391} (1997) 210
   [Conf.\ Proc.\ C {\bf 970512} (1997) 979].
\bibitem{FLAIR:2005}
  FLAIR collaboration (2005), FLAIR -- a facility for low-energy
  antiproton and ion research, Technical Proposal available from
  \verb\http://flairatfair.eu/\.
\bibitem{vonHahn:2016wjl}
  R.~von Hahn {\it et al.},
  Rev.\ Sci.\ Instrum.\  {\bf 87} (2016),  063115
\bibitem{Widmann:2015lna}
  E.~Widmann,
  Phys.\ Scripta T {\bf 166} (2015) 014074
\bibitem{Stoehlker:2015}
  T. St{\"o}hlker, V. Bagnoud, K. Blaum, {\it et al.},
  Nucl. Instrum. Methods {\bf B 365} (2015) 680.
\bibitem{Katayama:2015wsm}
  T.~Katayama {\it et al.},
  Phys.\ Scripta T {\bf 166} (2015) 014073.
\bibitem{Oelert:2017mud}
  W.~Oelert,
  Acta Phys.\ Polon.\ B {\bf 48} (2017) 1895.
\bibitem{Ahmadi:2016fir}
  M.~Ahmadi {\it et al.} [ALPHA Collaboration],
  Nature {\bf 541} (2016) 506.
\bibitem{Ahmadi:2017gwe}
  M.~Ahmadi {\it et al.} [ALPHA Collaboration],
  Nature {\bf 548} (2017)  66.
\bibitem{Welsch:2008zzd}
  C.~P.~Welsch, K.~U.~Kuhnel, C.~D.~Schroter and J.~Ullrich,
  AIP Conf.\ Proc.\  {\bf 1037} (2008) 318.
\bibitem{Trzcinska:2001sy}
  A.~Trzcinska, J.~Jastrzebski, P.~Lubinski, F.~J.~Hartmann, R.~Schmidt, T.~von Egidy and B.~Klos,
  Phys.\ Rev.\ Lett.\  {\bf 87} (2001) 082501.%
\bibitem{Bugg:1973zz}
  W.~M.~Bugg, G.~T.~Condo, E.~L.~Hart, H.~O.~Cohn and R.~D.~McCulloch,
  Phys.\ Rev.\ Lett.\  {\bf 31} (1973) 475.
\bibitem{Wada:2004ee}
  M.~Wada and Y.~Yamazaki,
  Nucl.\ Instrum.\ Meth.\ B {\bf 214} (2004) 196.
\bibitem{Obertelli:2017}
  A.~Obertelli {\it et al.},
  CERN-SPSC-2017-045, Letter of Intent, CERN, (2017).
\bibitem{Kienle:2005gn}
  P.~Kienle,
  AIP Conf.\ Proc.\  {\bf 796} (2005) 441.
\bibitem{Kienle:2007zza}
  P.~Kienle,
  Int.\ J.\ Mod.\ Phys.\ E {\bf 16} (2007) 905.
\bibitem{Akaishi:2002bg}
  Y.~Akaishi and T.~Yamazaki,
  Phys.\ Rev.\ C {\bf 65} (2002) 044005.
\bibitem{Bendiscioli:2007zzb}
  G.~Bendiscioli, L.~Lavezzi, A.~Panzarasa, P.~Salvini and T.~Bressani,
  Nucl.\ Phys.\ A {\bf 797} (2007) 109.
\bibitem{Zmeskal:2009jep}
  J.~Zmeskal {\it et al.},
  Hyperfine Interact.\  {\bf 194} (2009),  249.
\bibitem{Sakuma:2012wva}
  F.~Sakuma {\it et al.},
  Hyperfine Interact.\  {\bf 213} (2012),  51.


\bibitem{SHARKOV2013} B. Sharkov {\it et al.},  Nucl. \ Instr. \ Methods A {\bf 733}  (2013) 238.

\bibitem{VARENTSOV2013} D. Varentsov {\it et al.}, Physica Med. {\bf 29}  (2013) 208.

\bibitem{HOFFMANN2002} D.H.H. Hoffmann {\it et al.},  \ Phys.  \ Plasmas . {\bf 9} (2002) 3651.

\bibitem{TAHIR2000} N.A. Tahir, Phys. Rev. E {\bf 63}  (2000) 016402.

\bibitem{TAUSCHWITZ2007} An. Tauschwitz  {\it et al.},   \ High \ Energy \ Phys. {\bf 3}  (2007) 371. 

\bibitem{DURANTE2011} M. Durante, F. Cucinotta, Rev. \ Mod. \ Phys. {\bf 83} (2011) 1245. 

\bibitem{DURANTE2014} M. Durante, H. Stöcker, Combined ion irradiation and ion radiography device, EU EP 2 602 003 (B1) (2014).

\bibitem{TAHIR2017} N. A. Tahir {\it et al.}, The Astrophysical Journal Supplement Series {\bf 232}, (2017) 1.




 \bibitem{NuPECC} 	NuPECC: Nuclear Physics for Medicine (European Science Foundation; 2014.

 \bibitem{Durante2016}	M. Durante, H. Paganetti, Reports Prog Phys. {\bf 79} (2016) 96702.

 \bibitem{Durante2011} 	M. Durante, F.A. Cucinotta, Rev Mod Phys. {\bf 83} (2011)

 \bibitem{LaTessa2016}	C. La Tessa, M. Sivertz, I.H. Chiang {\it et al.},  Life Sci Sp Res. {\bf 11} (2016) 18

 \bibitem{Durante2010}  M. Durante, G. Reitz, O. Angerer, Radiat Environ Biophys. {\bf 49} (2010) 295

 \bibitem{Stoehlker2015} T. St\"ohlker, V. Bagnoud, K. Blaum  {\it et al.}, Nucl Instruments Methods Phys Res {\bf B365} (2015) 680

 \bibitem{Tommasino2015}  F. Tommasino, E. Scifoni, M. Durante, Int J Part Ther. {\bf 2} (2015) 428

 \bibitem{Durante2013} 	M. Durante, N. Reppingen, K.D. Held. Trends Mol Med. {\bf 19} (2013) 565

 \bibitem{Lehmann2016}	 H.I. Lehmann, C. Graeff, P. Simoniello {\it et al.}, Sci Rep. {\bf 20} (2016) 38895

 \bibitem{ISEGG} 	ISECG: The Global Exploration Roadmap. Vol. NP-2013-06. (NASA, Washington) (2013) 50

 \bibitem{Chancellor2014} J. Chancellor, G. Scott, J. Sutton, Life {\bf 4} (2014) 491

 \bibitem{Durante2005} 	M. Durante, A. Kronenberg, Adv Sp Res. {\bf 35} (2005) 180

 \bibitem{Wilson1997} J.W. Wilson, J. Miller, A. Konradi, F.A. Cucinotta, Shielding Strategies for Human Space Exploration (1997)

 \bibitem{Paganetti2012} H. Paganetti, Phys Med Biol. {\bf 57} (2012)  R99

 \bibitem{Poludniowski2015} G. Poludniowski, N.M. Allinson, P.M. Evans, Br J Radiol. {\bf 88} (2015) 20150134

 \bibitem{Parodi2014} K. Parodi, Phys Medica. {\bf 30} (2014)  539


 \bibitem{Varentsov2013} D. Varentsov D, A. Bogdanov, V.S. Demidov {\it et al.}, Phys Medica {\bf 29} (2013) 208

 \bibitem{Prall2016} M. Prall, M. Durante, T. Berger {\it et al.}, Sci Rep. {\bf 6} (2016) 27651

 \bibitem{Prall2015} M. Prall, P.M. Lang, C. LaTessa {\it et al.}, J Phys Conf Ser. {\bf 599} (2015) 12041

 \bibitem{Durante2012} M. Durante, H. St\"ocker, J Phys Conf Ser. {\bf 373} (2012) 12016

 \bibitem{Chu1993} W.T. Chu, B.A. Ludewigt, T.R. Renner, Rev Sci Instrum. {\bf 64} (1993)  2055

 \bibitem{Parodi2015} K. Parodi, Med Phys. {\bf 42} (2015) 7153

 \bibitem{Llacer1984} J. Llacer, A.  Chatterjee, E.L. Alpen {\it et al.}, IEEE Trans Med Imaging. {\bf 3} (1984) 80

 \bibitem{Sakurai2014} H. Sakurai, F. Itoh, Y. Hirano {\it et al.}, Phys Med Biol. {\bf 59} (2014)  7031

 \bibitem{Hirano2016} Y. Hirano, H. Takuwa, E. Yoshida {\it et al.}, Phys Med Biol. {\bf 61} (2016) 1875

 \bibitem{Augusto2016} R.S. Augusto, T.M. Mendonca, F. Wenander {\it et al.}, Nucl Instruments Methods Phys Res {\bf B376} (2016) 374

 \bibitem{Sellner2011} S. Sellner, C.P. Welsch, M. Holzscheiter, Radiat Meas. {\bf 46} (2011) 1770

 \bibitem{Weinrich2011} U. Weinrich, Nucl Instruments Methods Phys Res Sect {\bf B269} (2011) 2879

 \bibitem{Bert2011} C. Bert, M. Durante, Phys Med Biol. {\bf 56} (2011) R113

 \bibitem{Mizushima2014} K. Mizushima, K. Katagiri, Y. Iwata {\it et al.},  Nucl Instruments Methods Phys Res Sect {\bf B331} (2014) 243

 \bibitem{Favaudon2014} V. Favaudon, L. Caplier, V. Monceau {\it et al.}, Sci Transl Med. {\bf 6} (2014)  245

 \bibitem{Montay2017} P. Montay-Gruel, K. Petersson, M. Jaccard {\it et al.}, Radiother Oncol. {\bf 124} (2017)  3

 \bibitem{Durante2017} M. Durante, E. Br\"auer-Krisch, M. Hill,  Br J Radiol. (November 27) (2017) 20170628
 
 
\end{thebibliography}
\end{document}